\documentclass{aa}
\usepackage{natbib}
\usepackage{graphicx}
\usepackage{txfonts}
\begin{document} 

   \title{GALACTICNUCLEUS: A high angular resolution $JHK_s$ imaging survey of the Galactic Centre}
   \titlerunning{GALACTICNUCLEUS}
   \authorrunning{Nogueras-Lara et al.}

   \subtitle{I. Methodology, performance, and near-infrared extinction
   towards the Galactic Centre\thanks{Extinction maps as well as their corresponding uncertainty maps (Figs. \ref{ext_RC}, \ref{ext_all} and \ref{g1g2}) are available in electronic form at the CDS via anonymous ftp to cdsarc.u-strasbg.fr (130.79.128.5) or via http://cdsweb.u-strasbg.fr/cgi-bin/qcat?J/A+A/}}

  \author{F. Nogueras-Lara
          \inst{1}
          \and
          A. T. Gallego-Calvente
          \inst{1}
          \and
         H. Dong
          \inst{1}
          \and
           E. Gallego-Cano
          \inst{1}
          \and
          J.~H.~V. Girard
          \inst{2}
          \and
          M. Hilker
          \inst{3}
          \and
          P. T. de Zeeuw
          \inst{3,4}
         \and
          A. Feldmeier-Krause
          \inst{5}
         \and
          S. Nishiyama
          \inst{6}
         \and
          F. Najarro 
          \inst{7}   
         \and
          N. Neumayer 
          \inst{8}             
          \and      
          R. Sch\"odel 
          \inst{1}
          }

   \institute{
    Instituto de Astrof\'isica de Andaluc\'ia (CSIC),
     Glorieta de la Astronom\'ia s/n, 18008 Granada, Spain
              \email{fnoguer@iaa.es}
         \and
      European Southern Observatory (ESO), Casilla 19001, Vitacura, Santiago, Chile
         \and
      European Southern Observatory (ESO), Karl-Schwarzschild-Straße 2, 85748 Garching, Germany
         \and
      Leiden Observatory, Leiden University,  Postbus 9513, 2300 RA Leiden, The Netherlands
         \and
      The University of Chicago, The Department of Astronomy and Astrophysics, 5640 S. Ellis Ave, Chicago, IL 60637, USA
         \and
      Miyagi University of Education, Aoba-ku, 980-0845 Sendai, Japan
      \and
       Departamento de Astrof\'isica, Centro de Astrobiolog\'ia (CSIC-INTA), Ctra. Torrej\'on a Ajalvir km 4, E-28850 Torrej\'on de Ardoz, Spain
       \and 
       Max-Planck Institute for Astronomy, K\"onigstuhl 17, 69117 Heidelberg, Germany
           }

   \date{Received September 26, 2017; accepted November 14, 2017}

 
  \abstract
   {The Galactic Centre is of fundamental astrophysical interest, but
     existing near-infrared surveys fall short covering it adequately,
   either in terms of angular resolution, multi-wavelength coverage,
   or both. Here we introduce the GALACTICNUCLEUS survey, a
   $JHK_{s}$ imaging survey of the centre of the Milky Way with a
   $0.2''$ angular resolution.}
   {The purpose of this paper is to present the observations of
     Field\,1 of our survey, centred approximately on SgrA*  with an approximate size of 7.95$'$
  $\times$ 3.43$'$. We describe the observational set-up and
 data reduction pipeline and discuss the quality of the data. Finally,
we present the analysis of the data.}
   {The data were acquired with the near-infrared camera HAWK-I (High Acuity Wide field K-band Imager) at the
     ESO VLT (Very Large Telescope). Short readout times in combination with the speckle
     holography algorithm allowed us to produce final images with a
     stable, Gaussian PSF (point spread function) of $0.2''$ FWHM (full width at half maximum).  Astrometric calibration is  achieved via the VVV (VISTA Variables in the Via Lactea) survey and photometric calibration is based
     on the SIRIUS/IRSF (Infrared Survey Facility telescope) survey. The quality of the data is assessed by
   comparison between observations of the same field with different
   detectors of HAWK-I and at different times. }
   {We reach $5\,\sigma$ detection limits of approximately $J=22$,
     $H=21$, and $K_{s}=20$. The photometric uncertainties are less than $0.05$ at $J\lesssim20$, $H\lesssim17,$ and $K_{s}\lesssim16$. We can
     distinguish five stellar populations in the colour-magnitude
     diagrams; three of them appear to belong to foreground spiral
     arms, and the other two correspond to high- and low-extinction star groups
     at the Galactic Centre. We use our data to analyse the
     near-infrared extinction
   curve and find some evidence for a possible difference between the extinction index between $J-H$ and $H-K_s$. However, we conclude that it can be described very well by a power law
   with an index of $\alpha_{JHK_{s}}=2.30\pm0.08$. We do not find
   any evidence that this index depends on the position along the
   line of sight, or on the absolute value of the extinction. We produce extinction maps that show the clumpiness of
   the ISM (interstellar medium) at the Galactic Centre. Finally, we estimate that the majority of the stars have solar or super-solar metallicity by comparing our extinction-corrected colour-magnitude diagrams with isochrones  with different metallicities and a synthetic stellar model with a constant star formation.}
   {}


   \maketitle
%

\section{Introduction}

The centre of the Milky Way is the only galaxy nucleus in which we can
actually resolve the nuclear star cluster (NSC) observationally and
examine its properties and dynamics down to milli-parsec scales. The
Galactic Centre (GC) is therefore of fundamental interest for
astrophysics and is a crucial laboratory for studying stellar nuclei and
their role in the context of galaxy evolution \citep[e.g.][]{Genzel:2010fk,Schodel:2014bn}. It is a prime target
for the major ground-based and space-borne observatories and will be
so for future facilities, such as ALMA (Atacama Large Millimeter Array), SKA (Square Kilometer Array), the JWST (James Webb Space Telescope), the TMT (Thirty Meter Telescope), or the
E-ELT (European Extremely Large Telescope).

Surprisingly, in spite of its importance, only 1\% of the
projected area of the GC has been explored with sufficient angular and
wavelength resolution to allow an in-depth study of its stellar
population. The strong stellar crowding and the extremely high
interstellar extinction toward the GC \citep[$A_V\gtrsim30$,
$A_{K_{s}}\gtrsim2.5$, e.g.
][]{Scoville:2003la,Nishiyama:2008qa,Fritz:2011fk,Schodel:2010fk} require
an angular resolution of at least $0.2''$ and observations in at least
three bands. Moreover, the well-explored regions, the central parsec
around the massive black hole Sagittarius\,A* (Sgr\,A*) and the Arches
and the Quintuplet clusters, are extraordinary and we do not know
whether they can be considered as representative for the entire
GC. Accurate data are key to understanding the evolution of the GC and
to infer which physical processes shape our Galaxy. The aim of
obtaining a far more global view of the GC's stellar population,
structure, and history, and the methods to achieve this lie at the
heart of the new survey GALACTICNUCLEUS that will provide photometric
data in $JHK_{s}$ at an angular resolution of $0.2''$ for an area of a
few 1000 square parsecs. In this paper we present the
observations and analysis of the first, and central, field of
GALACTICNUCLEUS. We present our methodology, discuss the precision and
accuracy reached, and show an in-detail determination of the near-infrared (NIR) extinction curve toward the GC. To reach the desired angular resolution, we used the speckle
holography technique \citep{Schodel:2013fk}. 

Studying the extinction curve in the NIR is one of our goals. It is
generally thought that it can be approximated by a power law
\citep[e.g.][]{Nishiyama:2008qa,Fritz:2011fk} of the form
\begin{equation}
A_\lambda \propto \lambda^{-\alpha} \hspace{0.5cm},
\end{equation}
where $A_\lambda$ is the extinction at a given wavelength ($\lambda$) and $\alpha$
is the power-law index. While early work found values of
$\alpha\approx1.7$ \citep[e.g.][]{Rieke:1985fq,Draine:1989eq},
recently a larger number of studies has appeared, with particular
focus on the GC, where interstellar extinction reaches very high
values ($A_{K_s}\gtrsim 2.5$) that suggest steeper values of
$\alpha>2.0$ \citep[e.g.][]{Nishiyama:2006tx,Stead:2009uq,Gosling:2009kl,Schodel:2010fk,Fritz:2011fk}. For
a more complete discussion of the NIR extinction curve and
corresponding references, we refer the interested reader to the recent
work by \citet{Fritz:2011fk}. One of the limits of previous work
on the GC was that it was limited either to small fields
\citep[e.g.][]{Schodel:2010fk,Fritz:2011fk} or, because of crowding
and saturation issues, to bright stars or to fields at large offsets
from Sgr\,A* \citep[e.g.][]{Nishiyama:2006tx,Gosling:2009kl}. The
high angular resolution of the data presented in this work allows us
to study the extinction curve with accurate photometry in the $J$,
$H$, and $K_{s}$ bands with large numbers of stars and, in
particular, with stars with well-defined intrinsic properties (red
clump stars, \citealt[e.g.][]{Girardi:2016fk}). An accurate determination of the NIR extinction curve is
indispensable for any effort to classify stars at the GC through
multi-band photometry.  

This work constitutes the first paper of a series that will describe, make public, and exploit the GALACTICNUCLEUS survey. In the
following sections, we describe our methodology and the data reduction
pipeline that we have developed. We test the photometry and check its
accuracy comparing different observations. Finally, we study the
extinction law towards the first field of the survey and show how we
can use the known extinction curve in combination with $JHK_{s}$
photometry for a rough classification of the observed stars.


\section{Observations and methodology}

\subsection{Observations}

The imaging data were obtained with the NIR camera
HAWK-I (High Acuity Wide field K-band Imager, \citealt{Kissler-Patig:2008fr})  located at the ESO VLT (Very Large Telescope) unit
telescope\,4, using the broadband filters $J$, $H$, and $K_s$. HAWK-I has a
field of view (FOV) of 7$'$.5 $\times$ 7$'$.5 with a cross-shaped gap of
$15''$ between its four Hawaii-2RG detectors. The pixel scale is $0.106''$
per pixel. In order to be able to apply the speckle holography
algorithm described in \citet{Schodel:2013fk} to reach an angular
resolution of $0.2'' $ FWHM (full width at half maximum), we used the fast-photometry mode to take
a lot of series of short exposures. The necessary short readout times
required windowing of the detector. Here we present data of the
central field of our survey from 2013 and 2015 (D13 and D15, hereafter). The first epoch corresponds to a
pilot study. D15 data form part of the GALACTICNUCLEUS
survey that we are carrying out within the framework of an ESO Large
Programme.\footnote{Based on observations made with ESO telescopes at
  the La Silla Paranal Observatory under programme ID
  195.B-0283.} Table \ref{obs} summarises the relevant information of
the data.

\begin{itemize}

\item \textbf{D13 data}:\\

  The DIT (detector integration time) was set to $0.851$\,s, which
  restricted us to the use of the upper quarter of the lower two
  detectors and the lower quarters of the upper detectors.  The
  FOV of each of the four detectors was thus
  $2048\times512$ pixels.  We designed a four offset pointing pattern
  to cover the gap between the detectors. For each pointing we took
  four series of 480 exposures each.
  The observed region was centred on Sgr A* (17$^h$ 45$^m$ 40.05$^s$, -29$^\circ$ 00$'$ 27.9$''$) with a size of 8.2$'$ $\times$ 2.8$'$.\\

\item \textbf{D15 data}:\\

  In the 2015 observations, we used a random offset pattern with a
  jitter box width of 30$''$ to cover the detector gaps. We chose a
  longer DIT of $1.26$\,s, which allowed us to use larger detector
  windows ($3/8$ of each detector) and thus increase the efficiency of
  our observations. The FOV of each detector was thus $2048\times768$
  pixels.  We took 20 exposures each at 49 random offsets. The observed
  region was also centred on SgrA* (17$^h$ 45$^m$ 40.05$^s$,
  -29$^\circ$ 00$'$ 27.9$''$) with an approximate size of 7.95$'$
  $\times$ 3.43$'$.

\end{itemize}

\begin{table}
\begin{center}
\caption{Details of the imaging observations used in this
  work.}
\label{obs} 
\begin{tabular}{cccccc}
 &  &  &  &  & \tabularnewline
\hline 
\hline 
Date & HAWK-I & Seeing$^a$  & N$^b$ & NDIT$^c$ & DIT$^d$\tabularnewline
 & band   & (arcsec) &  &  & (s)\tabularnewline
\hline 
07 June 2013 & $J$ & 0.46 & 16 & 32 & 0.85\tabularnewline
07 June 2013 & $H$ & 0.42 & 16 & 32 & 0.85\tabularnewline
07 June 2013 & $K_s$ & 0.41 & 16 & 32 & 0.85\tabularnewline
08 June 2015 & $J$ & 0.37 & 49 & 20 & 1.26\tabularnewline
06 June 2015 & $H$ & 0.52 & 49 & 20 & 1.26\tabularnewline
06 June 2015 & $K_s$ & 0.57 & 49 & 20 & 1.26\tabularnewline
\hline 
\hline 
 &  &  &  &  & \tabularnewline

\end{tabular}
\end{center}
\footnotesize
\textbf{Notes.}
$^a$In-band seeing estimated from the PSF FWHM
measured in long exposure images. 
$^b$Number of pointings. 
$^c$Number of exposures per pointing. 
$^d$Integration time for each exposure. The total integration time of each observation is given by N$\times$NDIT$\times$DIT. 
 \end{table}

 HAWK-I was rotated to align the rectangular FOV with the Galactic
 Plane \citep[assuming an angle of $31.40^{\circ}$ east of north in
 J2000.0 coordinates,][]{Reid:2004ph}.  Each science observation was
 preceded or followed by randomly dithered observations, with the same filter, of a field centred on a dark cloud in the GC, located approximately at 17$^h$
 48$^m$ 01.55$^s$, -28$^\circ$ 59$'$ 20$''$, where the stellar density
 is very low. These observations were used to determine the sky
 background. The FOV was rotated by $70^{\circ}$ east of north to
 align it with the extension of the dark cloud.

\subsection{Data reduction}

 As HAWK-I has four independent detectors, all data reduction steps
 were applied independently to each of them. We followed a standard
 procedure (bad-pixel correction, flat fielding, and sky
 subtraction), paying special attention to the sky subtraction. Due to the
 extremely high stellar density in the GC, it was impossible to estimate the
 sky background from dithered observations of the target
 themselves. The dark cloud that we observed provided us with good
 estimates of the sky, but at intervals of about once per hour, which
 is far longer than the typical variability of the NIR sky (on the
 order of a few minutes). To optimise sky subtraction, we therefore
 scaled the sky image from the dark cloud to the level of the sky
 background of  each exposure. The latter was estimated from the
 median value of the 10\% of pixels with the lowest value in each
 exposure. A dark exposure was subtracted from both the sky image and
 from each reduced science frame before determining this scaling
 factor. A comparison of noise maps obtained with this strategy, or not, showed that this approach reduced the noise by a factor of about 10. 

\subsection{Image alignment}

For the images of D15, it was necessary to correct the dithering, which we did in a two-step procedure (it was not necessary for D13 images because the four offset pattern that we designed let us reduce every pointing independently). Firstly, we used the image headers to obtain the telescope offsets with respect to the initial pointing and shifted the images accordingly. Subsequently, we fine-aligned the frames by using a cross-correlation procedure on the long-exposures (merged image of all the corrected dithered frames) for each pointing. 

\subsection{Distortion solution}

Geometric distortion is significant in HAWK-I. When comparing the long-exposure image with the VVV (VISTA Variables in the Via Lactea) corresponding images, the position of a given star can deviate by as much as 1$''$ or about ten HAWK-I pixels between them. To correct that effect we used the VVV survey \citep{Minniti:2010fk,Saito:2012fk} as astrometric reference. We cross-identified  stars in both the VVV and a long exposure image for a given HAWK-I pointing (using as criterion a maximum distance of 0.1$''$). We iteratively matched the stellar positions by first using a polynomial of degree one and, subsequently, of degree two. Due to serious saturation problem in the VVV images at the $H$ and $K_s$ bands, as well as the lower stellar density in the VVV $J$ images, we used the $J$-band image of tile b333 to perform the distortion solution in all three bands. We found that the common stars were homogeneously distributed over the detectors, so that no region had an excessive influence over the derived distortion solution.

\begin{figure}
   \includegraphics[scale=0.35]{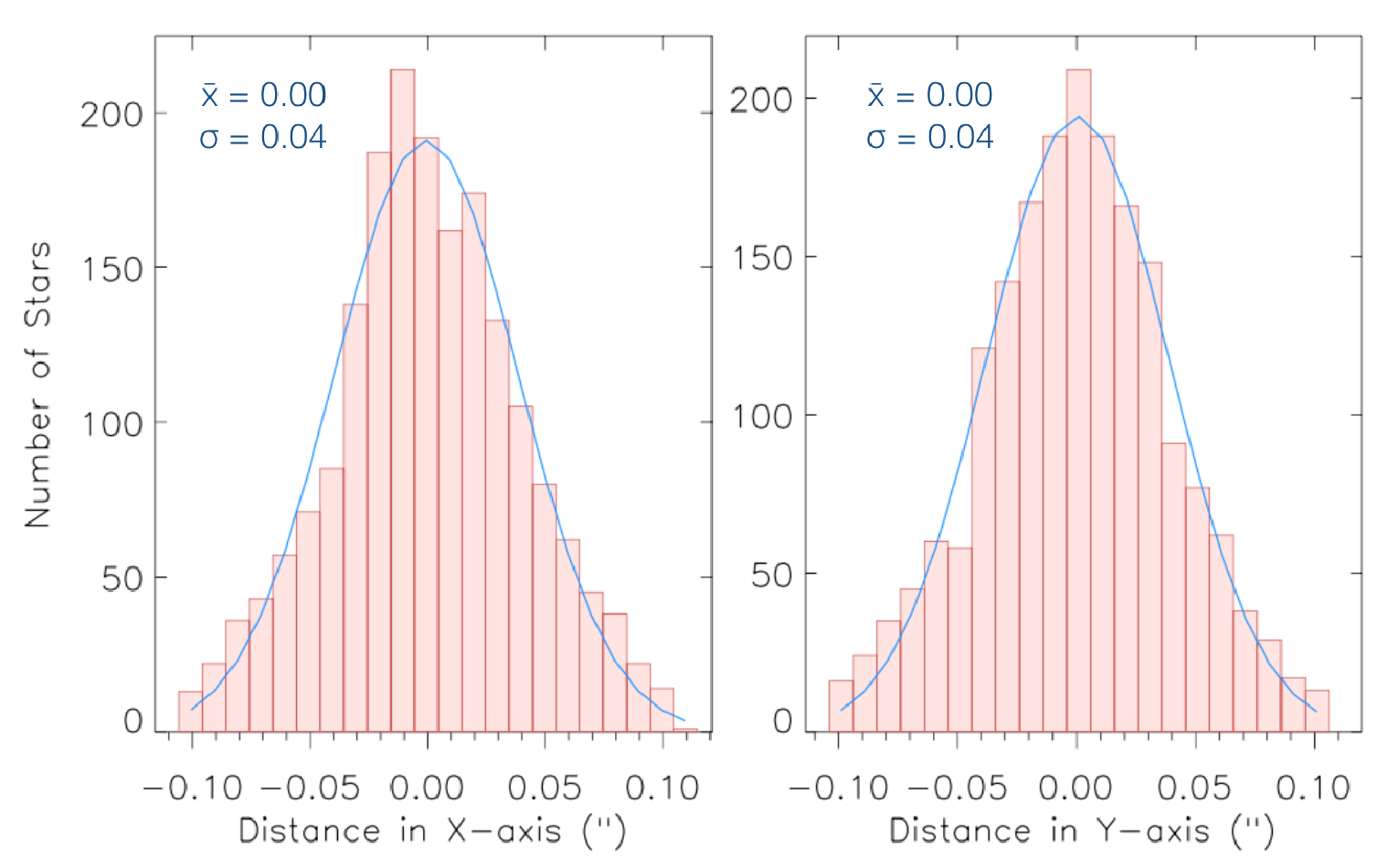}
   \caption{Goodness of the distortion solution (detector $\#$2 $H$-band D15 data). Differences in arcseconds between the relative positions of stars in the corrected HAWK-I frames and in the VVV reference image. Left panel X-axis and right panel Y-axis. The blue lines are Gaussian fits to the histograms. For both histograms, the centre of the fit lies at $0.0$ arcseconds, with  $\sigma=0.04$ arcseconds.}
   \label{positionVVV}
    \end{figure}

\begin{figure}
   \includegraphics[scale=0.47]{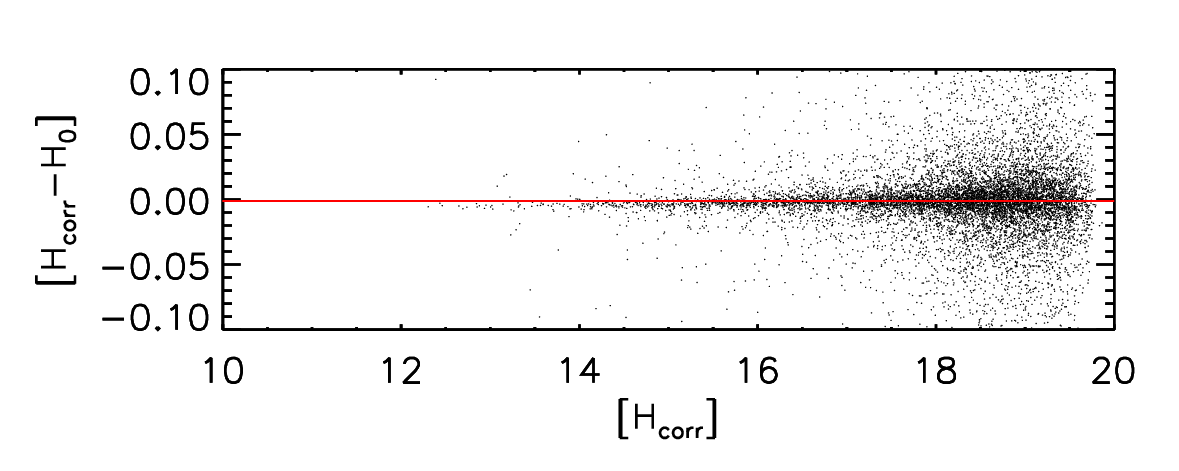}
   \caption{Difference in $H$-band photometry before and after applying the distortion solution. The results of the photometry computed before and after the application of the distortion solution are indicated by $H_0$ and $H_{corr}$, respectively. This plot corresponds to detector $\#$3 of D15 data. The units are magnitudes, with the zero point being the one specified in the HAWK-I user manual.}
   \label{photoVVV}
    \end{figure}

To check the quality of the distortion solution, we compared the relative  positions of the stars found in the corrected HAWK-I long exposure image with their positions in the VVV reference image. As can be seen in Fig. \ref{positionVVV}, the correction is quite satisfactory ($\sigma<0.05 ''$). Besides, we checked whether the application of the distortion solution had any systematic effect on the photometry. Figure \ref{photoVVV} shows a comparison between photometric measurements of stars on chip $\#$3 with and without distortion correction. As can be seen, applying the distortion solution has no significant effect on the photometry. The distortion solution was computed for each band and chip independently and then applied to each individual frame with a cubic interpolation method.

\subsection{Speckle holography}

\begin{figure}
   \includegraphics[scale=0.4]{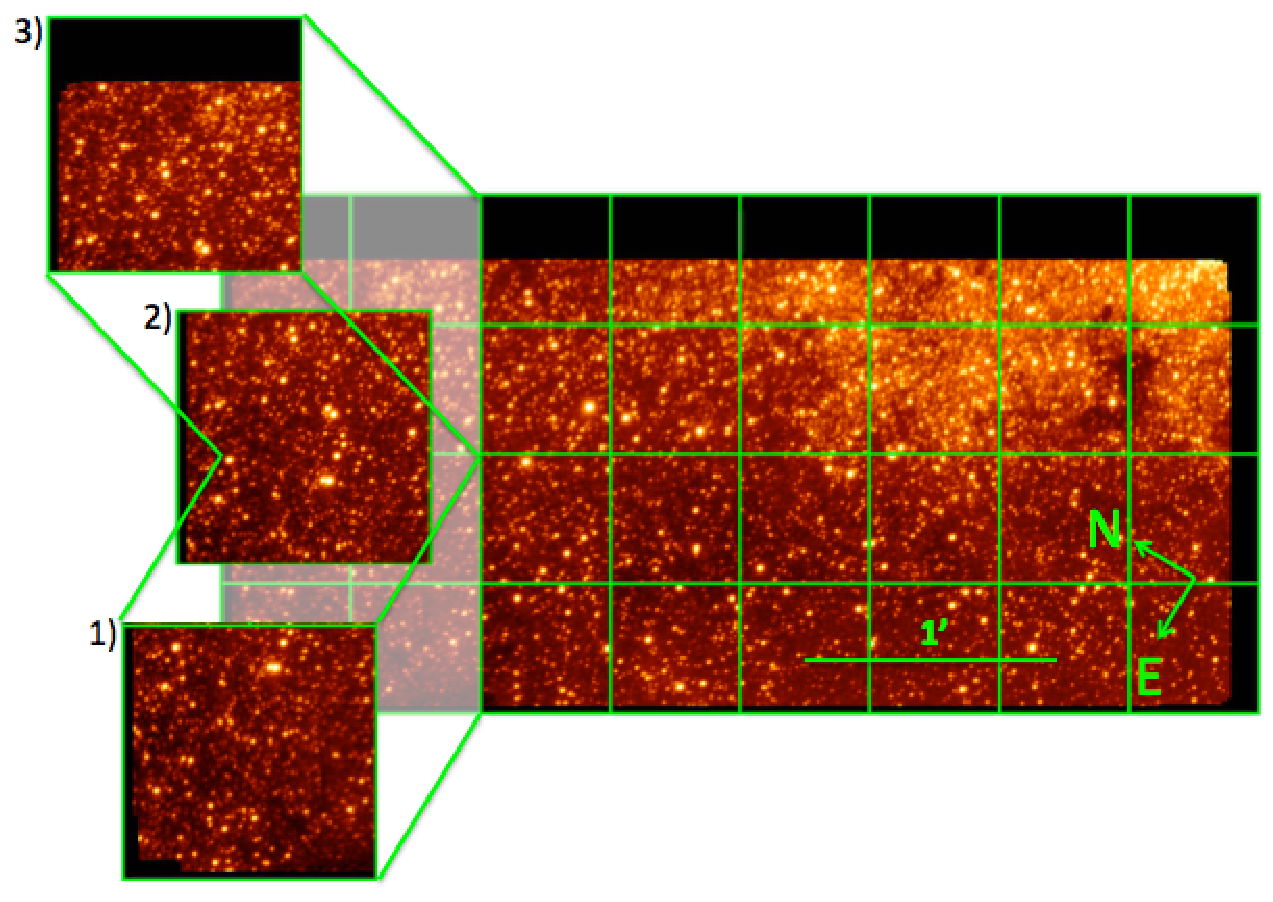}
   \caption{Long exposure image ($H$-band chip1, D15 data) divided to obtain the sub-regions used in the holographic procedure. Region 2) corresponds to an overlapping region between 1) and 3) showing the overlapping strategy described in the text. The green lines show the division of the rest of the detector.}
   \label{subcubes}
    \end{figure}

To produce high angular resolution images from the short exposure images, we used the speckle holography algorithm \citep[see e.g.][]{Primot:1990fk,Petr:1998vn}. This requires the knowledge of the point spread function (PSF) for each short exposure. For the latter purpose, we applied the methodology described by \citet{Schodel:2013fk}, which works very well in crowded fields such as the GC. In brief, this methodology consists in the superposition of multiple reference stars and iterative improvement to determine the PSF. The image resulting from the speckle holography algorithm was convolved with a Gaussian beam of 0.2$''$ FWHM in order to suppress noise at high spatial frequencies. In this way we overcome the image blurring imposed by
seeing. As the PSF not only varies with time but is also a
function of position, mainly due to anisoplanatic
effects, we divided the aligned frames into regions of 1 arcmin
$\times$ 1 arcmin (from now on referred to as
\textit{sub-regions}). Figure \ref{subcubes} shows the grid of the
sub-regions drawn on a long exposure image. Overlap between the
sub-regions corresponds to one half of their width. Speckle holography
was applied to every single sub-region independently.

    The PSF for each sub-region and exposure was extracted in an
    automatic way. First, we generated long exposures and
    corresponding noise maps from all the frames corresponding to a
    given pointing and filter. Then, we used the {\it StarFinder} software
    package \citep{Diolaiti:2000qo} for PSF fitting astrometry and
    photometry. From the list of detected stars we selected reference
    stars for PSF extraction in each exposure according to the
    following criteria:
\begin{itemize}
\item  Reference stars had to be fainter than $J\approx12$; $H\approx12$,  $K_{s}\approx11$ to avoid saturated stars. To compute these limits we performed PSF fitting photometry on the long exposure images and computed a preliminar ZP (zero point) using the SIRIUS/IRTF (Infrared Survey Facility telescope) GC survey \citep[e.g.][]{Nagayama:2003fk,Nishiyama:2006tx}. We obtained a systematic deviation for the brightest stars when we compared the ZP versus the magnitude in the SIRIUS/IRTF GC survey. Thus, the starting point of the deviation gave us the limit for the saturation.

\item Reference stars had to be brighter than $J=18$; $H=14$,  $K_{s}=13$. 
\item For a given exposure, the full PSF of a reference star needed to
  be visible, meaning\ stars close to the image edges were excluded.
\item Reference stars had to be isolated. Any neighbouring star within a distance corresponding to two times the FWHM of the seeing PSF had to be at least $2.5$ magnitudes fainter. No star brighter than the reference one was allowed within a distance corresponding approximately to the full radial extent of the PSF.
\end{itemize}

\begin{figure}
      \begin{center}
   \includegraphics[scale=0.4]{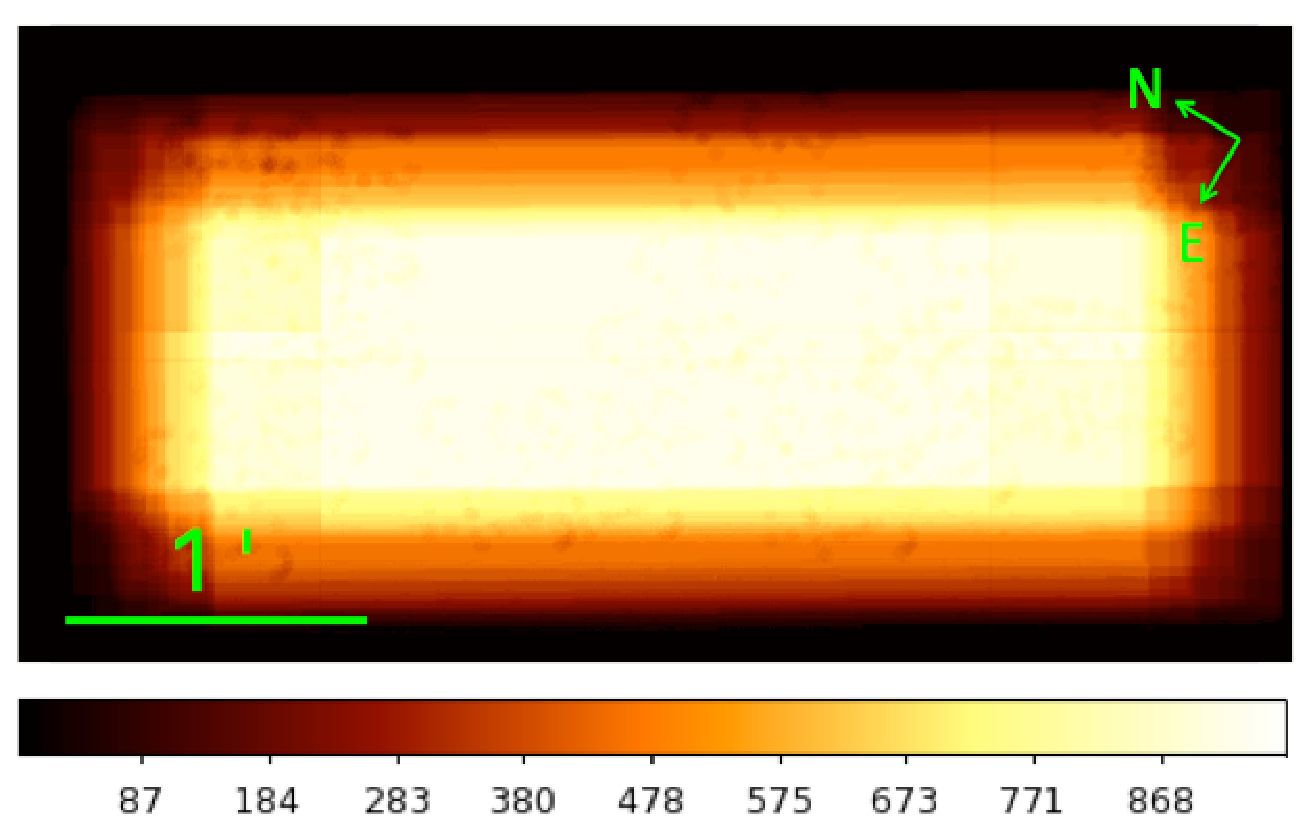}
   \caption{Example of an exposure map for a final holographic product (detector $\#$1 $H$-band, D15 data). The scale depicts the number of valid frames for each pixel.}
   \label{exp}
      \end{center} 
   \end{figure}

After having determined the PSF for each frame, we applied the speckle
holography algorithm. Once the process was finished for each
sub-region, we created the final image by combining the
holographically reduced sub-regions. To do that, we performed PSF
fitting astrometry and photometry (with {\it
  StarFinder}) for each
sub-region in each band and used the positions of the detected stars
to align all three bands with each other, taking as reference the
$H$-band image. This step was important to correct small relative
shifts between the sub-regions that may arise from the holography
algorithm. We also produced an exposure map that informs us about the
number of frames contributing to each pixel in the final image (see
Fig. \ref{exp}), and a noise map computed using for each pixel the error of the mean (the standard deviation of the mean divided by $\sqrt{N-1}$, where $N$ is the number of the measurements) of the frames that contribute to it. We produced a deep image from all the data and three
so-called \emph{sub-images} from three disjunct sub-sets of the data,
with each one containing $1/3$ of the frames, as well as the corresponding noise maps. The sub-images were used
to determine photometric and astrometric uncertainties of the detected
stars as described in the next section.

\subsection{Rebinning}
\label{rebinning}
Since we aim to obtain final images with 0.2$''$ angular resolution,
the sampling (0.106$''$ per pixel) is barely sufficient. The quality
of the reconstructed images can be improved by rebinning each input
frame by a factor > 1 (using cubic interpolation). The PSF fitting
algorithm then can fit and disentangle the stars in this crowded field
with higher accuracy. To test the optimum value of this factor and its
usefulness, we simulated several hundreds of images and then applied
exactly the same procedure that we followed when we reduced and
analysed our science images. To do so, we selected the most difficult
region that we studied in our data, namely, a square of 27$''$
centred on SgrA*, where the source density is the highest in the entire Galactic
Centre. We used a list of stars extracted from diffraction-limited
$K_{s}$-band observations of the GC with NACO (NAOS-CONICA) instrument installed at the ESO VLT (S27, camera, date 9
Sept 2012), consisting of 9840 stars with
magnitudes of $9\lesssim Ks\lesssim19$.  This allowed us to test also
the reliability of our procedure under the worst crowding conditions
possible.

To generate the images, we used cubes of real PSFs from HAWK-I and we
added readout and photon noise for stars and sky. Speckle holography
was applied in those images using rebinning factors of 1, 2, and
3. Then, we applied the procedure described in Section
\ref{photometry} to obtain the photometry and the uncertainties of the
stars. The obtained results are shown in Table \ref{rebinning}.

\begin{table}
\begin{center}
\caption{Results of simulations with different rebinning factors.}
\label{rebinning} 
\begin{tabular}{cccc}
 &  &   & \tabularnewline
\hline 
\hline 
Rebinning & Detections$^a$& Spurious$^b$  & Success$^c$\tabularnewline
 factor & &  & (\%) \tabularnewline
  &  &    &   \tabularnewline
\hline 
1 &  1571 & 3 & 76.4\tabularnewline
2 &  2041 & 11 & 81.6\tabularnewline
3 & 1918 & 10 & 77.4\tabularnewline
\hline 
  &  &  & \tabularnewline
\end{tabular}
\end{center}
\footnotesize
\textbf{Notes.} $^a$Detections that have a counterpart in the data used to simulate the images once we have removed detections with uncertainties above 10 $\%$. $^b$Stars without a counterpart after removal of detections with photometric uncertainties $>0.1$\,mag. $^c$Rate of valid identifications with $K_{s}\leq15$. 
 \end{table}

 As we can see, without rebinning the number of real detected sources was
 the lowest value obtained, whereas rebinning increased the number of
 detected sources significantly. To compare with the input data we
 discarded the outliers, removing all the stars with a photometric
 uncertainty $>0.1$\,mag. A rebinning factor of 2 turned out to be a
 reasonable choice. Higher rebinning factors do not improve the final
 product significantly and may lead to additional uncertainties from
 interpolation. Also, computing time increases quadratically with the
 image size. The completeness until magnitude 15 for a rebinning factor of 2 is
 above $81\%$. We note that the completeness of our actual data will be
 almost $100\%$ at $K_{s}\approx15$ in less crowded regions outside the central parsec.

\begin{figure}
   \includegraphics[scale=0.45]{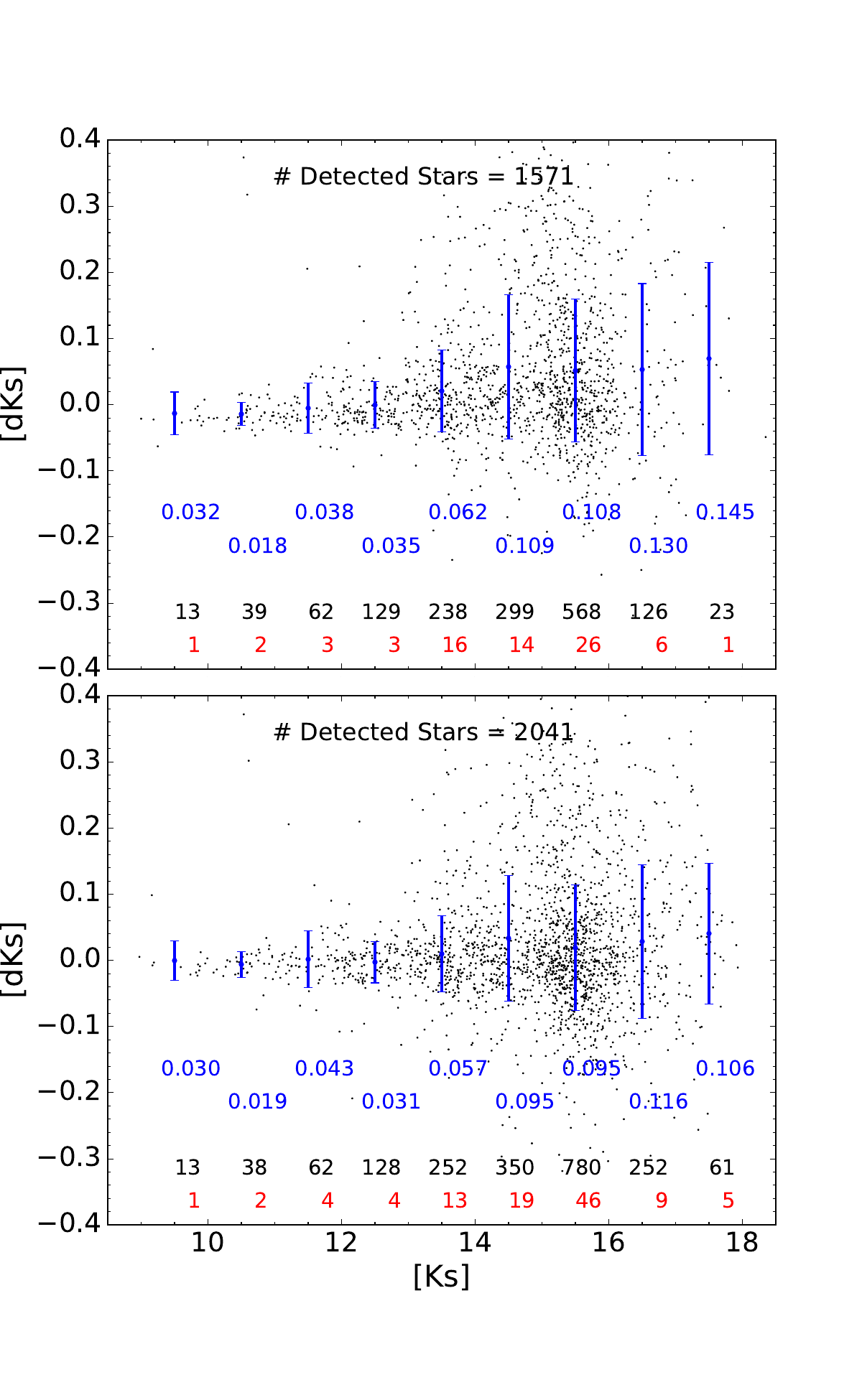}
   \caption{Photometric precision for simulated data with different
     rebinning factors. The upper and lower panels show  results for
     rebinning factors of 1 and 2 respectively. Blue error bars depict
     the standard deviation of the points in bins of one magnitude.  The
     first rows of numbers are the standard deviation in those
     bins. The second and third rows are the number of points used to
     compute the standard deviation and the number of rejected outliers.}
   \label{photo_sim}
    \end{figure}

With respect to the photometric accuracy obtained, we compared the
stellar fluxes measured in our simulations with their known input
fluxes. Then, we defined a mean flux, $f_{tot}$ , and its associated uncertainty, $df_{tot}$:

\begin{equation}
\label{compar}
\begin{split}
f_{tot} = \frac{f_{HAWK-I}+f_{input}}{2} ,
\\
df_{tot} = \frac{f_{HAWK-I}-f_{input}}{2} ,
\end{split}
\end{equation}

\noindent where $f_{HAWK-I}$ corresponds to the flux measured in the simulated
image and $f_{input}$ refers to the one in the input list. Once we
computed those fluxes, we expressed them in magnitudes, obtaining 
Fig. \ref{photo_sim}. As can be seen, the photometry is slightly more
accurate in the rebinned image, in particular at faint magnitudes.

\section{Photometry and astrometry}
\label{photometry}

Stellar fluxes in the final images were measured by means of PSF
fitting photometry with {\it StarFinder}. We used the noise map produced
previously for the deep image, which facilitates the
detection of stars and suppresses the detection of spurious sources.
Since the formal uncertainties given by the {\it StarFinder} package
tend to under-estimate the real uncertainties significantly (Emiliano Diolaiti, private communication), we
determined the error computing the photometry on the three independent sub-images.

We developed an automatic routine for PSF extraction that chooses the
reference stars taking into account isolation, saturation, brightness
limits (depending on the band), and weight of the stars (the number of
frames contributing to the final image at a star's position). We also
excluded stars near the image edges. We used the following {\it
  StarFinder}
parameters: A minimum correlation value of $min\_corr = 0.8$, no
diffuse background estimation, that is\ $ESTIM\_BG=0$, and a detection
threshold of $5\,\sigma$ with two iterations.

\subsection{Photometric uncertainties}

We took into account two different effects for the uncertainty: statistical uncertainties and the PSF variation across the detectors.

\subsubsection{Statistical uncertainties}

A star was accepted only if it was detected in all three sub-images
and in the deep image, using as a criterion a maximum distance of
one\,pixel between its relative positions (corresponding to about
$0.05''$ because of the rebinning factor used, or about one quarter of
the angular resolution). This is a conservative strategy as the deep
image has a higher signal to noise ratio than the sub-images. In this
way we can be certain that hardly any spurious detections will be
contained in the final lists. We used the flux of a star as measured
on the deep image and estimated the corresponding uncertainty, $\Delta f$, from the
measurements on the three sub-images according to the formula
\begin{equation}
\Delta f = \frac{f_{max}-f_{min}}{2 \sqrt{N}} ,
\end{equation}
\noindent where $f_{max}$ and $f_{min}$ correspond to the maximum and minimum flux obtained for each star in the measurements on the sub-images and $N$ is equal to 3, the number of measurements.

We compared the formal errors provided by {\it StarFinder} on the deep
image with the ones obtained with the procedure described above.  We confirm that {\it
  StarFinder} generally under-estimates the uncertainties
systematically. However, for some stars (mainly faint ones), the
uncertainty given by {\it StarFinder} can be larger than the one
obtained by the previous procedure. In those cases, we took the larger
value, to be conservative.

\subsubsection{PSF uncertainties}
\label{uncertainties}

\begin{figure}
      \begin{center}
   \includegraphics[scale=0.35]{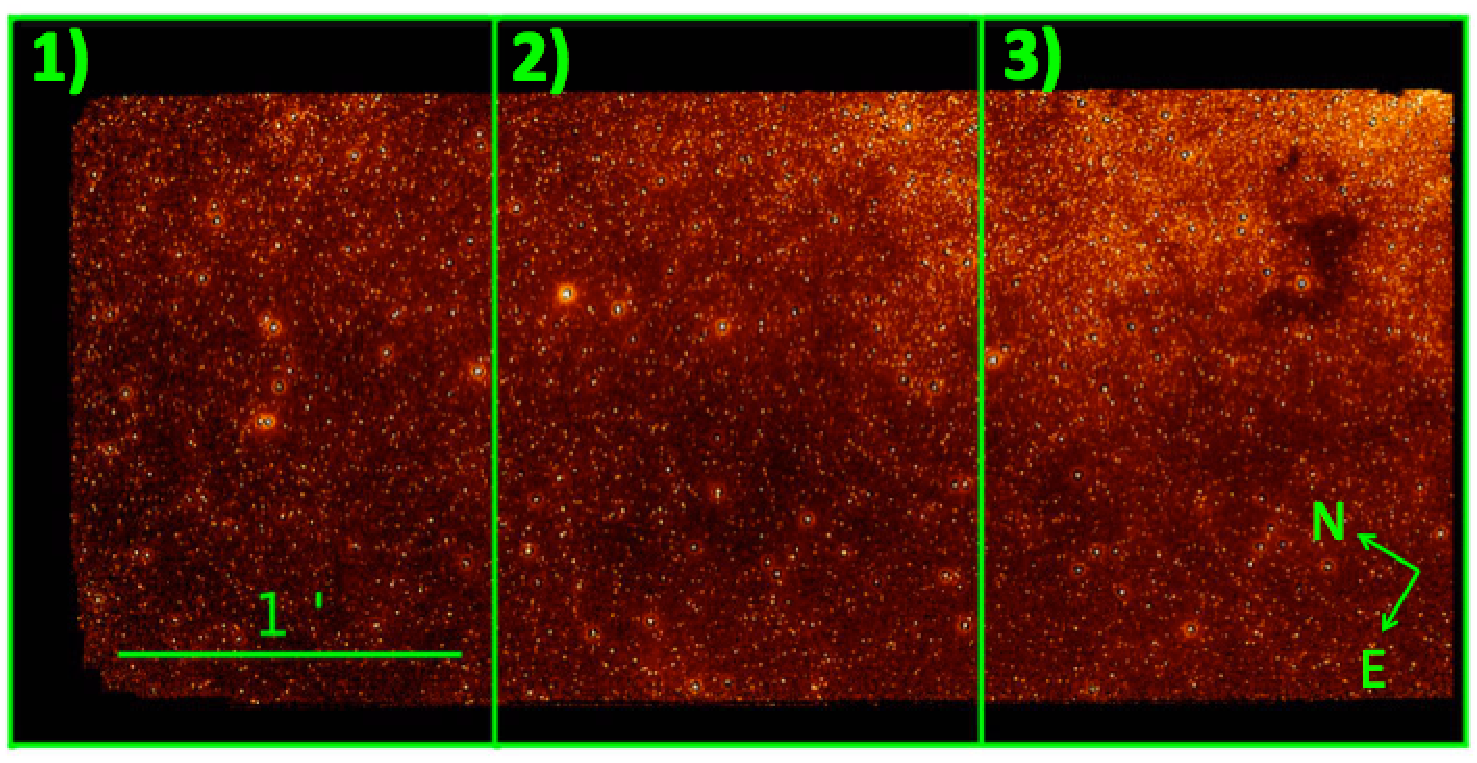}
   \caption{Image division of chip $\#$1 $H$-band (D15) to obtain three PSFs to quantify the PSF variation across the image. Each of the numbers indicates the region considered to extract the three PSFs.}
   \label{psf_division}
       \end{center}
    \end{figure}

The PSF can potentially vary across the field. To quantify this effect, we divided the
deep image horizontally into three equal regions as shown in
Fig. \ref{psf_division}. Subsequently, we obtained a PSF for each
region as described above. From a comparison between the PSFs
(essentially fitting the PSF from one sub-region with that from
another), we estimated the corresponding photometric uncertainty
and added it quadratically to the statistical uncertainties. The effect
of PSF variability is only of the order of $\lesssim$2\% and depends on the
observing conditions. 

This small effect of PSF variability highlights the excellent
performance of the speckle holography algorithm. It also justifies our
choice of relatively large sub-regions ($1'\times1'$ for the speckle
holographic reconstruction). The regions are considerably larger than
the size of the isoplanatic angle in the near-infrared, which is,
depending on the filter used, of the order of $10''-20''$. A possible
explanation why we can use such large regions is that we do not
reconstruct images at the diffraction limit of the telescope, but
rather at a less stringent $0.2''$ FWHM. We suspect that we are
therefore working in a "seeing-enhancer" regime, similar to
ground-layer adaptive optics (GLAO) systems. GLAO corrects image
degradation by turbulence in a layer close to the ground and leads
therefore to moderate corrections, but over large fields \citep[see
e.g. the description of HAWK-I's future GLAO system GRAAL (GRound layer Adaptive optics Assisted by Lasers) in][and
references therein]{Paufique:2010zl,Arsenault:2014fv}. We obtained the uncertainty for each individual star by adding quadratically the statistical and the PSF uncertainties. A plot of the
final, combined statistical and PSF uncertainties for chip \#1
is shown in Fig.\,\ref{uncertainties}.

   \begin{figure}
      \begin{center}
   \includegraphics[scale=0.7]{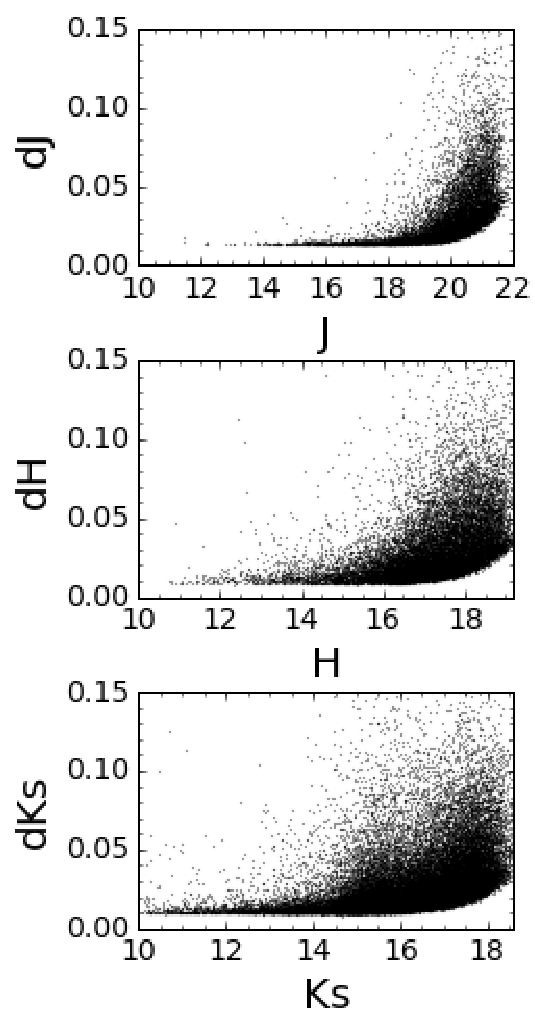}
   \caption{Combined statistical and PSF photometric uncertainties versus magnitude for $J$ (top), $H$
     (middle), and $K_s$ (bottom) for chip $\#1$ (D15 data).}
   \label{uncertainties}
       \end{center}
    \end{figure}

\subsection{Astrometric calibration}

We calibrated the astrometry by using VVV catalogue stars as reference
that we cross-identified with the stars detected in the images.  The
astrometric solution was computed with the IDL (Interactive Data Language) routine \textsc{solve\_astro}  \citep[see IDL Astronomy User's Library,][]{Landsman:1993aa}.

\begin{figure}
   \includegraphics[scale=0.47]{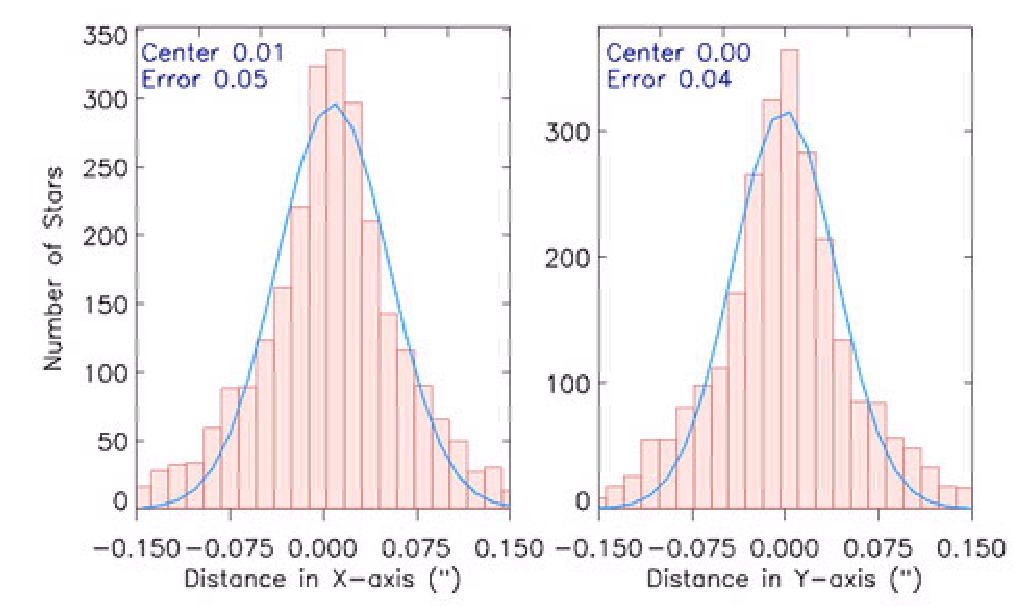}
   \caption{Position accuracy. Differences in arcseconds between the
     positions of stars in the final HAWK-I chip \#4 (D15 $H$-band) and
     in the VVV reference image. Left panel X-axis and right panel
     Y-axis. The mean and standard deviation are $0.01\pm0.05''$ in X
     and $0.00\pm0.04''$ in Y.}
   \label{pos_uncertainty}
    \end{figure}

To estimate the uncertainty of the astrometric solution, we compared
all stars common to our image and to the VVV
survey. Figure\,\ref{pos_uncertainty} shows the histograms of the
differences in right ascension and declination. For all bands and
chips the standard deviation of this distribution is
$\lesssim0.05$\,arcseconds.

\subsection{Zero point calibration}

Since the VVV catalogue uses aperture photometry, which will lead to
large uncertainties in the extremely crowded GC field, the zero point
calibration was carried out relying on the catalogue from the
SIRIUS/IRTF GC survey \citep[e.g.][]{Nagayama:2003fk,Nishiyama:2006tx},
which uses PSF fitting photometry. In that catalogue, the zero point
was computed with an uncertainty of 0.03\,mag in each band
\citep{Nishiyama:2006ai,Nishiyama:2008qa}. To select the reference stars, we took into account
several criteria:

\begin{itemize}

\item Only stars with an uncertainty $<5\, \%$ in both the  SIRIUS catalogue and our final list were accepted.
\item To avoid saturation or faint stars, we imposed brightness limits for all three magnitudes.
\item The reference stars should be as isolated as possible. For that reason, we excluded all the stars which have a neighbour within a radius of 0.5$''$ in our final list.
\item We did not use stars near the edges of the FOV or in regions  with a low number of exposures (see e.g. Fig.\,\ref{exp}).
\item Finally, we applied a two-sigma clipping algorithm to remove outliers.
\end{itemize}

\begin{figure}
       \begin{center}
   \includegraphics[scale=0.52]{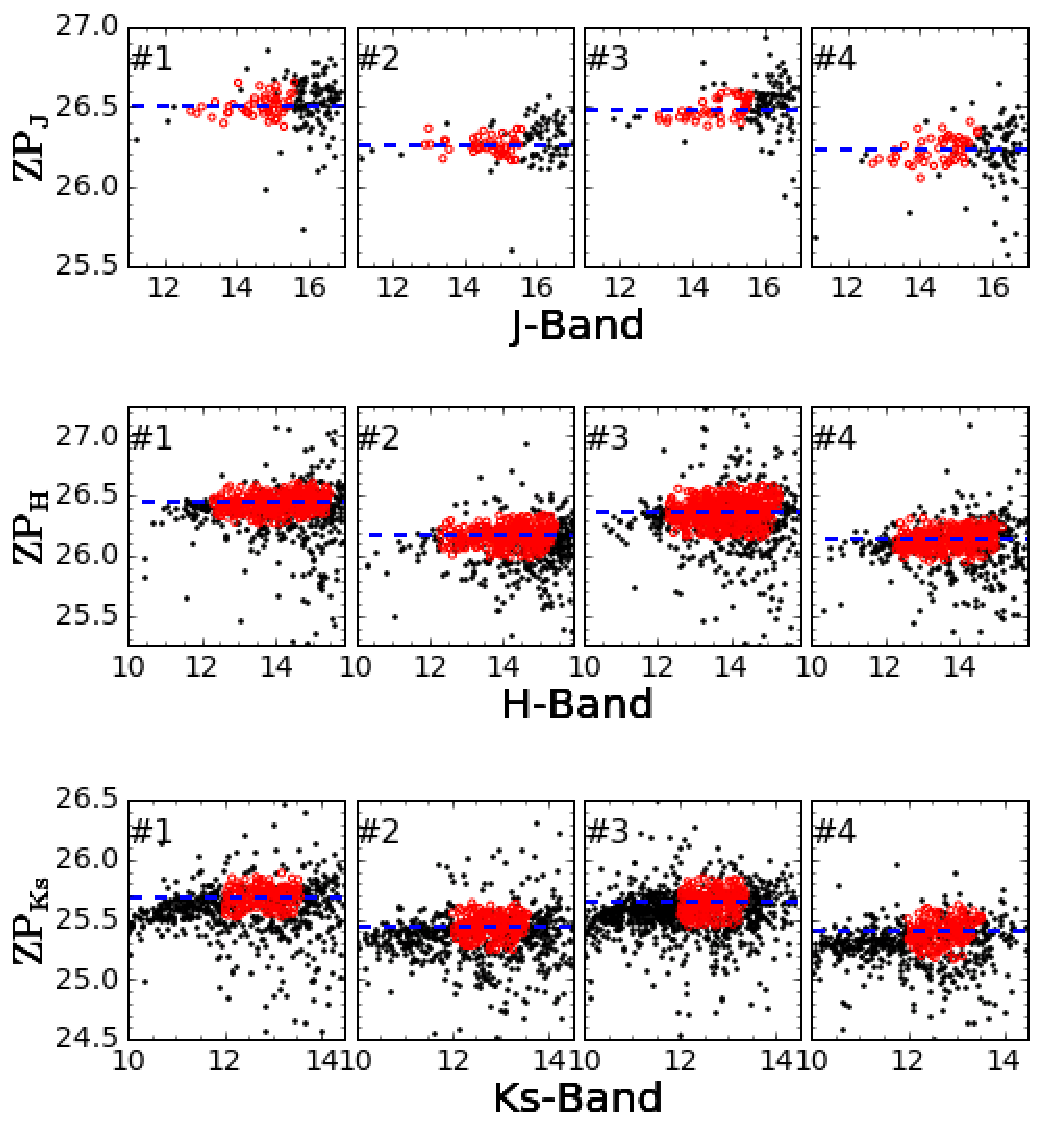}
   \caption{Zero point calibration for each chip and band in
     D15 data. Black points represent stars common to the SIRIUS catalogue
     and our final list. Red points mark the stars used to compute the
     zero points. The mean zero points are indicated by the blue
     lines. The systematic deviations for the brightest stars are due to
     saturation.}
   \label{zp_calculation}
    \end{center}   
    \end{figure}

Figure\,\ref{zp_calculation} shows the zero points computed for common
stars between the SIRIUS catalogue and our data in each band and
chip. In all cases the reference stars were well distributed across
each detector. There were also sufficient reference stars for a robust
calibration: $\sim$ 50 in $J$ band, $\sim$ 300 in $H$ band, and
$\sim$ 175 in $K_s$ band on each chip in the case of the 2015 data (about 30\%
less because of the smaller FOV in the case of the 2013 data). 

We also checked for spatial variability of the zero point across the
chips assuming a variable ZP and computing it with a slanting plane
(i.e. a one-degree polynomial). However, we did not find any
significant difference with the assumption of a constant ZP within the
uncertainties. This agrees with the findings of \citet{Massari:2016uq},
who also concluded that constant zero points could be used to
calibrate their HAWK-I imaging data.

\subsubsection{Pistoning correction}

Once the photometry obtained for every chip and band was calibrated,
we corrected the possible residual zero point offset that could have
remained between different chips (and pointings in the case of the 2013
data), known as "pistoning". To do that we used the technique
described in \citet{Dong:2011ff}, which consists in minimising a global
$\chi ^2$ that takes into account all the common stars for all chips
simultaneously. For that, we only used stars with less than 0.05 mag
of uncertainty. As expected, the variation in zero point between the
chips was quite low (less than 0.1 magnitude even in the worst cases
for all three bands and in both epochs).
Calibrating the photometry for every chip independently with the SIRIUS catalogue before applying the pistoning correction let us compare the overlapping region of the different calibrated chips and estimate the relative offset as described in Sect. \ref{quali}.

\subsubsection{Combined star lists}

\begin{figure}
    \begin{center}   
   \includegraphics[scale=0.5]{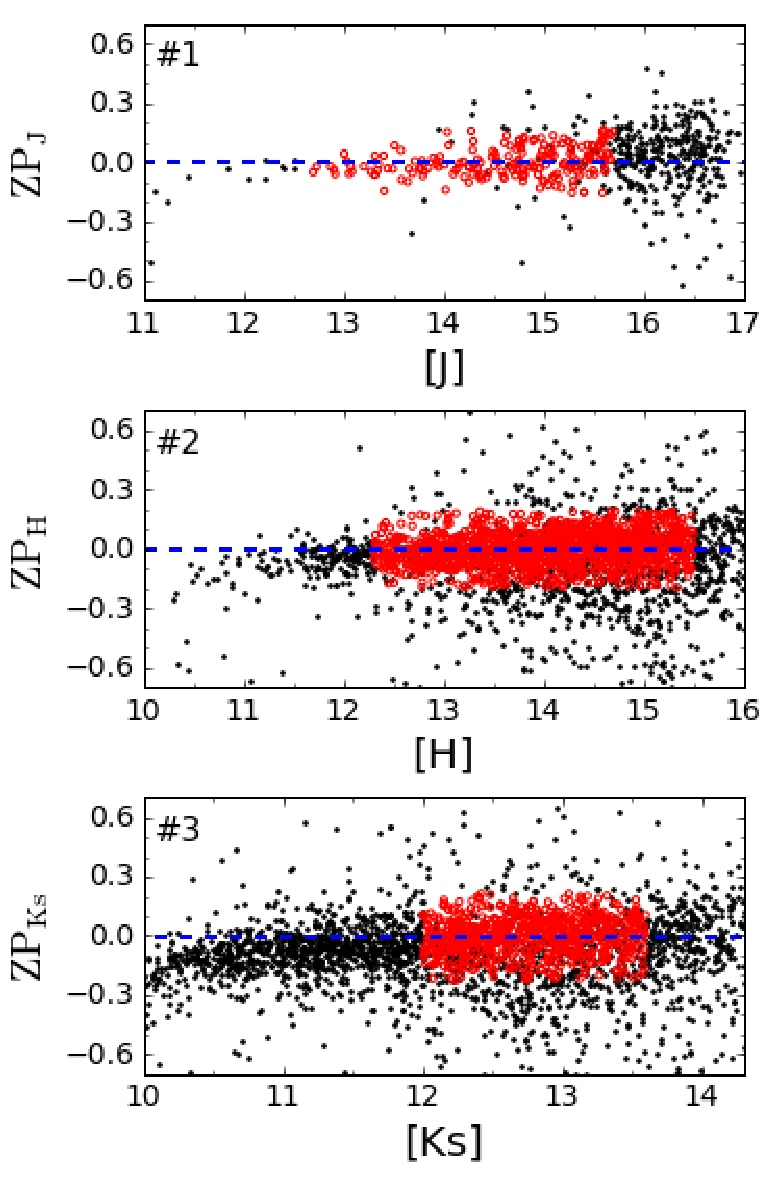}
   \caption{Zero points calculated in all three bands after merging
     the list for every chip and pointing (D15). Black points
     represent all the common stars between the SIRIUS catalogue and our
     data. Red points depict the stars used to compute the zero point
     and the blue line is its average. Systematic deviations for the
     brightest stars are due to saturation.}
   \label{zp-recal}
    \end{center}
    \end{figure}

To produce the final catalogue we merged all the photometric and
astrometric measurements. For the overlap regions between the chips,
the value for stars detected more than once was taken as the mean of
the individual measurements. In those cases, the uncertainty was computed
as the result of quadratically adding the individual
uncertainties of each measurement detection.
The pistoning corrections can result in minor zero point shifts
of the combined star lists. To avoid them, we
re-calibrated the zero point of the final, merged catalogue. This
procedure was completely analogous to the one described
above. Figure \ref{zp-recal} shows the final calibration. The deviations
at magnitudes $Ks\lesssim11$ are due to saturation.

\subsubsection{Zero point uncertainty}
\label{uncer_ZP}

\begin{figure}
    \begin{center}   
   \includegraphics[scale=0.42]{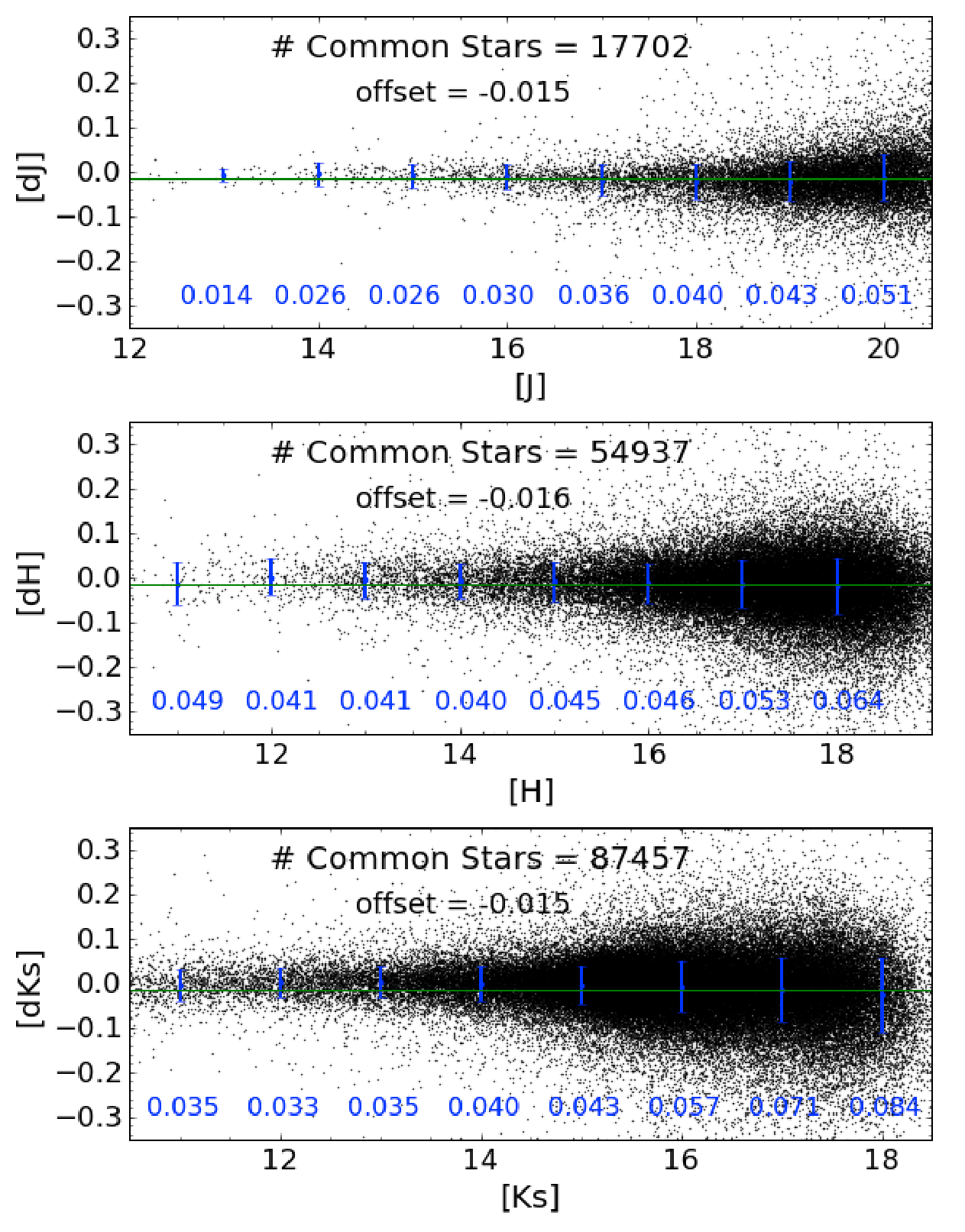}
   \caption{Photometric comparison of the common stars between the
     final lists of D13 and D15 data obtained using
     Eq. \ref{compar}. The green line depicts the mean of the
     difference between the points (we applied a two-sigma clipping
     algorithm to compute it). Blue lines and the number below them
     depict the standard deviation of the points in bins of one
     magnitude width.}
   \label{ZP_final_uncertainties}
    \end{center}   
    \end{figure}

The final ZP uncertainty was estimated by comparing the common stars
of the D13 and D15 combined lists
(Fig.\,\ref{ZP_final_uncertainties}). As they were calibrated
independently, the photometric offset that appears between them is a
measurement of the error associated with the calibration procedure. To compare the photometry, we used Eq. \ref{compar}, where the fluxes are calculated from the magnitudes in both epochs. This gives us an upper limit for the uncertainties that is shown in
Fig.\,\ref{ZP_final_uncertainties}. We obtained a rounded upper limit of
$0.02$\,mag for the ZP offset between the epochs.  This value takes
into account possible variations of the ZP across the detector, as
every star was located in a different position of the detector (or
even different detectors) in both epochs. It also takes into account
one of the main sources of uncertainty, namely that roughly 10\% of the
stars in the GC are variable \citep{Dong:2017ab}. Therefore, the absolute uncertainty of our catalogue results from
quadratically adding this uncertainty to that of the SIRIUS
catalogue. We thus obtain an absolute ZP uncertainty of $0.036$\,mag
for each band.

   \begin{figure}
    \begin{center}   
   \includegraphics[scale=0.4]{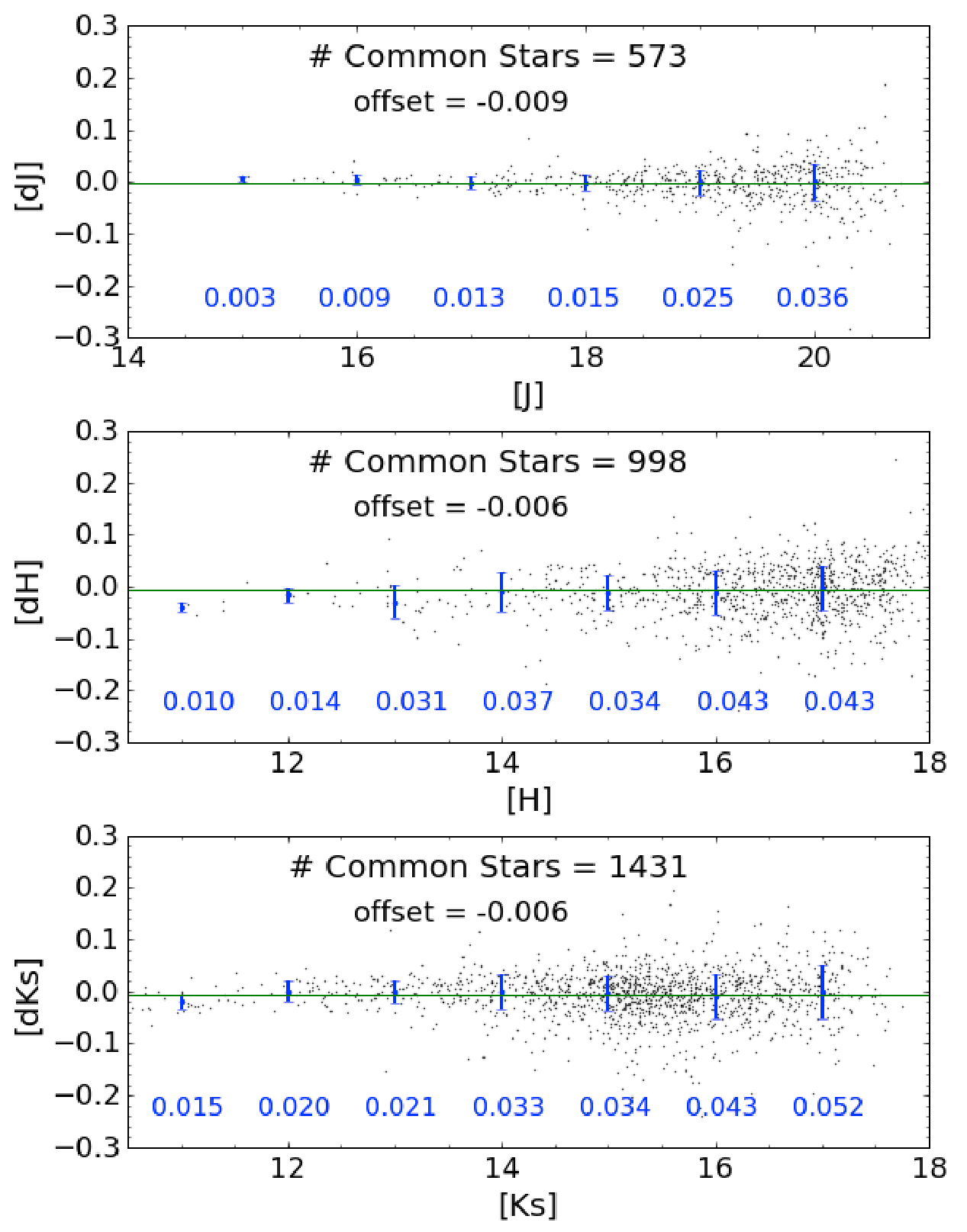}
   \caption{Photometric accuracy in the overlap regions between
     chips \#1 and \#4 (D15 data). Blue lines and the number below them
     depict the standard deviation of the points in bins of one
     magnitude width.}
   \label{overlap}
    \end{center}   
    \end{figure}

\section{Quality assessment}
\label{quali}
To check the accuracy of the photometry, we performed several tests:

(1) We compared the overlapping regions of all four chips in D15 data,
using Eq.\,\ref{compar}. We excluded the borders and compared the
inner regions of the overlap, where at least 100 frames contribute to
each pixel (see e.g. the weight map in
Fig.\,\ref{exp}). Figure\,\ref{overlap} shows the results for the overlap
between chips \#1 and \#4. As can be seen, the photometric accuracy is
satisfactory. The uncertainties are at most a few 0.01\,mag. These
independent estimates of the photometric uncertainties agree well with
the combined statistical and systematic uncertainties of our final
list.

(2) We compared the photometry from two different epochs, D13 and
D15.  In this case, we compared the common stars from each epoch. The
overlap region is much larger than the one that we had in the previous
test, so we found far more common stars, letting us improve on the
quality of the analysis and extend it to the central regions of the
chips.  We computed uncertainty upper limits using Eq. \ref{compar} and plotted the standard deviations in bins of one magnitude in
Fig.\,\ref{ZP_final_uncertainties}. The result demonstrates that the estimates
of the photometric uncertainties of our final lists are accurate.

(3) We took HAWK-I observations of several reference stars from the
2MASS (Two Micron All-Sky Survey) calibration Tile 92397 (ra(J2000) 170.45775 dec(J2000)
-13.22047), in all three bands, as a crosscheck to test the
variability of the zero point. The calibration field was observed in a
way that for each chip a reference star was positioned at nine
different locations.  We applied a standard reduction process to those
images taking into account the special sky subtraction that we
described above. Then, we performed aperture photometry for all nine
positions taking four different aperture radii to test the
uncertainties. All nine measurements for all the chips agree within
the uncertainties with the assumption of  a constant ZP across the chips.

(4) All the previous tests estimate the uncertainties supposing that only the detected sources are in the field. To complement the quality assessment, we used the simulations described in Sect. \ref{rebinning} and their corresponding uncertainties. They show the influence of extreme crowding in the photometry, as they simulated the inner parsec, which supposes the most crowed field. The photometric uncertainties that we obtained are lower because only the central region, the inner arcminute, suffers from extreme crowding. Therefore, the obtained results are consistent with this analysis.

\begin{figure}
    \begin{center}   
   \includegraphics[scale=0.5]{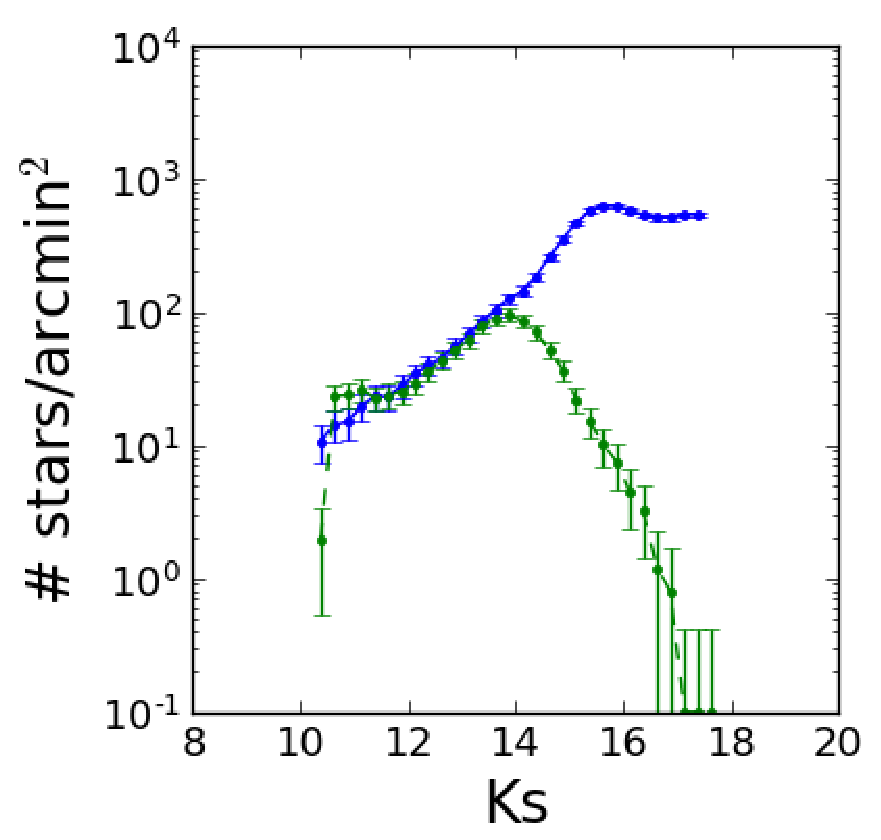}
   \caption{Luminosity functions in $K_s$ obtained with HAWK-I (blue line) and VVV (green dashed line).}
   \label{luminosity}
    \end{center}   
    \end{figure}

(5) We qualitatively compared the $K_s$ luminosity functions obtained with our HAWK-I data and the VVV survey. For that, we performed PSF fitting photometry with {\it StarFinder} on a final VVV mosaic centred on SgrA* observed in 2008. We produced luminosity functions for both VVV and HAWK-I data in a region of approximately $8'\times 2.5'$. Figure \ref{luminosity} shows the results.The HAWK-I data are roughly three magnitudes deeper than the VVV data. The brightest part of the luminosity function is slightly different as a consequence of the important saturation of the $K_s$ band in the central parsec in VVV.

\section{Colour-magnitude diagrams}
\label{RC_selection}

\begin{figure}
    \begin{center}   
   \includegraphics[scale=0.65]{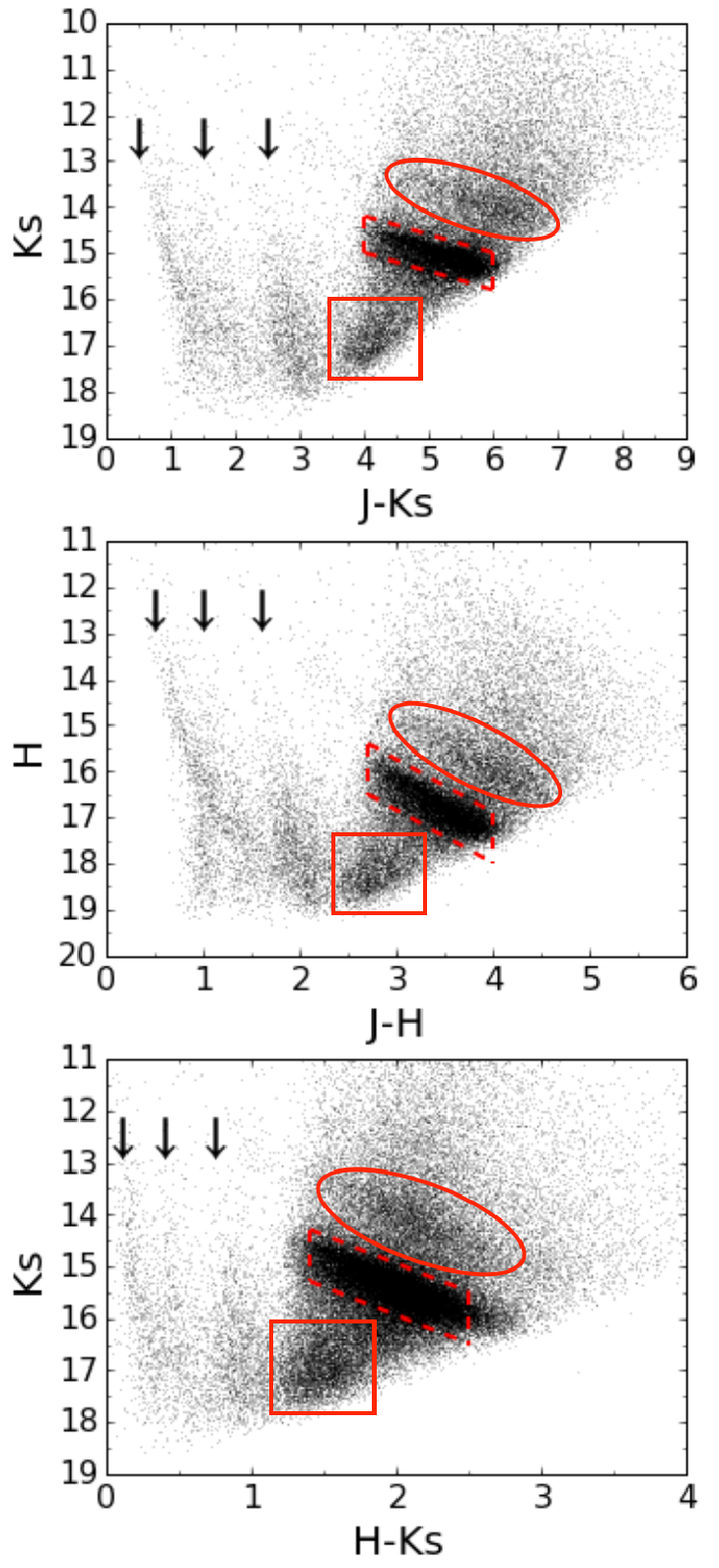}
   \caption{Colour-magnitude diagrams for $K_s$ versus $J-K_s$, $H$ versus $J-H,$ and $K_s$
     versus $H-K_s$ (D15 data). The red dashed parallelograms indicate the red clump, the red ellipse marks stars that belong to the asymptotic giant branch bump and the red square contains ascending giant branch and post-main sequence stars. The three black arrows indicate foreground stellar population probably tracing spiral arms.}
   
   \label{CMD}
    \end{center}   
    \end{figure}

Figure\,\ref{CMD} depicts the colour-magnitude diagrams (CMD) of the
D15 data. We can easily distinguish several features. The three black
arrows point to foreground stellar populations that probably trace
spiral arms.  The highly extinguished stars lie close to or in the
GC. The red ellipse indicates stars that belong to the asymptotic
giant branch (AGB) bump, the red square contains ascending giant branch and
post-main sequence stars, and the area marked with red dashed lines
corresponds to red clump (RC) stars. They are low-mass stars burning
helium in their core \citep{Girardi:2016fk}.  Their intrinsic colours and magnitudes depend weakly on age and metallicity, so that these stars are a good tracer population to study the extinction and determine
distances.

\begin{figure}
    \begin{center}   
   \includegraphics[scale=0.42]{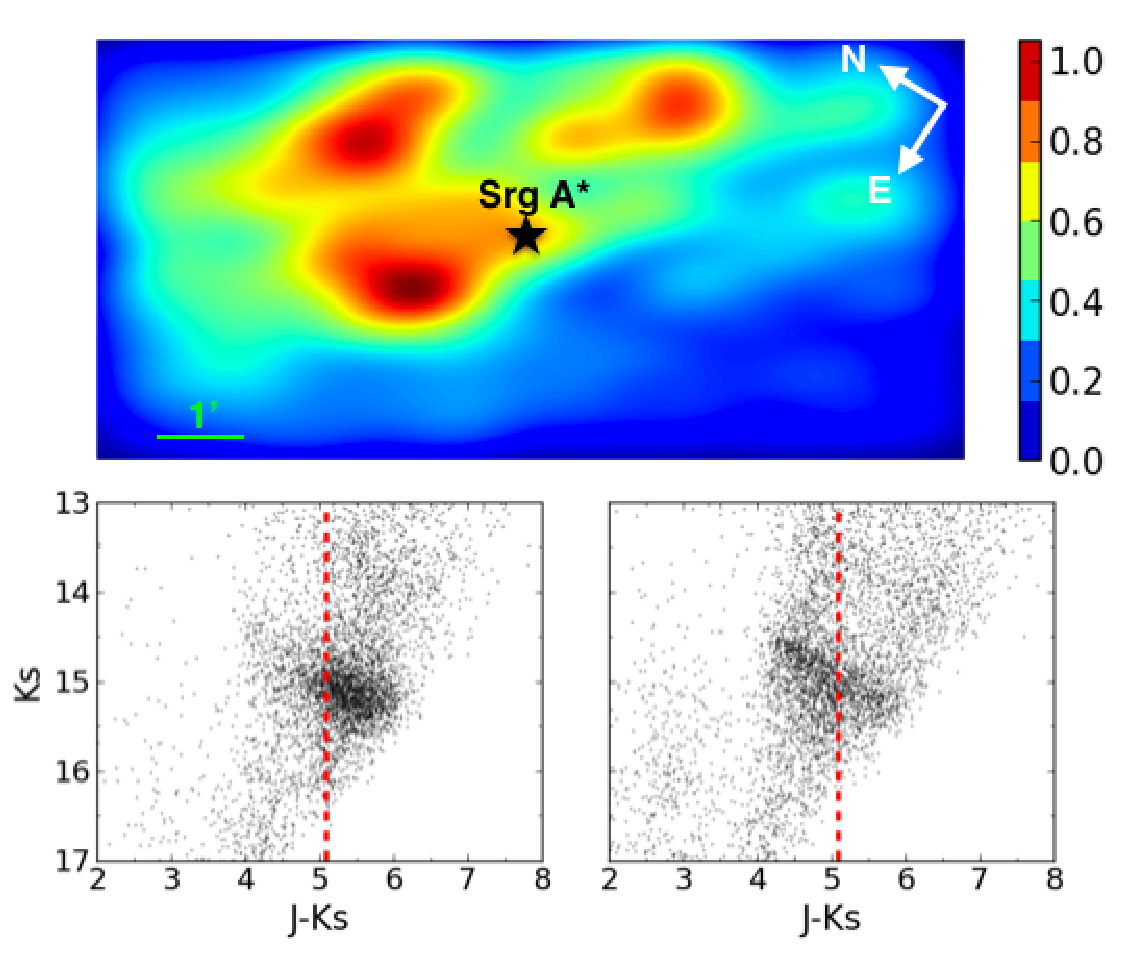}
   \caption{Upper panel: Plot of the $J$-band stellar density. Lower
     left panel: CMD for stars located in areas with high stellar
     density normalised to the same area and maximum density. Lower right panel: CMD of stars in regions dominated by
     dark clouds. The red dashed line approximately separates the foreground
     population from stars located at the Galactic Centre.}
   
   \label{density}
    \end{center}   
    \end{figure}

    The GC field studied here contains large dark
    clouds that affect significant parts of the area. In order to study the
    stellar population towards the GC, it therefore appears reasonable
    to separate areas with strong foreground extinction from those
    where we can look deep into the GC. Since the effect of extinction
    is a strong function of wavelength, dark clouds can be
    easily identified via the low $J$-band surface density of stars associated to them. This
    method works better than using stellar colours because those are
    biased towards the blue in front of dark clouds (dominated by
    foreground sources). The upper panel in Fig.\,\ref{density} depicts the $J$-band surface
    stellar density.

    To detect foreground stars and to avoid them in the
    subsequent analysis, we made CMDs ($K_s$ versus $J-K_s$) for regions
    dominated by dark clouds in the foreground of the GC,
    using as criterion a stellar density below $40 \%$ of the maximum density in Fig\,\ref{density}, and for the more transparent
    regions a stellar density $>75 \%$ of the maximum density. We normalised the resulting CMDs to the same area to compare them. For
    the transparent regions, the RC stellar population located at the
    GC forms a very clear clump on the right of the red dashed line in
    the left panel of Fig.\,\ref{density}. The density of
    foreground sources is much higher in the areas dominated by large
    dark clouds, as can be seen in the right panel of
    Fig.\,\ref{density}, where one of the potential spiral arms (at
    $J-K_{s}\approx3$) stands out clearly in the CMD. 

    The separation between the groups of stars in the foreground of
    dark clouds and in the deep, more transparent regions, is not
    unambiguous, however. Two important reasons for the overlap of the
    two CMDs are (1) the clumpiness of the dark clouds and (2) the
    large area dominated by the dark clouds. The angular size of the
    clumps is smaller than the smoothing length used in the stellar
    surface density map shown in Fig.\,\ref{density}. Also, we can see
    that the area with stellar density larger than $75\%$ of the maximum star density is roughly only one third of the area with $40\%$ of the maximum star density.

\begin{figure}
   \begin{center}   
   \includegraphics[scale=0.28]{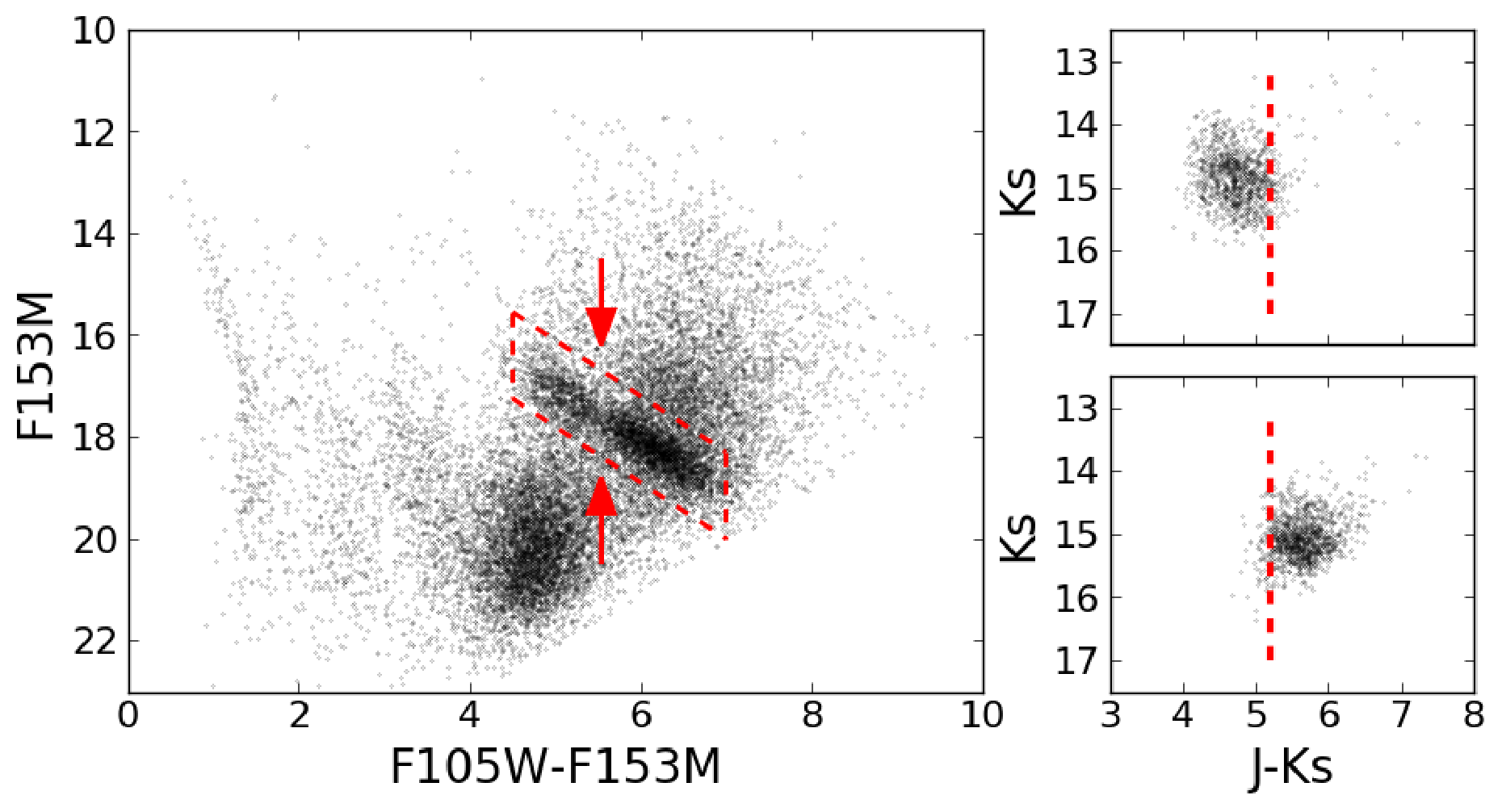}
   \caption{Left panel shows the colour-magnitude diagram F153M versus F105W-F153M. The red dashed parallelogram traces the RC and the red arrows mark an obvious gap in the distribution of the RC stars, which follows the reddening vector. The upper right panel depicts stars from our HAWK-I catalogue in $J$ and $K_s$ with counterpart in the HST data in the first detected bump. Analogously, the lower panel represents stars located in the second bump.}
   
   \label{HST_cut}
    \end{center}   
    \end{figure}

    To confirm the existence of two groups of stars in the field, with
    one group highly extinguished and the other group at lower extinction,
    we used data from HST WFC3 (Wide Field Camera 3) centred on SgrA* with an approximate size of 2.7' $\times$ 2.6', to produce a CMD F153M versus F105W-F153M
    (\citealt{Dong:2017aa}, MNRAS; Dong et al. in
    preparation). Figure\,\ref{HST_cut} shows that, when using these
    bands, there appears a clear gap between the two parts of the RC
    that we previously detected. If we select only stars that are
    bluer than $F105W-F153M=5.5$ in the HST CMD and identify those
    stars in our HAWK-I list, then we obtain the CMD shown in the
    upper right panel of Fig.\,\ref{HST_cut}. On the other hand, the
    stars on the red side of this colour produce the CMD in the lower
    right panel of Fig.\,\ref{HST_cut}.
Unfortunately, the HST data cover only a fraction of the HAWK-I
    FOV. However, as shown in the right panels of Fig.\,\ref{HST_cut},
    using a colour cut at $J-K_{s}=5.2$ we can separate fairly
    reliably the two giant populations with different mean
    extinctions.

\section{Determination of the extinction curve}
\label{alpha_sec}

As outlined in the introduction, here we assume that the NIR
extinction curve toward the GC can be described well by a power law of
the form $A_{\lambda}\propto\lambda^{-\alpha}$, where $\lambda$ is the
wavelength, $A_{\lambda}$ the extinction in magnitudes at a given
$\lambda$ , and $\alpha$ is the extinction index.
Here we use our data set to investigate whether $\alpha$ has the same
value across the $J$$H$$K_s$ filters, whether it can be considered independent
of the exact line of sight toward the GC, and whether it depends on the
absolute value of extinction. If $\alpha$ can be considered a constant
with respect to position, extinction, and wavelength, then we can
determine its mean value. We apply several different methods to compute $\alpha$
and perform tests of its variability.

\subsection{Stellar atmosphere models plus extinction grids}
\label{grid}

We selected RC stars and
assumed a Kurucz stellar atmosphere model \citep{Kurucz:1993fk} with
an effective temperature of 4750\,K, solar metallicity, and
$\log g =+2.5$  \citep{2014ApJ...790..127B}. We set the distance of the GC to 8.0 $\pm$ 0.25\,kpc
\citep{Malkin:2013fk} and assumed a radius of $10\pm 0.5\,R_\odot$ for
the RC stars \citep[see][with the uncertainty given in the former
reference]{Chaplin:2013kx,Girardi:2016fk}. Then, we computed the model
fluxes for the $J$, $H,$ and $K_{s}$ bands assuming a grid of different
values of $\alpha$ and the extinction at a fixed wavelength of
$\lambda=1.61\,\mu$m, $A_{1.61}$. The grid steps were 0.016 for both $A_{1.61}$ and $\alpha$.  To convert the fluxes into magnitudes, we
used a reference Vega model from Kurucz. We refer to this method as
the \emph{grid method} in the following text. We produced histograms to represent the distribution of the obtained parameters and fitted them with a Gaussian model.  We tested different ways of computing the bin widths of the histograms. Namely, we used the Freedman-Diaconis rule \citep{Freedman1981}, the Scott rule \citep{SCOTT:1979aa}, the Sturges rule \citep{doi:10.1080/01621459.1926.10502161}, or the Doane rule \citep{doi:10.1080/00031305.1976.10479172}, among others. We did not observe any significant difference in the result of the fits. We generally adopted the Sturges rule for the histograms presented in this paper, as this rule is appropriate when the distributions to be represented are Gaussian-like. Besides, it smooths the histograms, which is convenient to overcome the segregation that can appear due to the grid step. This bin width selection criterion is applied for all the histograms presented in the paper.

We applied this method to compute $\alpha$ between the $J$ and
$K_{s}$ bands, the $H$ and $K_{s}$ bands, the $J$ and $H$ bands, and
across all three bands together. We defined for each star a $\chi^2 = \sum ( (band_{measured}-band_{model})/\sigma_{band})^2$ and searched for the parameters that minimised it.

\subsubsection{RC stars in the high-extinction group}
\label{group1}

\begin{figure}
   \begin{center}   
   \includegraphics[scale=0.36]{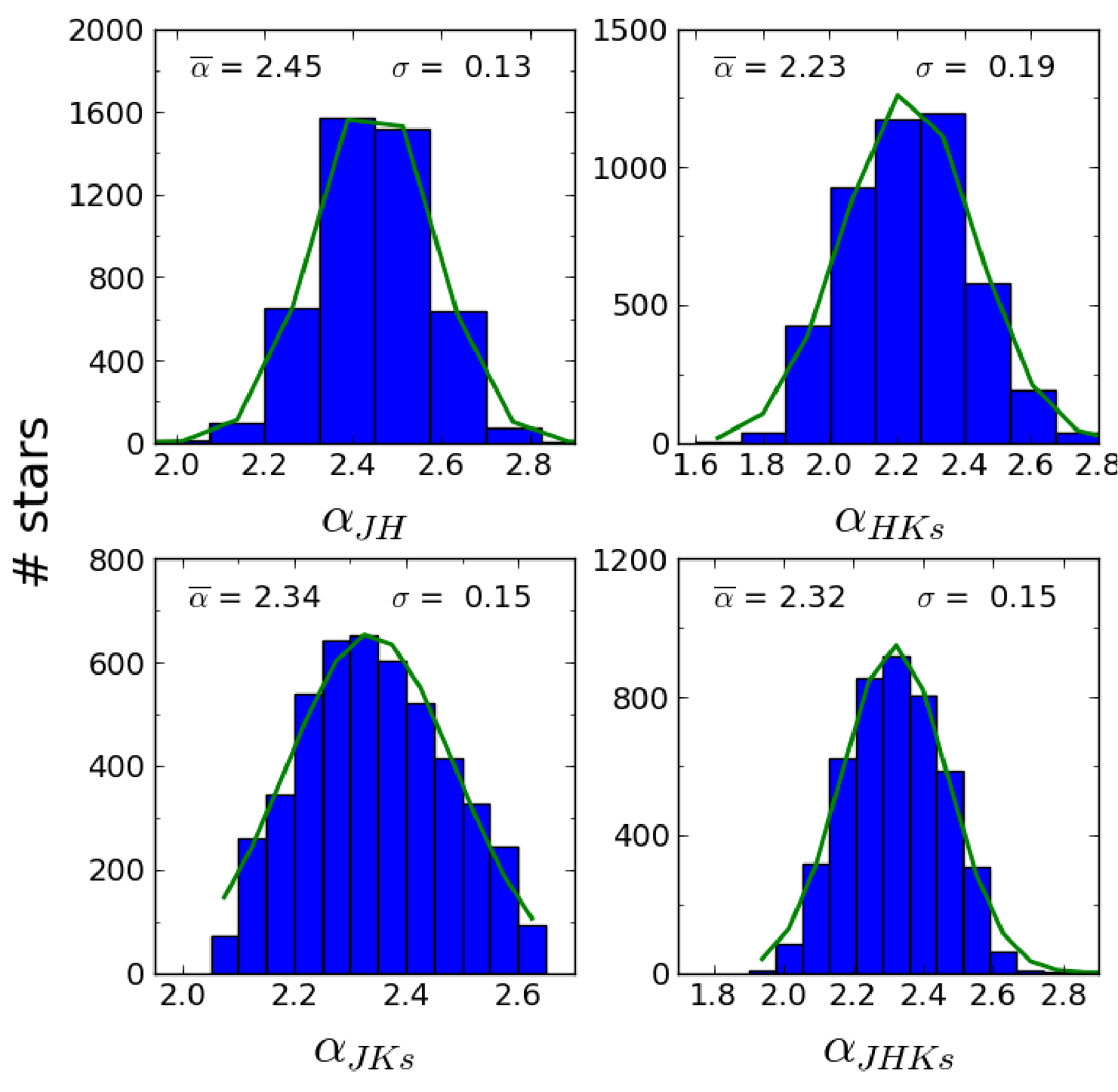}
   \caption{Histograms of $\alpha$ computed with the grid
     method for the RC stars in the high-extinction group. Gaussian fits are overplotted as green lines, with the mean and standard deviations annotated in the plots.}
   
   \label{alpha_grid_gaussian}
    \end{center}   
    \end{figure}

\begin{figure}
   \begin{center}   
   \includegraphics[scale=0.36]{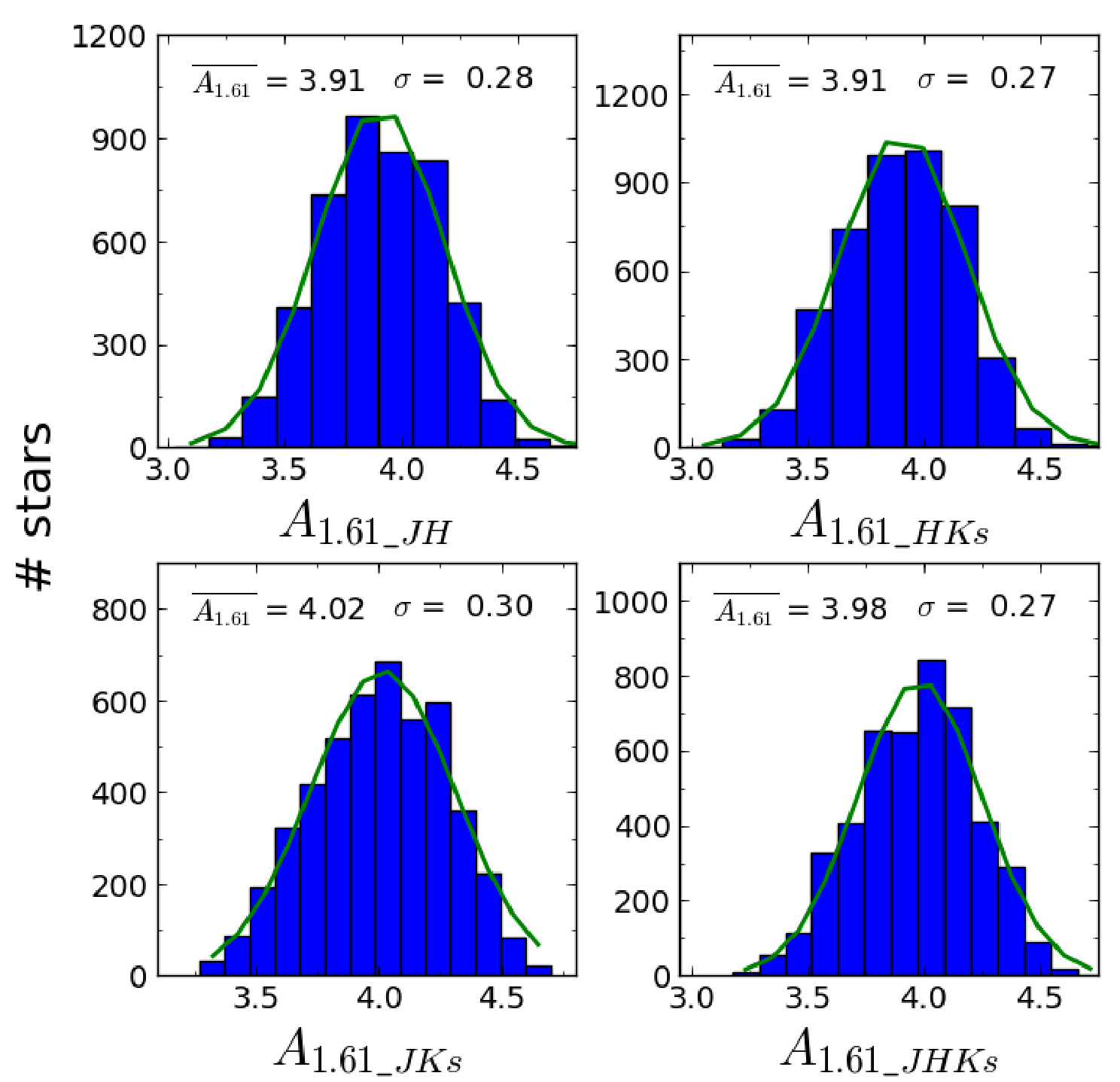}
   \caption{Histograms of $A_{1.61}$ computed with the grid
     method for the RC stars in the high-extinction group. Gaussian fits are overplotted as green lines, with the mean and standard deviations annotated in the plots.}
   
   \label{A_H_grid_gaussian}
    \end{center}   
    \end{figure}

We applied the method to the RC stars with observed colours between $J-K_{s}>5.2$ and $J-K_{s}<6$. The histograms of the optimal values of $\alpha$ and $A_{1.61}$, along with Gaussian fits, are shown in Figs.\,\ref{alpha_grid_gaussian} and
\ref{A_H_grid_gaussian}, respectively.
The statistical uncertainty is given by the error of the mean of the
approximately Gaussian distributions and is negligibly small. We
considered the systematic uncertainties due to the uncertainties in
the temperature of the assumed model, its metallicity, the atmospheric
humidity, the distance to the GC, the radius of the RC stars, and the
systematics of the photometry.  We re-computed the values of
$A_{1.61}$ and $\alpha_{wavelength\_range}$ varying individually the
values of each of these factors:

\begin{itemize}
\item For the model temperature we took two models with  \mbox{4500 K} and
  5000 K. That range takes into account the possible temperature variation for the RC \citep{2014ApJ...790..127B}.

\item For the metallicity we took five values in steps of 0.5 from -1 to +1 dex.
\item For the humidity we varied the amount of precipitable water vapour between 1.0, 1.6, and 3.0 mm.
\item For the distance to the GC, we used $7.75$, $8.0$ and $8.25$ kpc.
\item We varied the stellar radius between  $9.5$, $10$ and $10.5$.
\item We took three different values for the $\log g$  used in the Kurucz model: $2.0$, $2.5,$ and $3.0$.
\item Finally, we tested the effect of the variation of the systematics of the photometry for every band independently, subtracting and adding the systematic uncertainty to all the measured values.
\end{itemize}

The largest errors arise from the uncertainty in the radius of the RC
stars and the temperature of the models. The final systematic
uncertainty was computed by summing quadratically all the individual
uncertainties. Table\,\ref{grid_table} lists the values of $\alpha$
and $A_{1.61}$ we obtained along with their uncertainties. We can see
that $\alpha_{JH}$, $\alpha_{HK_s}$, $\alpha_{JK_s}$ , and
$\alpha_{JHK_s}$ are consistent within their uncertainties. On the other hand, we obtained very similar values for $A_{1.61\_JH}$, $A_{1.61\_HK_s}$, $A_{1.61\_JK_s}$ , and
$A_{1.61\_JHK_s}$ to what we expected because we used the same stars to compute them.

\begin{table}
\begin{center}
\caption{Values of $\alpha$ and $A_{1.61}$ obtained with the grid
  method.}
\label{grid_table} 
\begin{tabular}{cccc}
 &  & \tabularnewline
\hline 
\hline 
&Bands & $\alpha$ & $A_{1.61}$\tabularnewline
\hline 
&$JH$ & $2.45 \pm 0.09$  & $3.91 \pm 0.16$\tabularnewline
High&$HK_s$ & $2.23\pm 0.14$ & $3.91 \pm 0.15$\tabularnewline
extinction&$JK_s$ & $2.34 \pm 0.09$ & $4.02 \pm 0.20$\tabularnewline
&$JHK_s$ & $2.32 \pm 0.09$ & $3.98 \pm 0.18$\tabularnewline
\hline 
&$JH$ & $2.41 \pm 0.11$  & $3.30 \pm 0.16$\tabularnewline
Low&$HK_s$ & $2.19\pm 0.16$ & $3.30 \pm 0.15$\tabularnewline
extinction&$JK_s$ & $2.28 \pm 0.11$ & $3.40 \pm 0.19$\tabularnewline
&$JHK_s$ & $2.33 \pm 0.10$ & $3.36 \pm 0.17$\tabularnewline
\hline 
&$JH$ & $2.44 \pm 0.10$ &  $3.62 \pm 0.16$ \tabularnewline
All&$HK_s$ & $2.21 \pm 0.14$ & $3.61 \pm 0.16$ \tabularnewline
RC stars&$JK_s$ & $2.31 \pm 0.10$  & $3.72 \pm 0.20$ \tabularnewline
&$JHK_s$ & $2.34 \pm 0.09 $  & $3.67 \pm 0.17$ \tabularnewline
\hline 
\end{tabular}
\end{center}
\vspace{1cm}
\end{table}

\subsubsection{RC stars in the low-extinction group}
\label{low_ext}
We also applied this method to the RC stars located in front of the dark clouds, $J-K_{s}>4.0$ and $J-K_{s}<5.2$. Analogously, the main source of error is the systematic uncertainty, and statistical uncertainties given by the error of the mean are negligible. The Gaussian fits for the extinction index and $A_{1.61}$ for $JH$, $HK_s$, $JK_s$ , and $JHK_s$ are presented in Appendix\,\ref{plots_low_ext}. Table\,\ref{grid_table} shows the results for the fits and the corresponding uncertainties.

\subsubsection{All RC stars, $JHK_{s}$}
\label{2gaussians}

We also applied the previously described method to all the RC stars ($J-K_s > 4$ and $J -K_s < 6$ ),
using $JHK_{s}$ measurements. We thus obtain the extinction index
simultaneously for low-extinction regions and for regions dominated by
dark clouds. The results are shown in Fig.\,\ref{all_grid} and in Appendix\,\ref{plots_all_ext}. The mean values of the Gaussian fit and their uncertainties (dominated by systematics) are presented in Table \ref{grid_table}. It can be
seen that the mean and standard deviation of the histogram of
the extinction indices agree very well with the values obtained for the
two RC populations analysed before (Table\,\ref{grid_table}). In the case of $\alpha_{JHK_s}$ and $A_{1.61\_JHK_s}$, a Gaussian fit for both gives a mean value of $\alpha_{JHK_s} = 2.34 \pm 0.09$ and $A_{1.61\_JHK_s} = 3.67 \pm 0.17$.

On the other hand, the extinction shows a broader distribution that extends to values
$A_{1.61}<3.4$.  We fitted a combination of two Gaussians and compared it with the single Gaussian fit. For that, we used the SCIKIT-LEARN python object GaussianMixture, which implements the  expectation-maximisation algorithm to fit a mixture of Gaussian models \citep{Pedregosa:2011aa}. Computing the Bayesian Information Criterion \citep{Schwarz:1978aa} and the Akaike Information Criterion \citep{Akaike:1974aa}, we checked that a combination of two Gaussians provides a
significantly better fit than using a single Gaussian. In the case of $A_{1.61\_JHK_s}$ , the mean values of the two Gaussians are $A_{1} = 3.32 \pm 0.20$ and $A_{2} = 3.88 \pm 0.20,$ with a standard deviation of $\sigma_1 = 0.37$ and $\sigma_2 = 0.29$, respectively. Again, the uncertainties are dominated by the systematics. It is remarkable that the systematic uncertainties do not change the relative position of the two mean extinctions detected, since they affect both in the same direction. The mean extinctions $A_1$ and $A_2$ correspond to the low and highly reddened population. The results agree with their 
previously determined values based exclusively on each population independently. This indicates that the method is able to estimate $\alpha_{bands}$ and $A_{1.61\_bands}$ independently of the selected range of colour or extinction, distinguishing stars with different extinction. As shown in Appendix\,\ref{plots_all_ext}, we repeated the procedure described above for $JH$, $HK_s$ , and $JK_s$ and found that the extinctions are compatible with a two-Gaussian model and the extinction index value is in agreement with one single $\alpha_{JHK_s}$ within the uncertainties of our study.

   \begin{figure}
   \begin{center}   
   \includegraphics[scale=0.36]{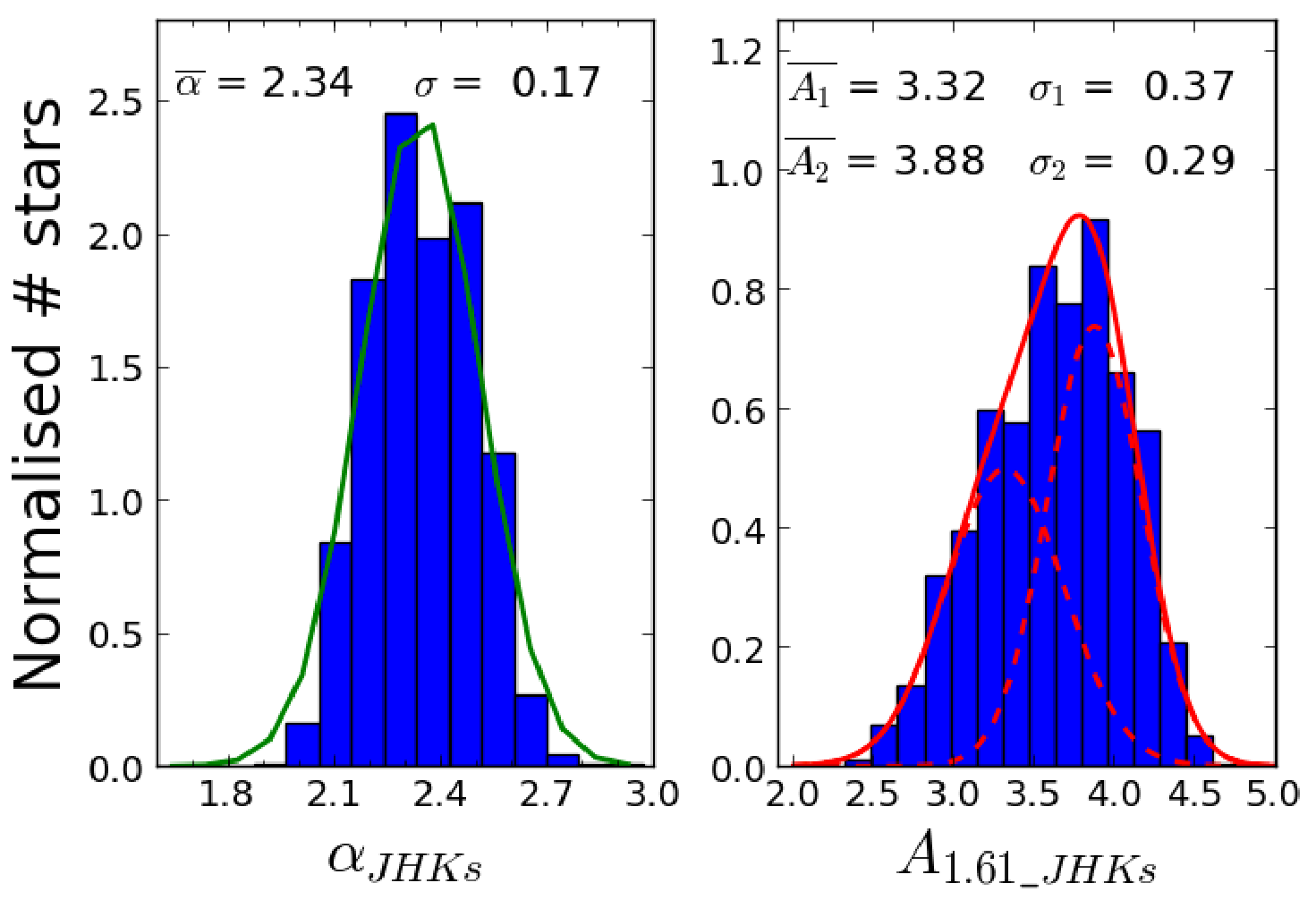}
   \caption{Normalised histograms of $\alpha_{JHK_s}$ and $A_{1.61}$ computed with all the RC stars using the grid method. The green line shows a Gaussian fit for the extinction index. The red continuous line depicts a two Gaussian fit (with the individual Gaussians marked by the red dashed lines).}
   \label{all_grid}
    \end{center}   
    \end{figure}

The values of the extinction indices and extinctions agree within their uncertainties. However, it is noticeable that we have found a slightly steeper value for $\alpha_{JH}$ than for $\alpha_{HK_s}$. Although the difference between those values is covered by the given uncertainties, the relative uncertainty between them is slightly lower. Namely, the variations in the temperature of the model, $\log g$, the radius of the RC stars, their metallicity, and the distance to the GC  produce a systematic uncertainty in the same direction for all the values computed. Therefore, we found a small difference between the extinction index depending on the wavelength in the studied ranges. This difference will be studied in detail in the next sections.

\subsubsection{Spatial distribution of $\alpha$}

We studied the spatial variability of the extinction index, using the individual values obtained for all the RC stars in the previous section. To do that, we produced a map calculating for every pixel the corresponding value of $\alpha_{JHK_s}$. To save computational time, we defined a pixel size of 100 real pixels ($\sim 5 ''$). The value for every pixel was obtained averaging (with a two-sigma criterion) the extinction indices obtained for all the stars located within a radius of $1'$ from its centre. The obtained map presents very small variations depending on the region, varying between $\alpha_{JHK_s} = 2.30$ and 2.38. Fitting with a Gaussian, we obtained a mean value of $\alpha_{JHK_s} = 2.35$ for all the pixels with a standard deviation of 0.03. We concluded that if there exists variation across the studied field, it is negligible. Therefore, we can assume that the extinction index does not vary with the position.

   

\subsection{Fixed extinction}
\label{individual_method}

As a first estimation, a constant $\alpha$ from $\lambda_J$ to
$\lambda_{K_{s}}$ can be assumed. Here, we use RC stars identified in all three
bands  to compute for
each of them their corresponding $\alpha$. To do that, we use the following expression:

\begin{equation}
\label{individual}
\frac{\left(\frac{\lambda_H}{\lambda_J}\right)^{\alpha}-1}{1-\left(\frac{\lambda_H}{\lambda_{K_s}}\right)^\alpha} = \frac{J-H-(J-H)_0}{H-K_s-(H-K_s)_0},
\end{equation}

where $\lambda_i$ refers to the effective wavelengths in each band,
$\alpha$ is the extinction index, $J$, $H,$ and $K_s$ are the observed
magnitudes in the corresponding bands, and the sub-index 0 indicates
intrinsic colours (see Appendix\,\ref{intrinsic}).

To compute the effective wavelength, $\lambda_{\rm eff}$, and the intrinsic colours, we used as
starting values the $\alpha$ and the $A_{1.61}$ calculated in
Section\,\ref{grid} (as explained in Appendix \ref{wav}). We
kept $A_{1.61}$ constant, but updated iteratively the values of
$\lambda_i$ and, subsequently, of $\alpha$. After several iterations,
the results converged. The systematic
uncertainties were estimated via MonteCarlo (MC) simulations
taking into account the uncertainties of the intrinsic
colours, the zero points, and the effective wavelengths.

\subsubsection{RC stars in the high-extinction group}

\begin{figure}
   \begin{center}   
   \includegraphics[scale=0.36]{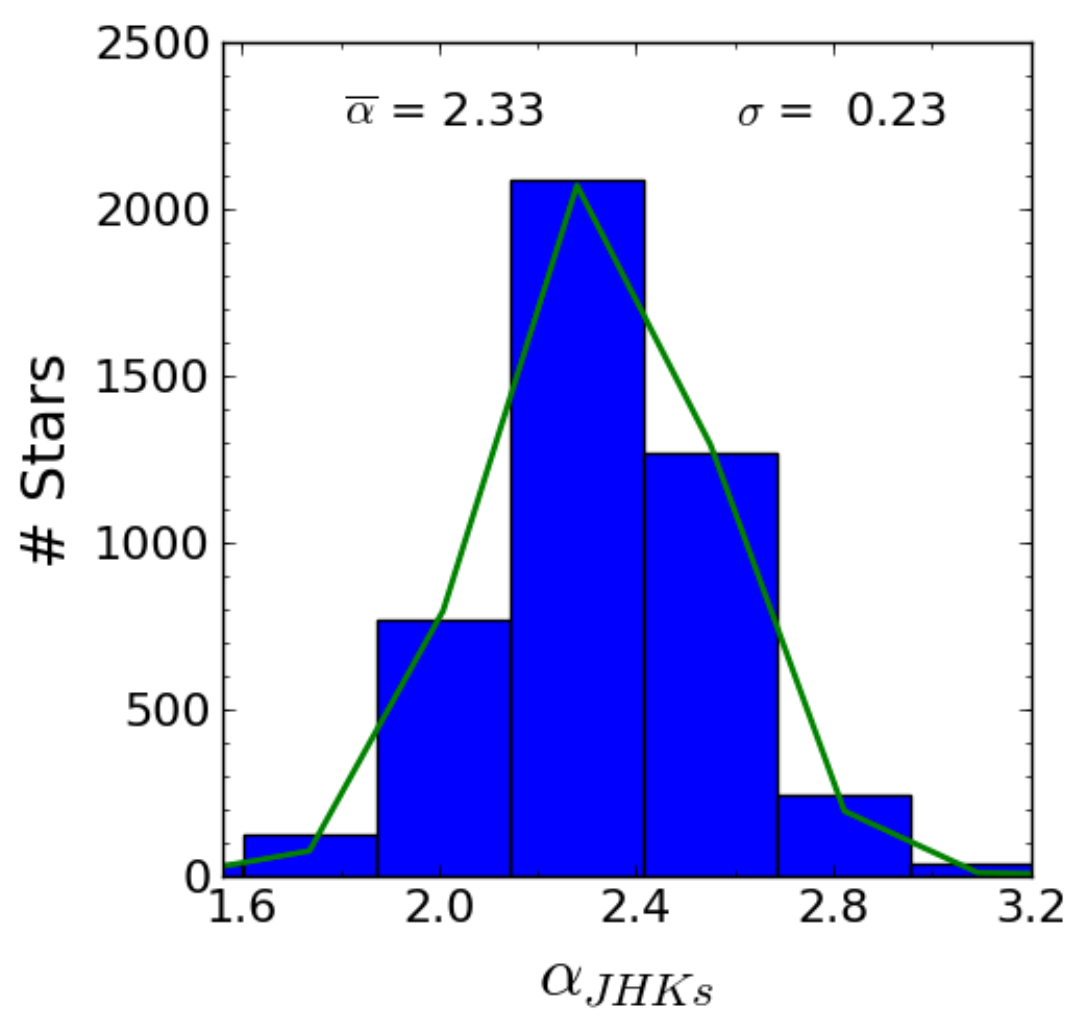}
   \caption{Distribution of the extinction index for the RC stars in the high-extinction group ($5.2<J-K_s<6$) computed using Eq. \ref{individual}. The green line shows a Gaussian fit, with the mean and standard deviation indicated in the legend.}
   
   \label{individual_alpha}
    \end{center}   
    \end{figure}

We used the RC stars identified in the high-extinction group with $J-K_{s}>5.2$ to compute the extinction index. The resulting histogram of  $\alpha$  for all the stars is shown in Fig. \ref{individual_alpha}. We obtained a value of  $\alpha_{JHK_s}=2.33 \pm 0.17$, where the 
systematic error is the main source of uncertainty and the statistical uncertainty is negligible.

\subsubsection{Spatial distribution of $\alpha$}

    As the RC stars that we employed were well distributed across the
    field, we used this method to test again the constancy of the
    extinction index for different positions. We selected
    several thousands of random regions of $1.5 '$  radius to cover
    the entire field. We computed the mean $\alpha$ for every region
    and then studied the resulting distribution of values. This distribution can be
    described by a quasi-Gaussian distribution with a mean value of $2.35$ and a small
    standard deviation of $0.03$. This supports the notion that
    $\alpha$ can be considered independent from the position in the
    field.

\subsubsection{Extinction towards dark clouds}
\label{individual_all}

\begin{figure}
   \begin{center}   
   \includegraphics[scale=0.36]{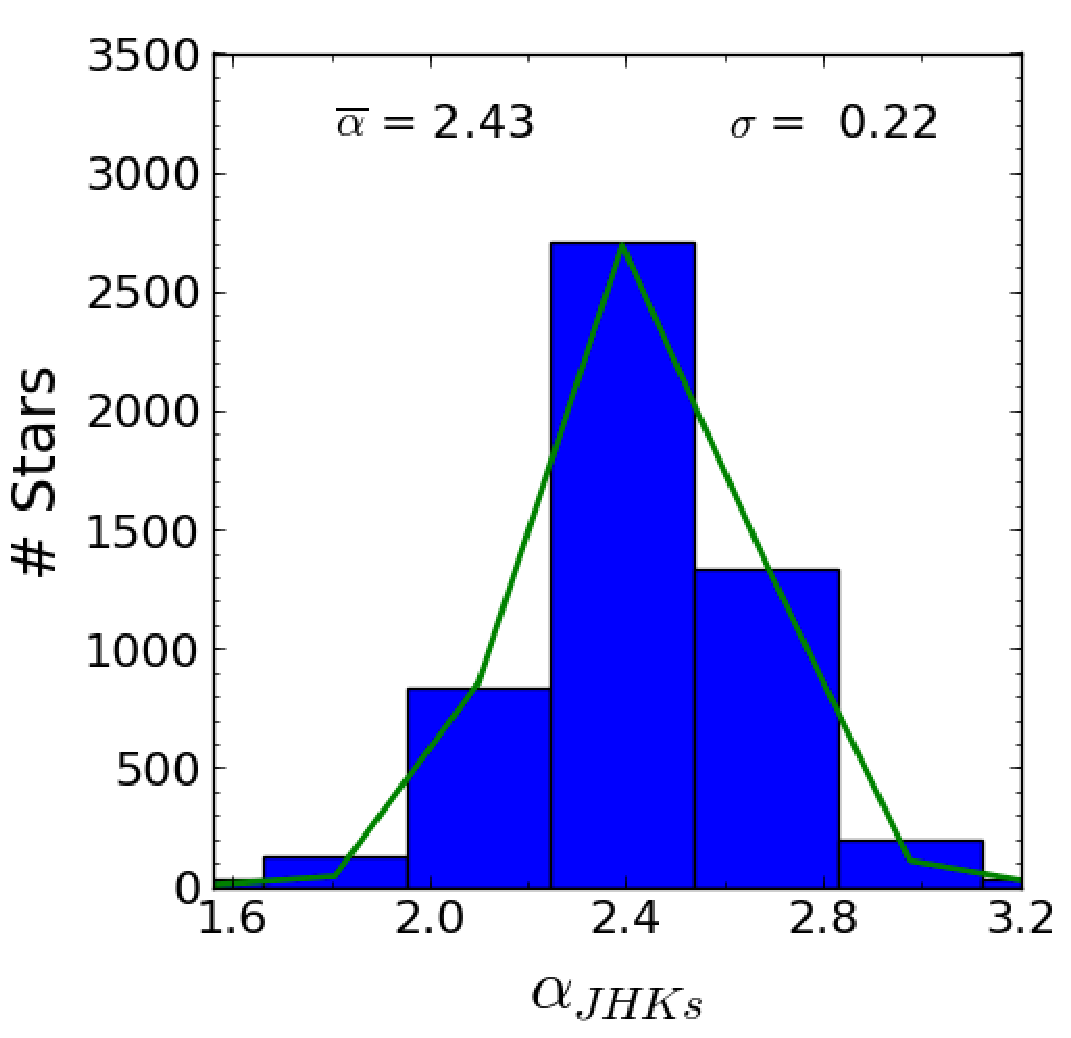}
   \caption{Distribution of the extinction index for the RC stars in the low-extinction group ($4.0<J-K_s<5.2$) computed using Eq. \ref{individual}. The green line shows a Gaussian fit, with the mean and standard deviation indicated in the legend.}
   \label{individual_alpha_dark}
    \end{center}   
    \end{figure}

Since this method computes the extinction index based on individual
stars, it is appropriate to study the stars that appear in front of
dark clouds. To do that, we used the RC stars with $J-K_{s}<5.2$ (the
ones on the left part of the red line in Fig.\,\ref{density}) and
assumed the extinction derived for this group in Section\,\ref{low_ext}.  We
obtained a value of $\alpha=2.43\pm 0.22$, where the statistical uncertainty given by the error of the mean is negligible and the systematics uncertainties are the main source of error.
Figure\,\ref{individual_alpha_dark} shows the corresponding
distribution. The slightly higher extinction index for the low-extinction stellar group can be explained by the assumption of a constant $A_{1.61}$ that suffers from systematic uncertainties. That was taken into account for the error estimation and both values agree within their uncertainties.

\subsection{Colour-colour diagram}

Studying the extinction index using a colour-colour diagram (CCD)
removes the distance effects and arranges the RC stars in a line that
follows the reddening vector. For this method, we assume that the
extinction curve does not depend on extinction, which is supported by
the tests in the preceding sections.  By using RC stars across a broad
range of extinction, a wider colour range can be used that reduces
the uncertainty of the fit. Therefore, we selected RC stars with $H-K_s\in [1.4,2.0]$ and $J-H\in [2.8,4.0]$. To compute the extinction index, we used
the slope of the RC stars's distribution in a $J-H $ versus\,$ H-K_{s}$
CCD. We divided the cloud of points in the CCD into bins of $0.03$\,mag
width on the x-axis. In every bin, we used a Gaussian fit to estimate
the corresponding density peak on the y-axis. The slope of the RC line
and its uncertainty were subsequently computed using a Jackknife resampling
method. Finally, we applied Eq.\,\ref{individual}  to calculate the
value of $\alpha_{JHK_s}$. To compute the effective wavelengths, which depend on
the absolute value of extinction, we assumed the value obtained with the Gaussian
fit in Fig.\,\ref{all_grid}.

\begin{figure}
   \begin{center}   
   \includegraphics[scale=0.3]{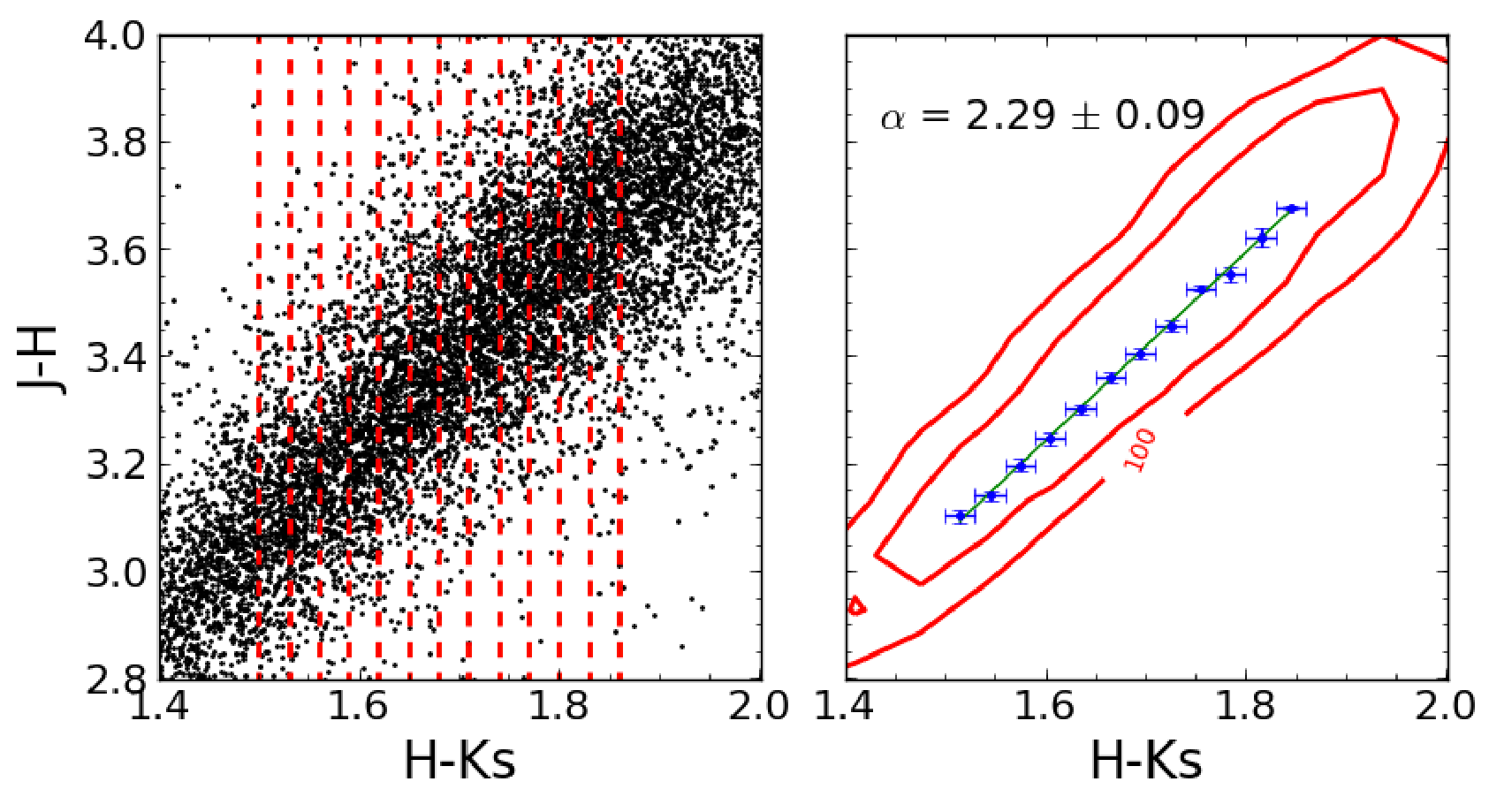}
   \caption{Calculation of $\alpha_{JHK_s}$ using the distribution of the RC stars in the CCD. Left panel shows the cloud of points and the bins used to computed the slope.  Right panel depicts the obtained points using the bins. The green line is the best fit and red contours depict the density distribution of the cloud of points.}
   
   \label{color_color}
    \end{center}   
    \end{figure}

The final result was obtained after reaching convergence through
several iterations that updated the values of $\lambda_{i}$ and
$\alpha_{JHK_s}$.  We obtained
$\alpha_{JHK_s} = 2.23 \pm 0.09$, where the uncertainty takes into account the formal
uncertainty of the fit and the error due to the effective uncertainty of the wavelength. If we omitted the bluest points, where the
number of stars was lower, and repeated the fit, then we obtained a
value of $\alpha_{JHK_s} = 2.29 \pm 0.09$ as shown in Fig.\,\ref{color_color}. In both cases, the agreement with the other methods is good.

\subsection{Obtaining $\alpha$ using known late-type stars}
\label{late_cool_stars}

We computed the extinction index using
late-type stars whose near-infrared $K$-band spectra, metallicities,
and temperatures are known \citep{Feldmeier-Krause:2017kq}. Those
stars are distributed in the central 4 $pc^2$ of the Milky Way nuclear
star cluster, corresponding to the central region of our catalogue. We
cross-identified those stars with our catalogue and excluded stars
with a photometric uncertainty $<0.05$ in any single band. We also
excluded variable stars. For the latter purpose, we compared our D15
HAWK-I $K_{s}$ -band photometry with
the photometry from the D13 HAWK-I data. In that way we excluded
several tens of stars that are possible variable candidates. Finally,
we found 367 stars for the subsequent analysis.

\subsubsection{Variable extinction}
\label{cold_var}

Stars whose stellar type is known let us study in detail the dependance of the extinction index on the wavelength. We used a slightly modified version of the grid method described in Section \ref{grid} to overcome the unknown radius of each used star. In this case, we defined a new $\chi^2 = \sum(colour_{measured}-colour_{model})^2/\sigma_{colour}^2$. With this approach, we do not need to know either the distance to the star or its radius. 

For each star we assumed the appropriate stellar atmosphere model.  We used Kurucz models because of their wide range of metallicities and
temperatures, which was necessary to analyse properly the data. We
varied the metallicity in steps of 0.5 dex from -1 to 1
dex. The temperature of the models was  3500\,K, 4000\,K, and 4500\,K, consistent with the uncertainty in temperature for
each star ($\sim 200$\,K in average). Because the lower limit for the
model temperatures was 3500\,K, we deleted ten stars that were not
covered by any model.

\begin{figure}
   \begin{center}   
   \includegraphics[scale=0.3]{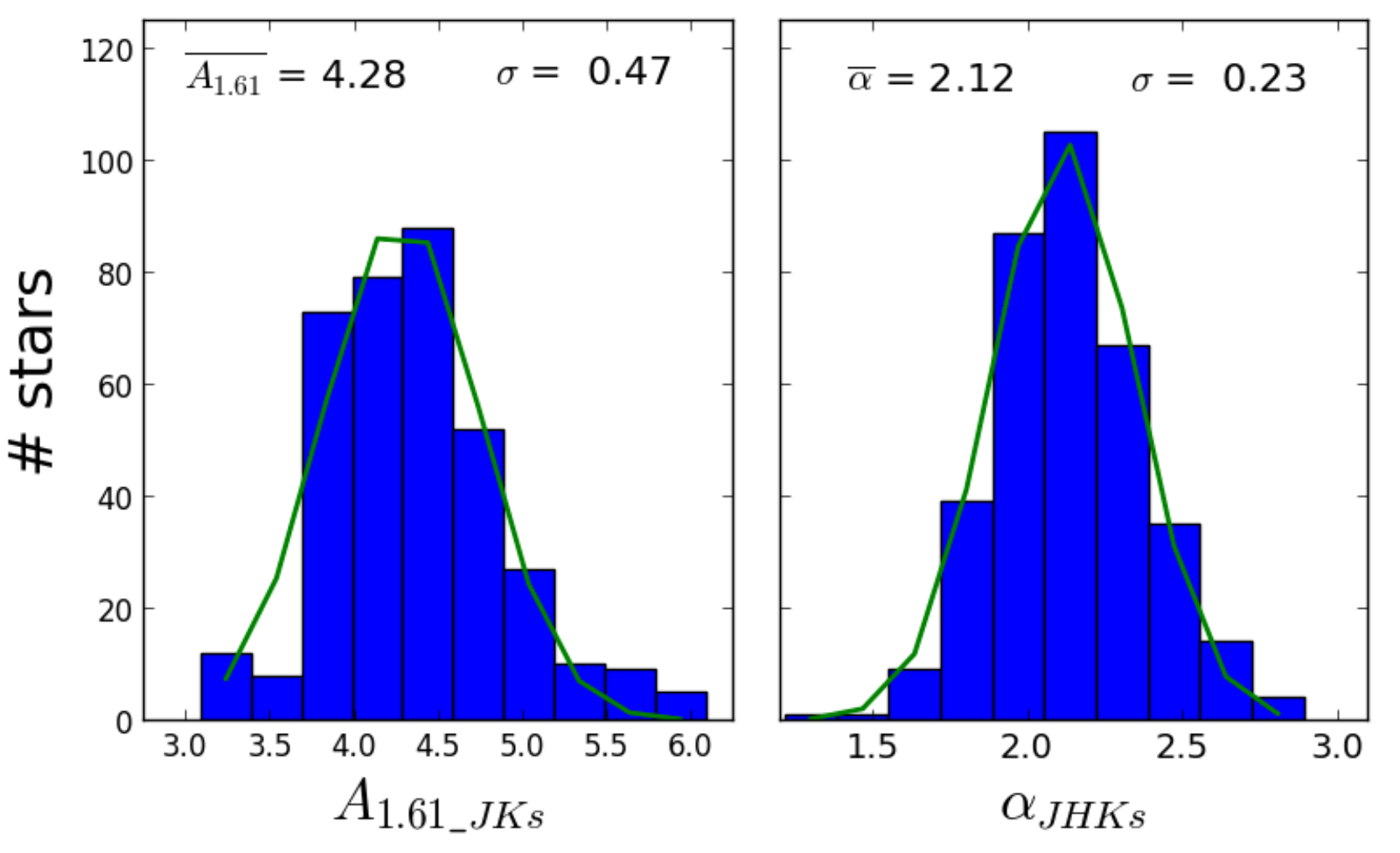}
   \caption{Left panel: $A_{1.61}$ distribution computed individually for each spectroscopically studied late-type star. Right panel: $\alpha$ estimated using the fixed extinction method for the same stars. The green line shows a Gaussian fit,
     with the mean and standard deviation indicated in the legend.}
   
   \label{late_type}
    \end{center}   
    \end{figure}

\begin{figure}
   \begin{center}   
   \includegraphics[scale=0.35]{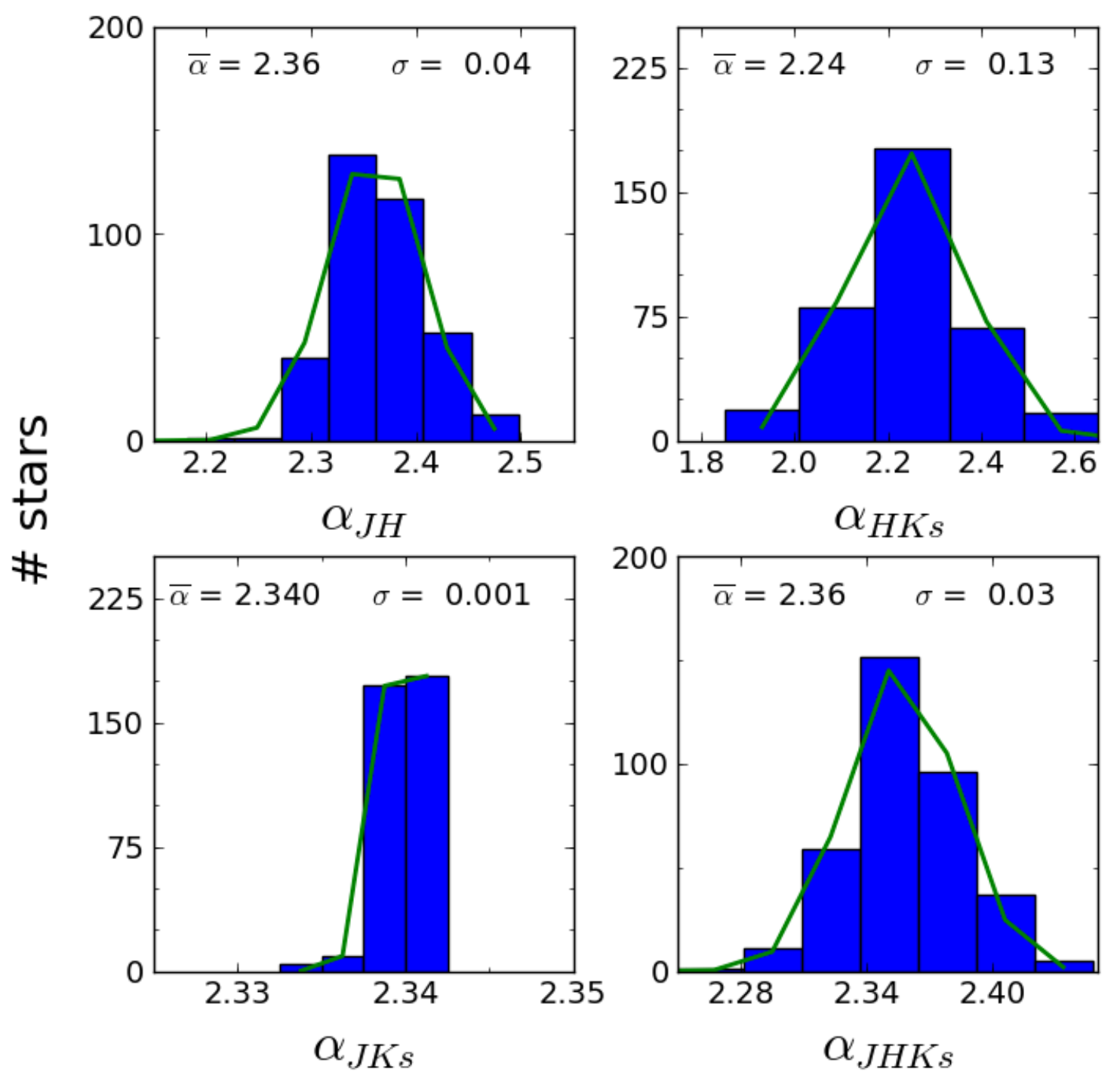}
   \caption{Histograms of $\alpha$ computed with the modified grid
     method for known late-type stars. Gaussian fits are overplotted as green lines, with the mean and standard deviations annotated in the plots.}
   
   \label{late_type_var}
    \end{center}   
    \end{figure}

Because this method has only one known variable, the colour, we cannot compute simultaneously the extinction and the extinction index. Thus, we computed the individual extinction of each star using the measured colour $J-K_s$. We minimised the difference between the data and the theoretical colour obtained using the appropriate Kurucz model (taking into account the metallicity and the temperature) and a grid of extinctions, $A_{1.61}$ , with a step of 0.01 mag. For that, we used the extinction index derived in Sect. \ref{grid}. We obtained a quasi-Gaussian distribution for the extinction that is shown in the left panel of Fig. \ref{late_type}. The systematic uncertainty dominates the errors whereas the statistical uncertainty is negligible. We computed the systematics taking into account the variation of the extinction index (using the uncertainties computed in Sect. \ref{grid}), the systematic uncertainty of the ZP, and different amounts of precipitable water vapour (1.0, 1.6 and 3.0\,mm). We ended up with a mean extinction of $A_{1.61}= 4.28 \pm 0.18$. We used $J-K_s$ because to convert colour into extinction, we need to assume a value of $\alpha$ to compute the grid of individual extinctions and, for that wavelength range, the value is similar to the one obtained assuming a constant extinction index for all the three bands (see Sect. \ref{grid}) and we do not need to assume different values for the ranges $J-H$ and $H-K_s$. In this way, we computed the extinction for each star and fixed it to compute the extinction index. Figure \ref{late_type_var} and Table \ref{grid_table_cold_hot} summarise the obtained results. The statistical uncertainties, given by the error of the mean, are negligible, and the systematic uncertainties are the main source of error. To compute them, we took into account the systematics introduced by the zero points, the initial $\alpha$ used to translate the colour $J-K_s$ into extinction, and the humidity of the atmosphere.

\begin{table}
\begin{center}
\caption{Values of $\alpha$ obtained with the modified grid
  method for known late-type stars.}
\label{grid_table_cold_hot} 
\begin{tabular}{ccc}
 &  & \tabularnewline
\hline 
\hline 
&Bands & $\alpha$ \tabularnewline
\hline 
&$JH$ & $2.36 \pm 0.08$   \tabularnewline
Known&$HK_s$ & $2.24\pm 0.16$ \tabularnewline
late-type&$JK_s$ & $2.34 \pm 0.09$  \tabularnewline
&$JHK_s$ & $2.35 \pm 0.08$  \tabularnewline
\hline 
&$JH$ & $2.36 \pm 0.08$  \tabularnewline
Known&$HK_s$ & $2.24\pm 0.13$ \tabularnewline
early-type&$JK_s$ & $2.34 \pm 0.09$ \tabularnewline
&$JHK_s$ & $2.36 \pm 0.08$ \tabularnewline
\hline 
\end{tabular}
\vspace{1cm}
\end{center}
\end{table}

\subsubsection{Fixed extinction}

We employed the same approach described above
(Sect. \ref{individual_method}). In this case, we used the fixed extinction given by the Gaussian fit shown in Fig \ref{late_type} (left panel). To obtain the final $\alpha_{JHK_s}$, we used an iterative approach updating the value of the extinction index obtained in every step until reaching convergence. To start the iterations we used the extinction index derived in Sect. \ref{2gaussians}. The resulting distribution of $\alpha_{JHK_s}$ is shown in Fig.\,\ref{late_type} (right panel). The estimation of the systematic uncertainty was carried out considering the systematics of the ZP, the possible variation of the fixed extinction, and different amounts of precipitable water vapour (1.0, 1.6 and 3.0\,mm).  The final result was $\alpha_{JHK_s}=2.12\pm 0.14$, which is in agreement with all the previous estimates. The slightly lower extinction index obtained can be explained by the assumption of a constant $A_{1.61}$, which suffers from systematic uncertainties. That was taken into account in the uncertainty estimation.

\subsection{Computing the extinction index using early-type stars}

We used again the same methods described in
Sect. \ref{late_cool_stars} to compute the extinction index towards
known hot, massive stars near Sgr\,A* \citep{Do:2009tg}. We used the
D13 data, which cover the central region far better than the D15
data. We excluded known Wolf-Rayet stars
\citep{Paumard:2006xd,Do:2009tg,Feldmeier-Krause:2015} because they
are frequently dusty and therefore intrinsically reddened. To avoid
spurious identifications because of the high crowding of the region,
we only used stars with $11.2\leq K_{s}\leq 13$. We used the published NACO $K_{s}$ magnitudes (approximately equivalent to the HAWK-I $K_s$ band) and
compared them with our HAWK-I data applying a $3-\sigma$ exclusion
criterion to delete any possible variable stars. Finally, we excluded
stars with photometric uncertainties larger than $0.05$\,mag in any single band. We ended up with 23 accepted stars for the analysis.

\subsubsection{Variable extinction}

\begin{figure}
   \begin{center}   
   \includegraphics[scale=0.3]{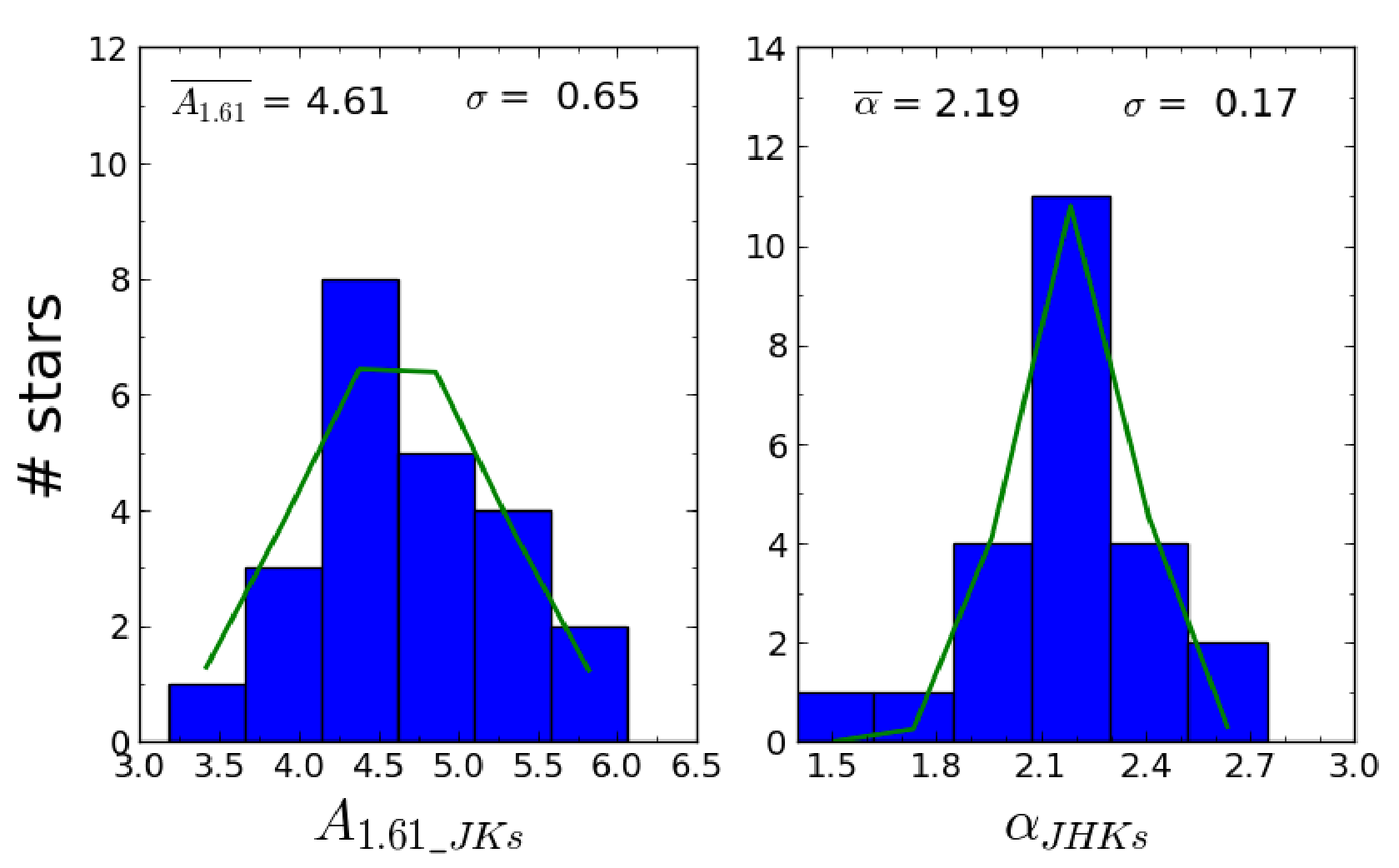}
   \caption{Left panel: $A_{1.61}$ distribution computed individually for each spectroscopically studied late-type star. Right panel: $\alpha$ estimated using the fixed extinction method for the same stars. The green line shows a Gaussian fit,
     with the mean and standard deviation indicated in the legend.}
   
   \label{young_stars}
    \end{center}   
    \end{figure}

\begin{figure}
   \begin{center}   
   \includegraphics[scale=0.35]{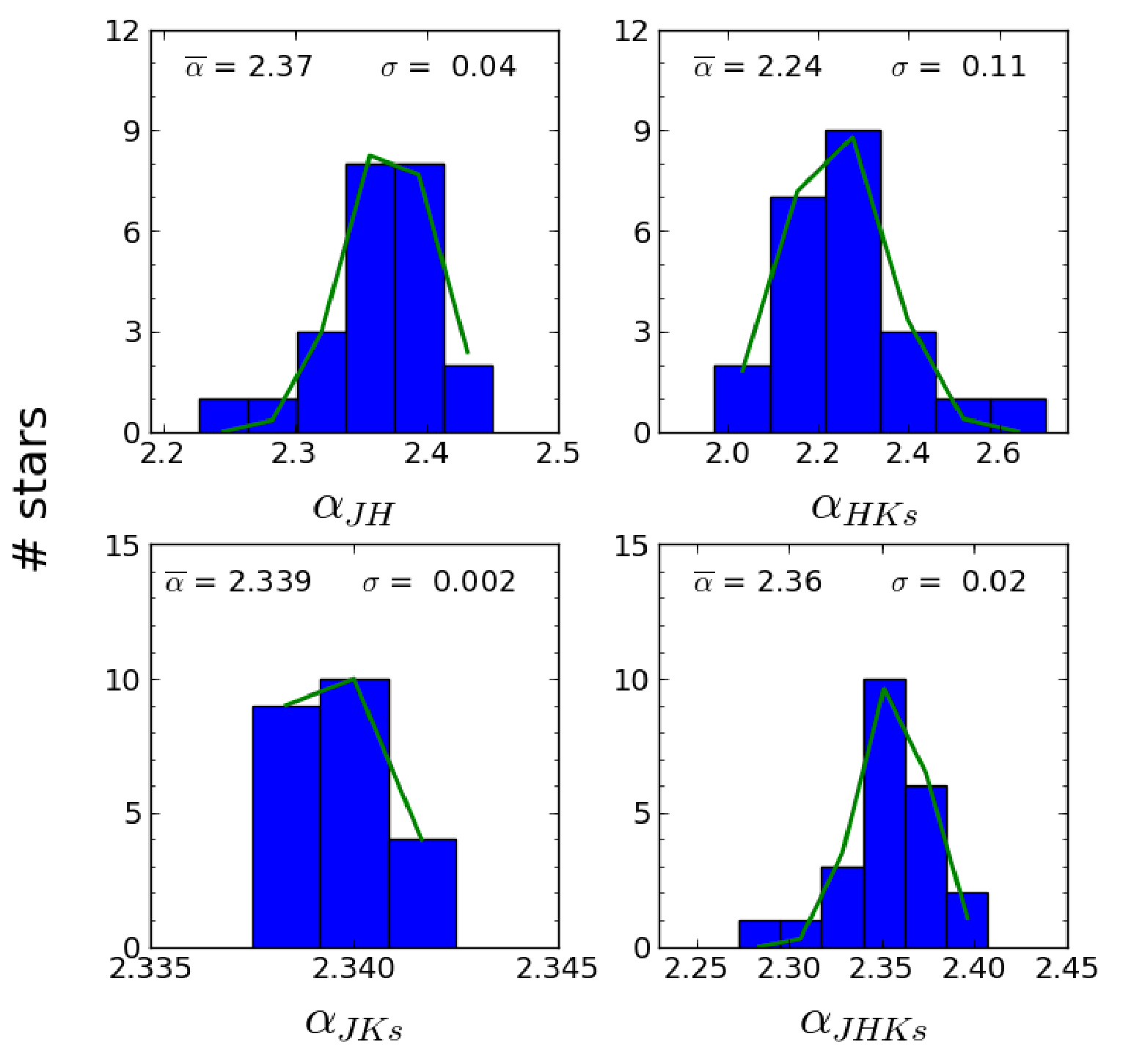}
   \caption{Histograms of $\alpha$ computed with the modified grid
     method for known early-type stars. Gaussian fits are overplotted as green lines, with the mean and standard deviations annotated in the plots.}
   
   \label{early_type_var}
    \end{center}   
    \end{figure}

As described in Sect. \ref{cold_var}, we applied the modified grid method and computed the individual extinction to each star to analyse the variation of the extinction index with the wavelength. We used a 30000 $K$ model, a solar metallicity, a $\log g = 4.0,$ and a humidity of 1.6 mm of precipitable water vapour. The uncertainties were estimated using the error of the mean of the quasi-Gaussian distribution and varying the parameters previously described to obtain the systematics. The final mean value is $A_{1.61\_JK_s}=4.61 \pm 0.20$, where 0.13 and 0.16 correspond to the statistical and systematic uncertainties. Figure \ref{young_stars} (left panel) shows the obtained results. Figure \ref{early_type_var} and Table \ref{grid_table_cold_hot} present the obtained results for the modified grid method.

We found again the variation in the extinction index that we already noticed in Sect. \ref{2gaussians}. This supports the evidence of having a steeper extinction index between $J$ and $H$ than between $H$ and $K_s$. However, the difference between the extinction indices and the uncertainties that we found is not enough to clearly distinguish two different values. Therefore, we estimate that, in spite of being different, their close values make necessary a deeper analysis with better spectral resolution or wide wavelength coverage, to clearly distinguish them. Within the limits of the current study, we assume that a constant extinction index is enough to describe the extinction curve between the analysed bands.

\subsubsection{Fixed extinction}

A Kurucz model with a temperature of 30000\,K and solar metallicity was
taken to compute $\alpha_{JHK_s}$. We used the mean value obtained above and presented in Fig. \ref{young_stars} (left panel). The result is shown in Fig. \ref{young_stars} (right panel). The systematic
uncertainty dominates the error and was computed using MC simulations considering the
uncertainties in the ZP calibration of all three bands, the effective wavelength, and the
intrinsic colour of $J-H$ and $H-K_s$. The final value was
$\alpha_{JHK_s} = 2.19\pm 0.13$, which is consistent with the values
determined by the other methods.

\subsection{Final extinction index value and discussion}

Given our results from all the different methods to estimate the
extinction index on our data, we can conclude that all the derived values
of $\alpha$ agree within their uncertainties and that there is no
evidence - within the limits of our study - for any variation of $\alpha$
 with position, or absolute value of interstellar extinction. On the other hand, we observed a small dependence with the wavelength when we considered different values for $\alpha_{JH}$ and $\alpha_{HK_s}$. However, a constant extinction index, $\alpha_{JHK_s}$ , seems to be sufficient to describe the extinction curve. We average all the values obtained with the different methods explained above and obtain a final value $\alpha_{JHK_s} = 2.30 \pm 0.08$, where the uncertainty is given by the  standard deviation of the measurements, which takes into account the dispersion of the values obtained using the different methods.\\

\begin{figure}
   \begin{center}   
   \includegraphics[scale=0.31]{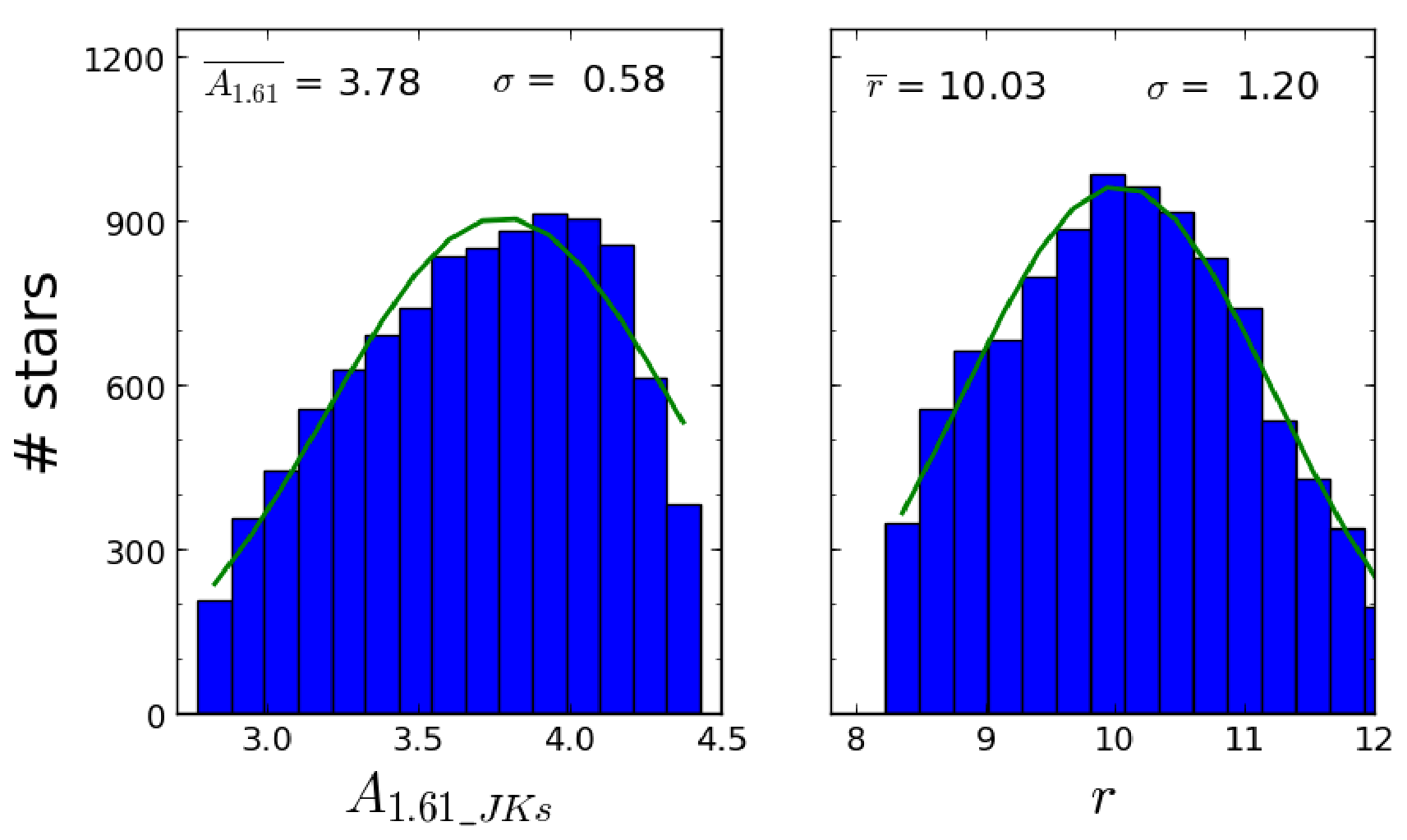}
   \caption{Left panel: $A_{1.61\_JK_s}$ distribution computed for all the RC stars. Right panel: radii distribution of the RC stars (in solar radii). The green line shows a Gaussian fit, with mean and standard deviation indicated in the legend.}
   
   \label{check}
    \end{center}   
    \end{figure}

To check the reliability of the obtained value for the extinction index, we used all the RC stars identified in Fig. \ref{CMD} and computed their radii using the individual extinction for each star and the derived extinction index. We used the colour $J-K_s$ to calculate the individual extinctions employing the grid approach described in Sect. \ref{cold_var}. Then, we computed the corresponding radius for each star assuming the final extinction index, the GC distance (8.0 kpc, \citep[]{Malkin:2013fk}) and a Kurucz model for Vega to convert the fluxes into magnitudes. We obtained the radius by comparing the obtained value with the measured $K_s$ of each star. Figure \ref{check} depicts the obtained individual extinctions (left panel) and the derived radius (right panel). We obtained a value of $A_{1.61} = 3.78 \pm 0.15$ and $r = 10.03 \pm 0.57$ $R_\odot$, where the error is dominated by the systematic uncertainty. The comparison between the obtained extinction and the one computed in Sect. \ref{grid}, and shown in Fig. \ref{all_grid}, is consistent. Moreover, we have obtained a mean radius for the RC stars that agrees perfectly with the standard value of $10.0 \pm 0.5$ R$_\odot$ \citep{Chaplin:2013kx,Girardi:2016fk}. Therefore, we conclude that the computed values for the extinction and the extinction index are consistent and let us calculate the radius of the RC stars obtaining an accurate value.

\section{A high angular resolution extinction map}

 To produce an $A_{K_s}$ extinction map, we removed all the foreground
 population and kept only the RC stars indicated by the parallelogram in Fig. \ref{CMD} and detected in $H$ and $K_s$ bands. The  observed colours were converted into extinction using the colour $H-K_s$ (as we have far more stars detected in $H$ and $K_s$ than in $J$) and the following equation:

 \begin{equation}
\label{ext_map}
ext = \frac{H-K_s-(H-K_s)_0}{\left(\frac{\lambda_H}{\lambda_{K_s}}\right)^{-\alpha}-1} \hspace{0.5cm},
\end{equation}

\noindent where the subindex 0 refers to the intrinsic colour and $\lambda_i$ are the effective wavelengths.

To save computational time, we defined a pixel scale of 0.5''/pixel  for the extinction map. For every pixel, we computed
the extinction using the colour of the ten closest stars, limiting the
maximum distance to 12'' from its centre. If less than ten stars were
found for a pixel, we did not assign any extinction value to avoid obtaining strongly biased values due to too distant stars. To take into account the different distances of the stars to every pixel, we
computed the colour using an inverse distance weight (IDW)
method. Namely, we weighted every star based on the distance to the
corresponding pixel following the expression:
\begin{equation}
v = \frac{\sum^n_{i=1}\frac{1}{d^p_i}v_i}{\sum^n_{i=1}\frac{1}{d^p_i}} \hspace{0.5cm},
\end{equation}
where $v$ is the magnitude to be computed in the target pixel, $d$ is
the distance to that pixel, $v_i$ are the known values, and $p$ is the
weight factor. We explored different values for $p$ and ended up with $p=0.25$ as the best estimate to avoid giving too much weight to the star closest to the pixel centre.

\begin{figure}
      \begin{center}
   \includegraphics[scale=0.37]{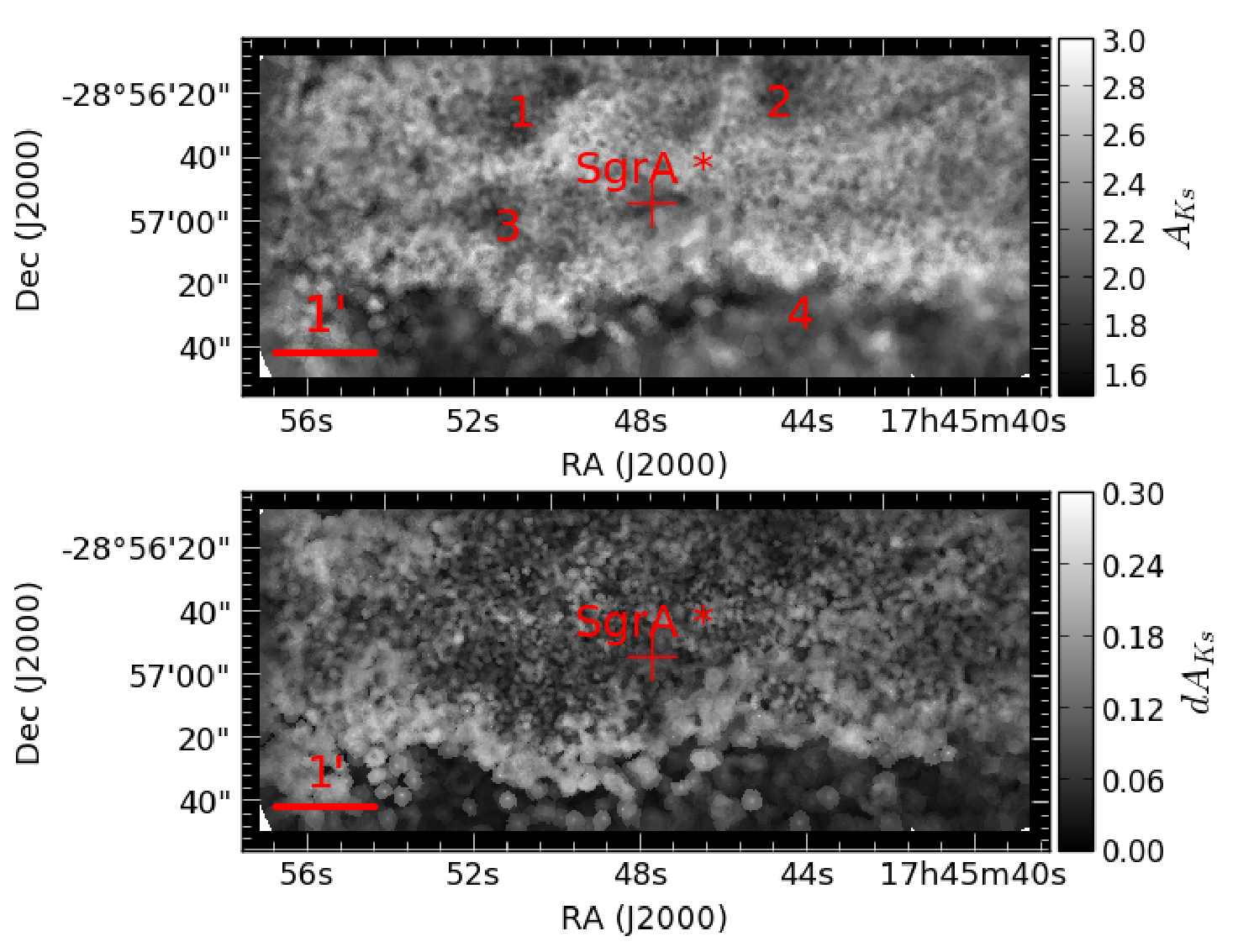}
   \caption{Upper panel: Extinction map $A_{K_s}$ obtained using all
     the RC stars shown in Fig.\,\ref{CMD}. Lower panel: Corresponding
     uncertainty map. The numbers in red indicate regions to be compared between the different extinction maps (see main text).}
   \label{ext_RC}
       \end{center}
    \end{figure}

The uncertainty of the extinction map was computed estimating the colour
uncertainty of every pixel, using a Jackknife algorithm in the calculation of the IDW. The resulting extinction map and its corresponding uncertainty map are shown in Fig. \ref{ext_RC}. The mean uncertainty is $\sim 5\%$. Besides, we estimate a systematic error of $\sim 5$\%, which takes into account the uncertainties of the effective wavelengths and the extinction index.

\begin{figure}
   \begin{center}   
   \includegraphics[scale=0.28]{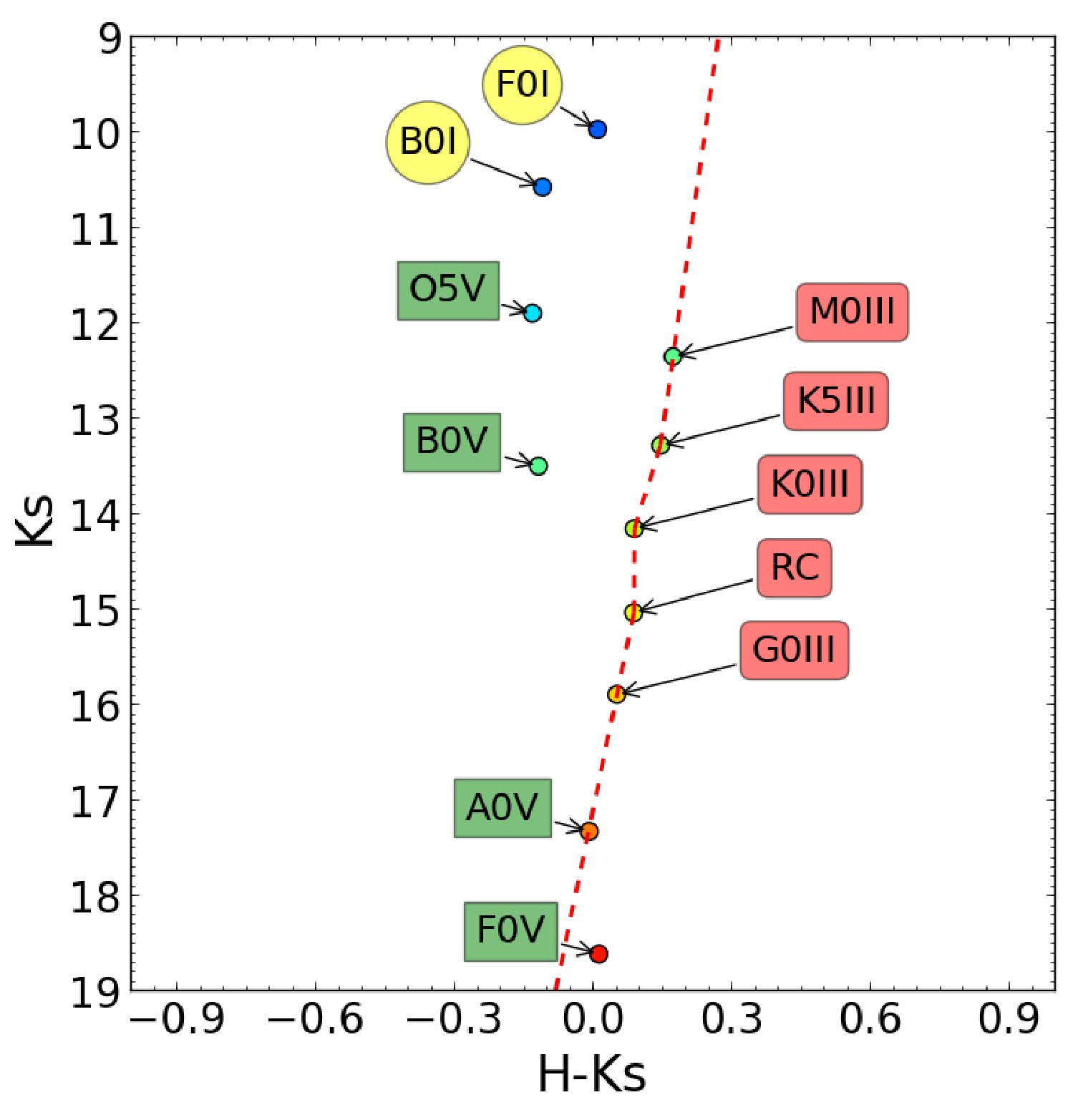}
   \caption{Reddened observed magnitude (assuming a distance modulus
     of $14.51$ and a mean extinction $A_{1.61}=3.67$) versus intrinsic
     colour for several types of stars computed using Kurucz models. The
     red dashed line indicates the intrinsic colours used for the stars
     selected to compute the extinction map.}
   \label{function}
    \end{center}   
    \end{figure}

\begin{figure}
      \begin{center}
   \includegraphics[scale=0.37]{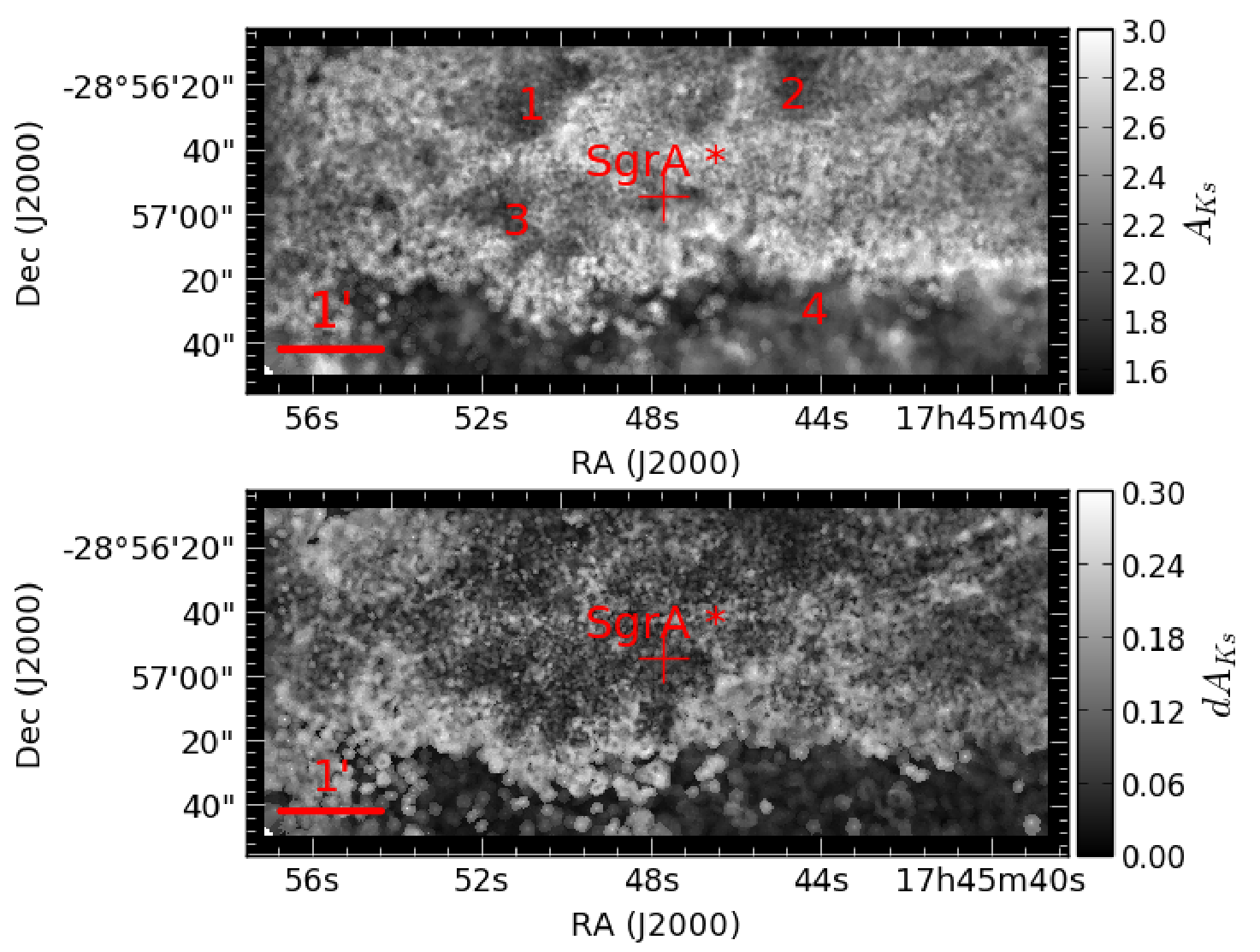}
   \caption{Upper panel: Extinction map $A_{K_s}$ obtained using stars
     with $12<K_{s}<17$ and $1.4< H-K_s < 3.0$. Lower
     panel: Corresponding uncertainty map. The numbers in red indicate regions to be compared between the different extinction maps (see main text).}
   \label{ext_all}
       \end{center}
    \end{figure}

    Due to the fact that the extinction can vary on scales of arc
    seconds, we tried to improve the extinction map using not only the
    RC stars, but all stars with $K_s$ between 12 and 17, and $H-K_s$ between 1.4
    and 3.0. In this way, we almost duplicate the number of selected
    stars, improving significantly the angular resolution of the
    map. To compute the intrinsic colour of every star, we assumed that
    the majority of the used stars can be considered as giants. We
    calculated the $K_s$ reddened magnitude of several Kurucz models of
    giants (assuming an average extinction of 3.67 magnitudes at 1.61
    $\mu$m, see Sect. \ref{grid}) and their intrinsic colour. We
    interpolated the points to build a smooth function that gives the
    intrinsic colour based on the measured $K_s$ magnitude (see
    Fig.\,\ref{function}). The resulting extinction map and its
    corresponding uncertainty map are shown in Fig.\,\ref{ext_all}. It
    can be seen that it is very similar to Fig. \ref{ext_RC}, but
    with a higher angular resolution. The statistical and systematic uncertainties are similar as well.

\begin{figure*}
      \begin{center}
   \includegraphics[scale=0.6]{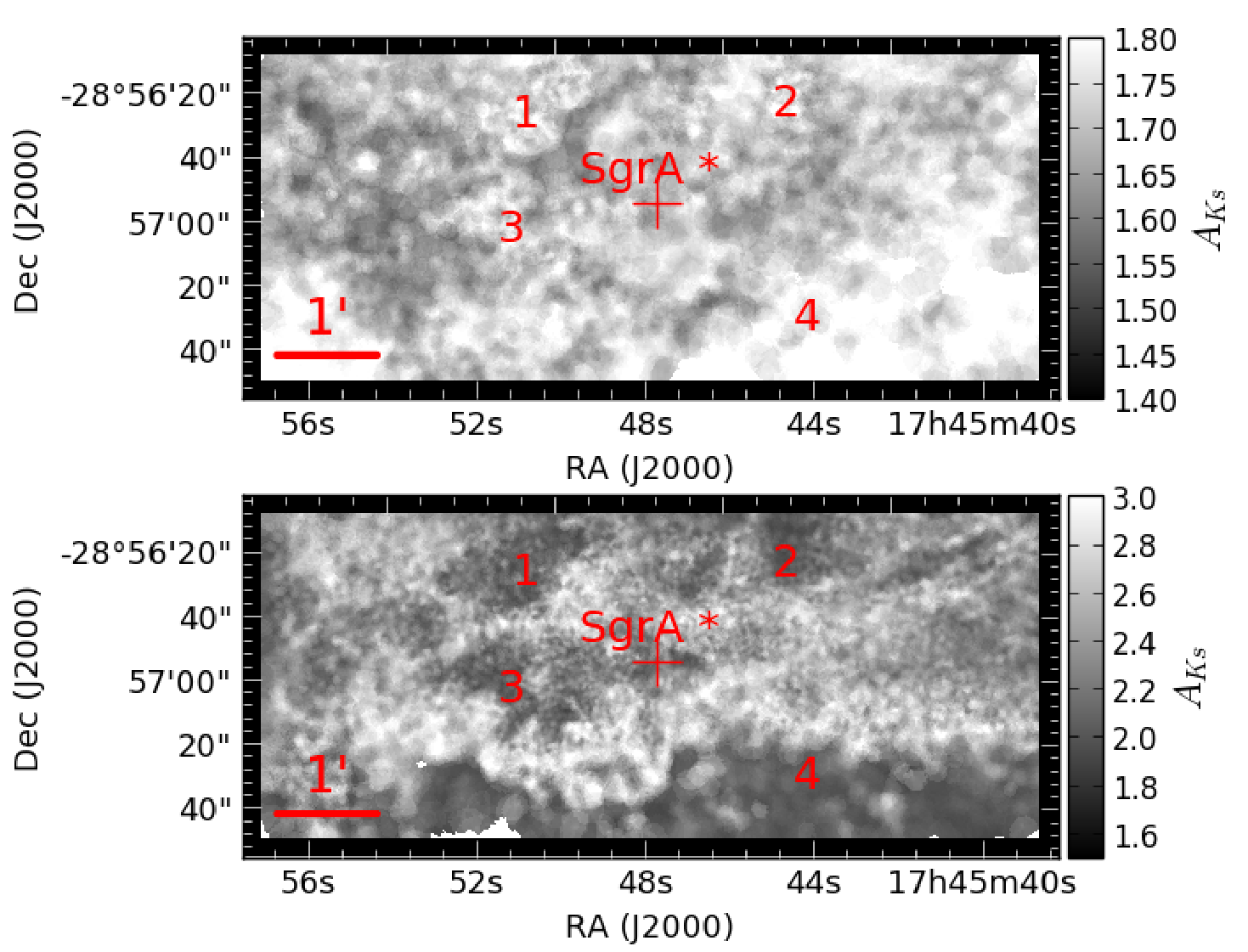}
   \caption{Upper panel: Extinction map for stars with observed
colours in the interval $1.4< H-K_{s}<1.7$. Lower panel:
Extinction map for stars with observed colours in the interval
$1.7< H-K_{s}<3$. The numbers in red indicate regions to be compared between the different extinction maps (see main text).}
   \label{g1g2}
       \end{center}
    \end{figure*}

    With this approach, we generated, in addition, two more extinction
    maps taking into account the two groups of stars that we
    identified in Sect.\,\ref{RC_selection}. Namely, we produced the
    first map using stars with $1.4< H-K_{s}<1.7$
    (Fig.\,\ref{g1g2}, upper panel) and the second one with stars
    between $1.7< H-K_{s}<3.0$ (Fig.\,\ref{g1g2}, lower
    panel). In this way, the first map represents the extinction
    screen toward the foreground population (excluding stars in the
    spiral arms), whereas the second one includes the interstellar
    extinction all the way toward GC stars.  Although the first map
    does show some correlation with the second map (e.g. in regions
    1, 2, 3) because it is impossible to separate the two stellar
    populations without any overlap, it shows a rather constant
    extinction, with a mean of 1.72 and a standard deviation of
     0.06. This demonstrates that this foreground screen can be
    considered to vary on very large scales. The statistical and systematic uncertainties for this first extinction map are $\sim 2$ \% and  $\sim 6$ $\%$ respectively. In this case the uncertainty is lower because we consider a narrower range in $H-K_s$ and all the stars are closer without mixing stars with different extinctions.
On the other hand, the second map contains fine structure on scales
  of just a few arcseconds, which is consistent with the extinction
  being caused by a clumpy medium close to the GC. The statistical and systematic uncertainties of the map are $\sim 4\%$ and $\sim 5$ $\%$ respectively.

\begin{figure*}
   \begin{center}   
   \includegraphics[width=\textwidth]{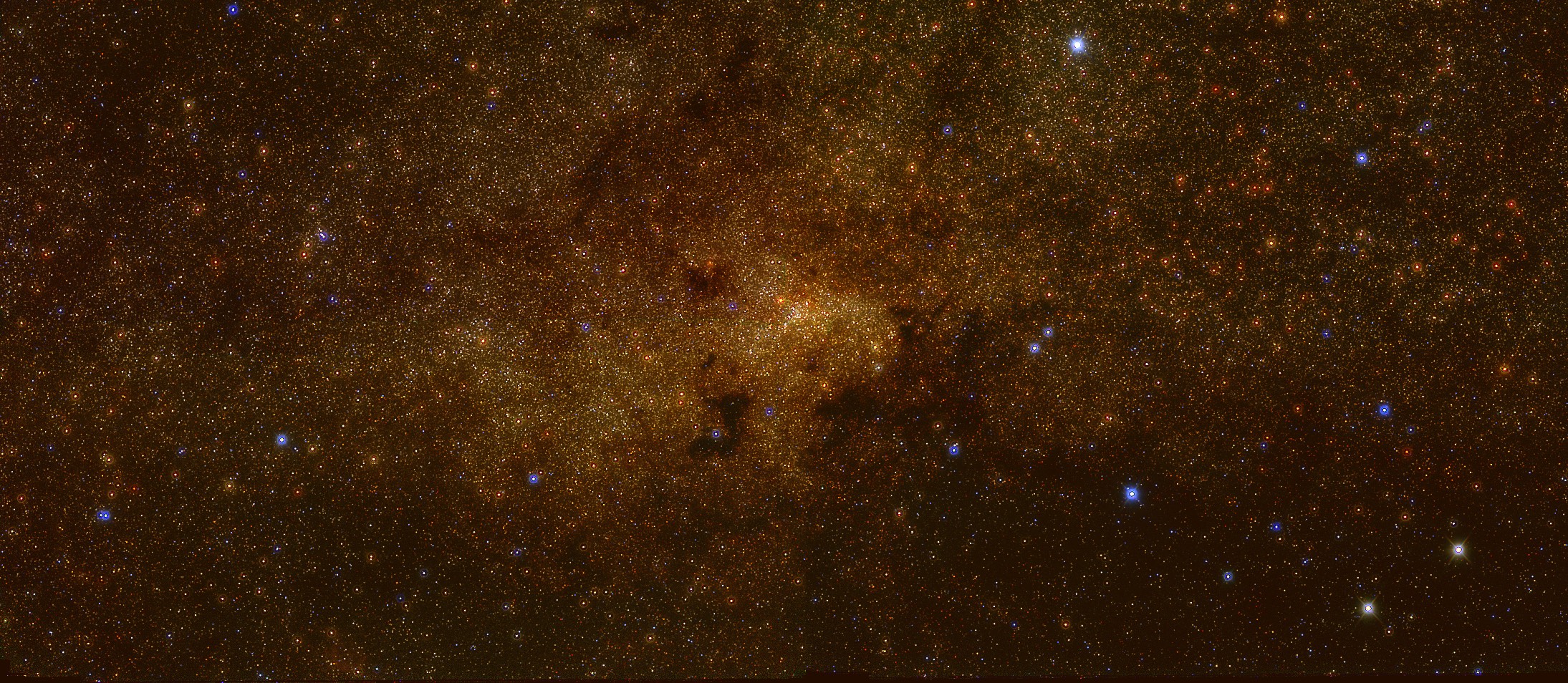}
   \caption{RGB image of Field 1 (red for $K_{s}$, green for $H$, blue
     for $J$.)}
   \label{rgb}
    \end{center}   
    \end{figure*}

    When comparing the extinction map in the lower panel of
    Fig.\,\ref{g1g2} with the $J$-band density plot
    (Fig.\,\ref{density}), we can see that the high density regions,
    marked with numbers 1, 2, and 3, correspond to low extinction as we
    expected. However, while region\,4 can be identified as a high-extinction region in the density plot, it appears as a
    low-extinction one in the lower panel of \ref{g1g2}, but as a
    relatively high-extinction region in the upper panel. We interpret
    this as evidence for a large dark cloud with very high density
    that blocks most of the light from the stars behind it. This is
    also rather evident from inspecting a $JHK_{s}$ colour image of our
    field (Fig.\,\ref{rgb}). We conclude that our method cannot produce reliable
    extinction maps for regions marked by the presence of highly
    opaque foreground clouds.

\section{Stellar populations}

\subsection{CMDs}

\begin{figure*}
   \begin{center}   
   \includegraphics[scale=0.36]{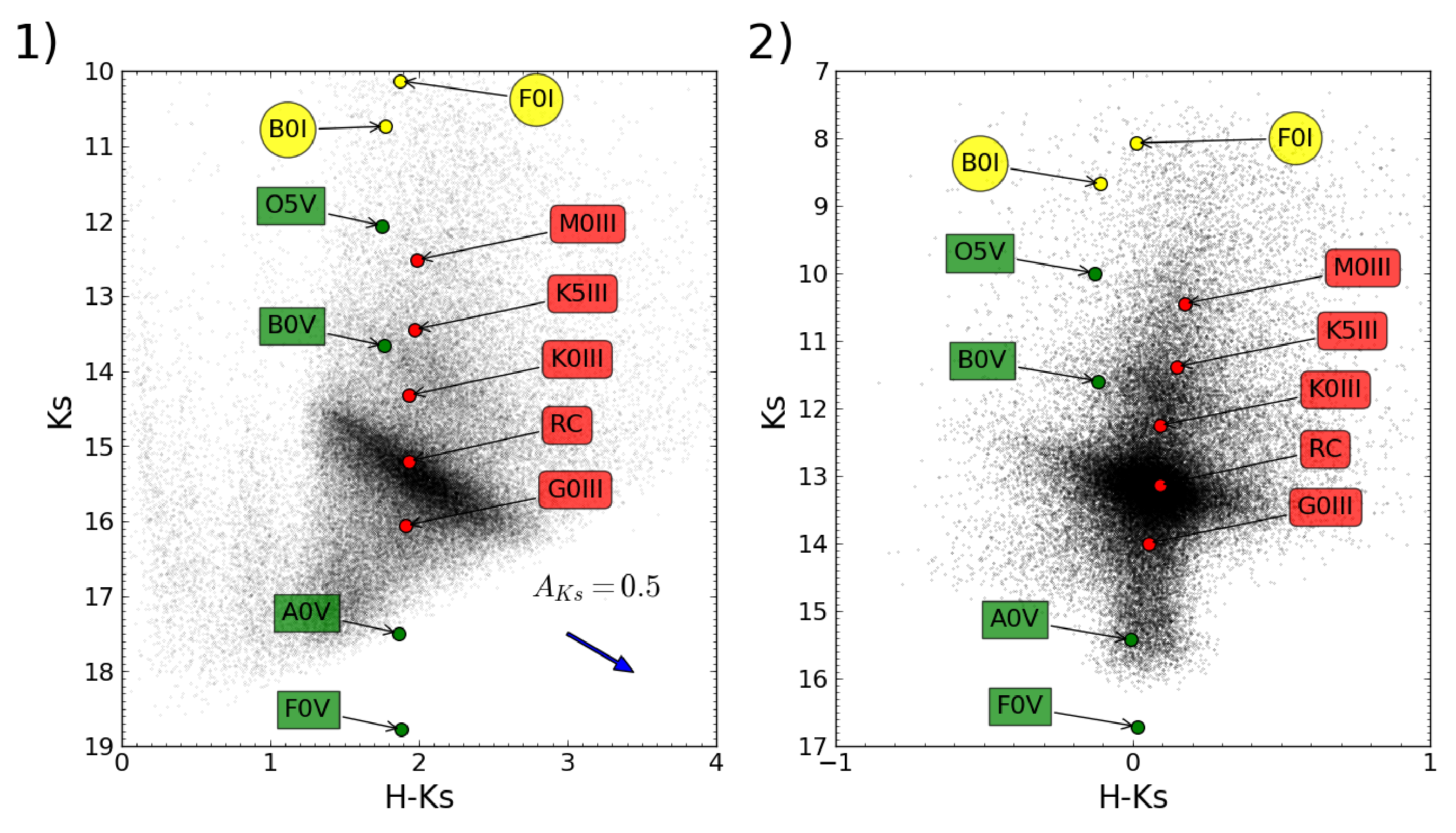}
   \caption{Panel 1 shows the CMD $K_s$ versus $H-K_s$. We have overplotted several Kurucz stellar models (assuming a mean extinction $A_{1.61}=3.97$, see Sect. \ref{grid}) to identify the expected position of those stars. Panel 2 depicts the dereddened map once we have applied the extinction maps. Both panels have different scales on the X-axis.}
   
   \label{HK_der}
    \end{center}   
    \end{figure*}

\begin{figure*}
   \begin{center}   
   \includegraphics[scale=0.36]{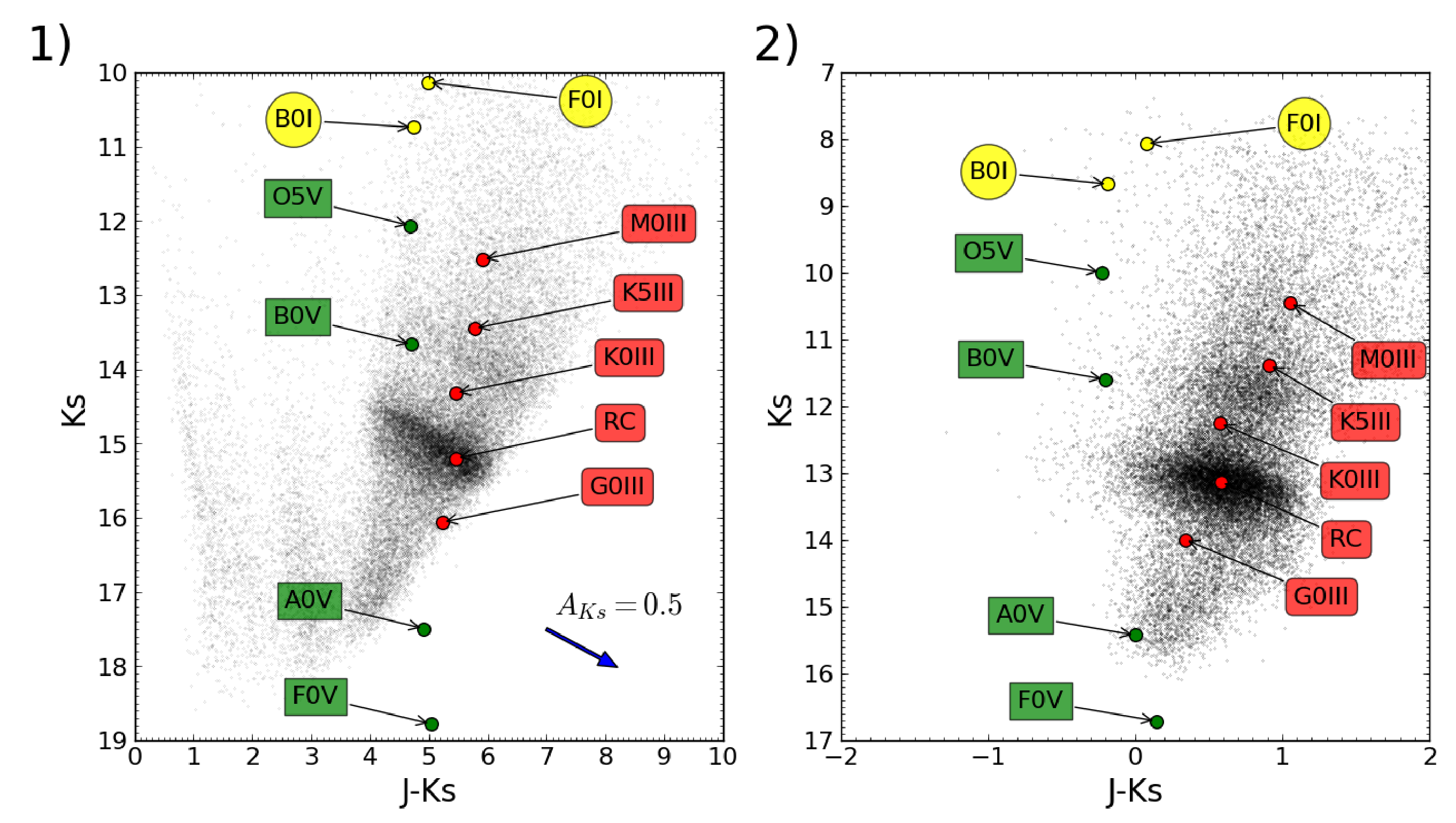}
   \caption{Panel 1 shows the CMD $K_s$ versus $J-K_s$. We have overplotted several Kurucz stellar models (assuming a mean extinction $A_{1.61}=3.97$, see Sect. \ref{grid}) to identify the expected position of those stars. Panel 2 depicts the dereddened map once we have applied the extinction maps. Both panels have different scales on the X-axis.}
   
   \label{JK_der}
    \end{center}   
    \end{figure*}

To obtain a rough idea of the stellar populations in Field\,1 of our
survey, we attempted to deredden the CMD.  We transformed the $A_{K_s}$
extinction maps into $A_H$ using the RC effective wavelength
and the extinction index of $\alpha_{JHK_s} = 2.30$, as determined by
us. We used the two extinction maps computed previously according to
the different $H-K_{s}$ colours. This enabled us to improve the dereddening
process as we considered two different maps that correspond to two
different layers in the line of sight. We used either one or the other,
depending on the $H-K_{s}$ colour of a given star. The divisions were $1.4<H-K_s<1.7$ and $1.7<H-K_s<3.0$, which is similar to the $J-K_s$ cut in Fig. \ref{HST_cut}. Figure\,\ref{HK_der}
 shows the CMDs for $K_{s}$ versus $H-K_{s}$ before and after de-reddening. 
 Analogously, we repeated the process for $J$ band but, in this case, we produced the extinction maps taking into account only stars identified in all three bands, $J$, $H,$ and $K_s$. By doing so, we obtain a consistent extinction map and avoid producing extinction maps biased towards too high extinction, as the completeness is higher in $H$ and $K_s$ than in $J$. Figure \ref{JK_der} depicts the CMD $K_s$ versus $J-K_s$ before and after de-reddening. It can be seen that the extinction correction clearly reduces the scatter in the CMDs. On the other hand, we overplotted different Kurucz stellar models. The giant sequence follows clearly
the main density ridge in both dereddened CMDs. In both cases, $K_{s}$ versus $H-K_{s}$ and $K_{s}$ versus $J-K_{s}$, the standard
deviation of the dereddened colours around the giant branch is of the order of
$\sigma=0.2$.

\subsection{CMD modelling}

We overplotted three 5 Gyr isochrones \citep{Marigo:2017aa} with different metallicities (0.1,1 and 2.5 solar metallicity) to the CMD $K_s$ versus $J-K_s$. In this way we are able to compare them with the stellar population. We can clearly see how a super-solar metallicity population fits better the data on the right hand side of the CMD. This is in particular indicated by the centring of the AGB bump. The RC bump appears to be centred at lower metallicities, but that is an effect of extinction and completeness, because the RC clump is cut off at the red end of the $JK_s$ CMD. Therefore, a range of solar metallicities around the solar metallicity can explain the observed data. This is in agreement with the low number of low metallicity stars found in the central 4 $pc^2$ in \citet{Feldmeier-Krause:2017kq}. On the other hand, the scatter of the stars in the upper part of the diagram gives us an idea of the goodness of the extinction correction $\sim 0.2$, explaining the difference in comparison with the isochrones. This is the first study that infers the GC metallicity by using only photometric data.  

\begin{figure*}
   \begin{center}   
   \includegraphics[scale=0.5]{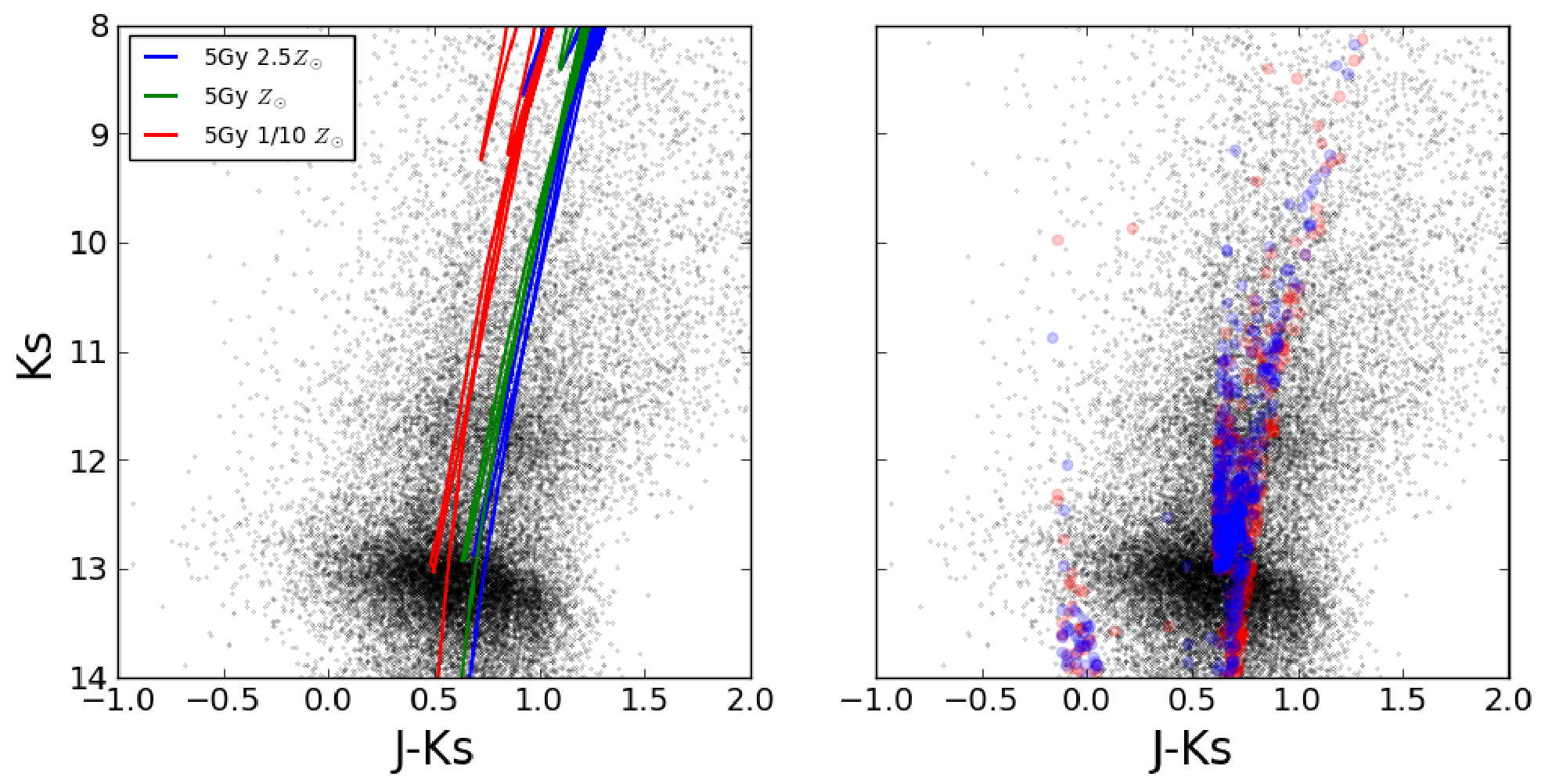}
   \caption{Left panel depicts 5 Gyr stellar isochrones with different metallicities overlaid on the CMD $K_s$ versus $J-K_s$. Right panel shows the comparison of the real data in the CMD $K_s$ versus $J-K_s$ with a synthetic stellar model produced using the code of \citet{Aparicio:2004fk}, a metallicity of 2.5 solar metallicity, and a constant stellar formation between 1 Gyr < t < 5 Gyr (blue circles)  and 5 Gyr < t < 12 Gyr (red diamonds).}
   
   \label{models}
    \end{center}   
    \end{figure*}

The right panel of Fig. \ref{models} shows a synthetic model population in comparison with the real data. The model assumes 2.5 solar metallicity and two different age intervals, from 1 Gyr < t < 5 Gyr and from 5 Gyr < t < 12 Gyr. In both cases the star formation rate is constant over the time. The synthetic population was created analogously as described in \citet{Pfuhl:2011uq}, using the code of \citet{Aparicio:2004fk} that used the stellar evolution library from \citet{Bertelli:1994aa} and the bolometric correction library from \citet{Lejeune:1997aa}. It can be seen that taking into account the uncertainty that produces the scatter of around 0.2 mag, we are able to reproduce several features observed in the data like the RC bump or the AGB branch. Therefore, within the limits of the current study, we suggest that a solar or super-solar metallicity explains well the observed stellar distribution.

\section{Discussion and conclusions}

We have presented the observations, data reduction, and the analysis for Field\,1 of our GALACTICNUCLEUS survey. Using
HAWK-I at the ESO VLT with short readout times, we applied the speckle
holography algorithm to create $JHK_{s}$ images with a PSF FWHM of
$0.2''$. This allows us to obtain accurate photometry in the crowded
field of the Galactic Centre. Our $5\,\sigma$ sensitivity limits are
approximately $J=22$, $H=21$, and $K_{s}=20$. The photometric
uncertainty is $<0.05$\,mag at $J\lesssim20$, $H\lesssim17$, and
$K_{s}\lesssim16$. The uncertainty of the zero points is $0.036$\,mag
in each band. The angular resolution and sensitivity of our data
allows us to reach well below the important tracer population of red
clump stars in all three bands, except in highly reddened areas in $J$.

We identify three stellar populations in the foreground, at low
extinction, with mean observed $H-K_{s}\approx0.2,0.5,0.8$, which we
interpret to belong to spiral arms along the line of sight. At the
distance of the GC, we identify a high-extinction population, with a
mean $H-K_{s}\approx2.0$, and a population at somewhat lower
extinction at $H-K_{s}\approx1.5$. Both populations appear clearly
distinct in an HST/WFC3 $F105W-F153M$ versus\ $F153M$ CMD and can be
separated relatively cleanly using a colour threshold of
$J-K_{s}=5.2$. Since the RC of the low-extinction population lies
along the reddening line of the more extinguished RC, we believe that
the low-extinction population lies approximately at the distance of the
GC. As a tentative explanation, we hypothesise that we are seeing stars
in front of the dust and molecular gas of the central molecular zone
(CMZ). This is in agreement with the findings of
\citet{Launhardt:2002nx} that about 80\% of the molecular gas in the
CMZ appears to be concentrated in a torus with radii between 120 to
220\,pc. The clumpiness of the ISM in the CMZ \citep[see
also][]{Launhardt:2002nx} will, of course, complicate this simple
picture.

Our deep and sensitive data allow us to study the NIR extinction curve
during the GC in unprecedented detail.  We found some evidence for a possible difference in the extinction index between $J-H$ and $H-K_s$. However, it is small and the extinction index can be assumed to be roughly constant within the limits of our study. Moreover, the value found here, $\alpha_{JHK_s}= 2.30 \pm 0.08$, is in excellent
agreement with the latest literature values on the NIR extinction
curve towards the GC. \citet{Nishiyama:2006tx} found
$\alpha_{JHK_s}=2.23\pm0.23$ \citep[value taken from Table\,5
in][]{Fritz:2011fk}, \citet{Schodel:2010fk}
 found $\alpha_{JHK_s}=2.21\pm0.24$, and \citet{Fritz:2011fk}
found $\alpha_{JHK_s}=2.11\pm0.06$.  The study carried out by \citet{Stead:2009uq} found $\alpha_{JHK_s}=2.14\pm0.05$ not towards
the GC but towards eight regions of the Galaxy between l $\sim 27^\circ$ and $\sim 100^\circ$.
\citet{Zasowski:2009ys} also found a compatible
$\alpha_{JHK_s}=2.26\pm0.17$ \citep[value taken from Table\,5
in][]{Fritz:2011fk} towards different regions of the Milky Way (over nearly $150^\circ$ of contiguous Milky Way midplane longitude).  This
may indicate a universal value for $\alpha_{JHK_s}$ that is steeper
than assumed in older work
\citep[e.g.][]{Rieke:1985fq,Draine:1989eq,Cardelli:1989kx}. Besides, \citet{Stead:2009uq} also point out that great care must be
taken in the choice of the filter wavelength when using broadband
filters. This has been done in the current work by computing carefully the effective wavelengths accordingly, as shown in Appendix \ref{wav}.

We studied possible variations of $\alpha_{JHK_s}$ with position in the
field and with the absolute value of extinction, but
cannot find any evidence for any such dependence, as suggested, for example,
by \citet{Gosling:2009kl}, who found a broad range for $\alpha_{JHK_s}$
in the GC with a strong dependency on the line of sight. We believe
that the latter work suffered from crowding and low angular
resolution, which limited the authors to make assumptions on the
median properties of their observed stellar populations. They could
not select specifically RC stars and had to rely on median colours of
stars in their fields. Hence, we claim that the NIR extinction curve
toward the GC can be approximated by a power law with constant index
$\alpha_{JHK_s}$.

Such a well-behaved extinction curve can facilitate the next step of
our analysis: the attempt to identify hot, massive stars in the GC,
which can provide evidence for the star formation history in the past
 few 100 Myrs. On the other hand, we have created several extinction maps and have identified a first layer of low varying extinction and a second one closer to the GC where the extinction varies in arc-second scales. Therefore, it appears that a significant part of the extinction toward the GC is originated from a highly clumpy medium near the GC. The clumpiness is in agreement with previous studies of the ISM in the CMZ \citep[e.g.][]{Launhardt:2002nx}. This division enables us to improve the de-reddening process to obtain more accurate de-reddened CMDs. Using them, we have employed isochrones with different metallicities \citep{Marigo:2017aa} and a synthetic stellar model \citep{Aparicio:2004fk} to infer the metallicity of the stellar population in the studied region. We found that a solar to super-solar metallicity fits the data well. This study supposes the first time that metallicity has been guess using only photometry in the near infrared for the GC.

This is the first of a series of papers that will present and exploit
the GALACTICNUCLEUS survey. It is supported by an ESO Large Programme
and all images and source catalogues will be made publicly available.

\begin{acknowledgements}
      The research leading to these results has received funding from
      the European Research Council under the European Union's Seventh
      Framework Programme (FP7/2007-2013) / ERC grant agreement
      n$^{\circ}$ [614922]. This work is based on observations made with ESO
      Telescopes at the La Silla Paranal Observatory under programmes
      IDs 195.B-0283 and 091.B-0418. We thank the staff of
      ESO for their great efforts and helpfulness. F
N-L acknowledges financial support from a MECD pre-doctoral contract, code FPU14/01700. F N acknowledges Spanish grants FIS2012-39162-C06-01, ESP2013-47809-C3-1-R and ESP2015-65597-C4-1-R. This work has made use of the IAC-STAR Synthetic CMD computation code. IAC-STAR is supported and maintained by the computer division of the Instituto de Astrof\'isica de Canarias.
\end{acknowledgements}

\appendix

  \section{Extinction and extinction index analysis. Plots and fits.  \label{plots}}
  
  \subsection{Grid method. RC stars. Low extinction.}
  \label{plots_low_ext}
  
\begin{figure}
   \begin{center}   
   \includegraphics[scale=0.36]{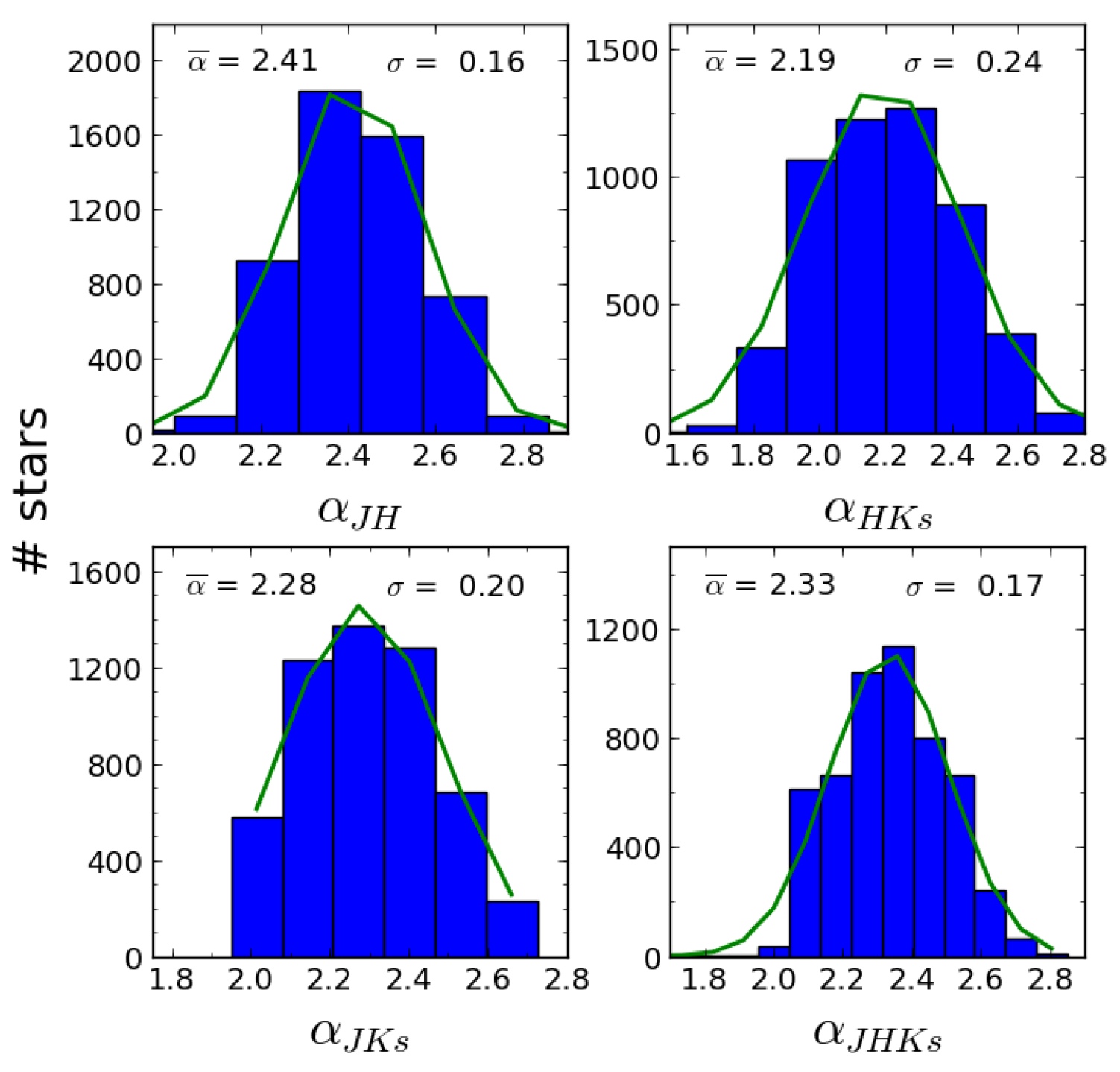}
   \caption{Histograms of $\alpha$ computed with the grid
     method for the low-extinction group of RC stars. Gaussian fits are overplotted as green lines, with the mean
     and standard deviations annotated in the plots.}
   
   \label{alpha_grid_1}
    \end{center}   
    \end{figure}

\begin{figure}
   \begin{center}   
   \includegraphics[scale=0.36]{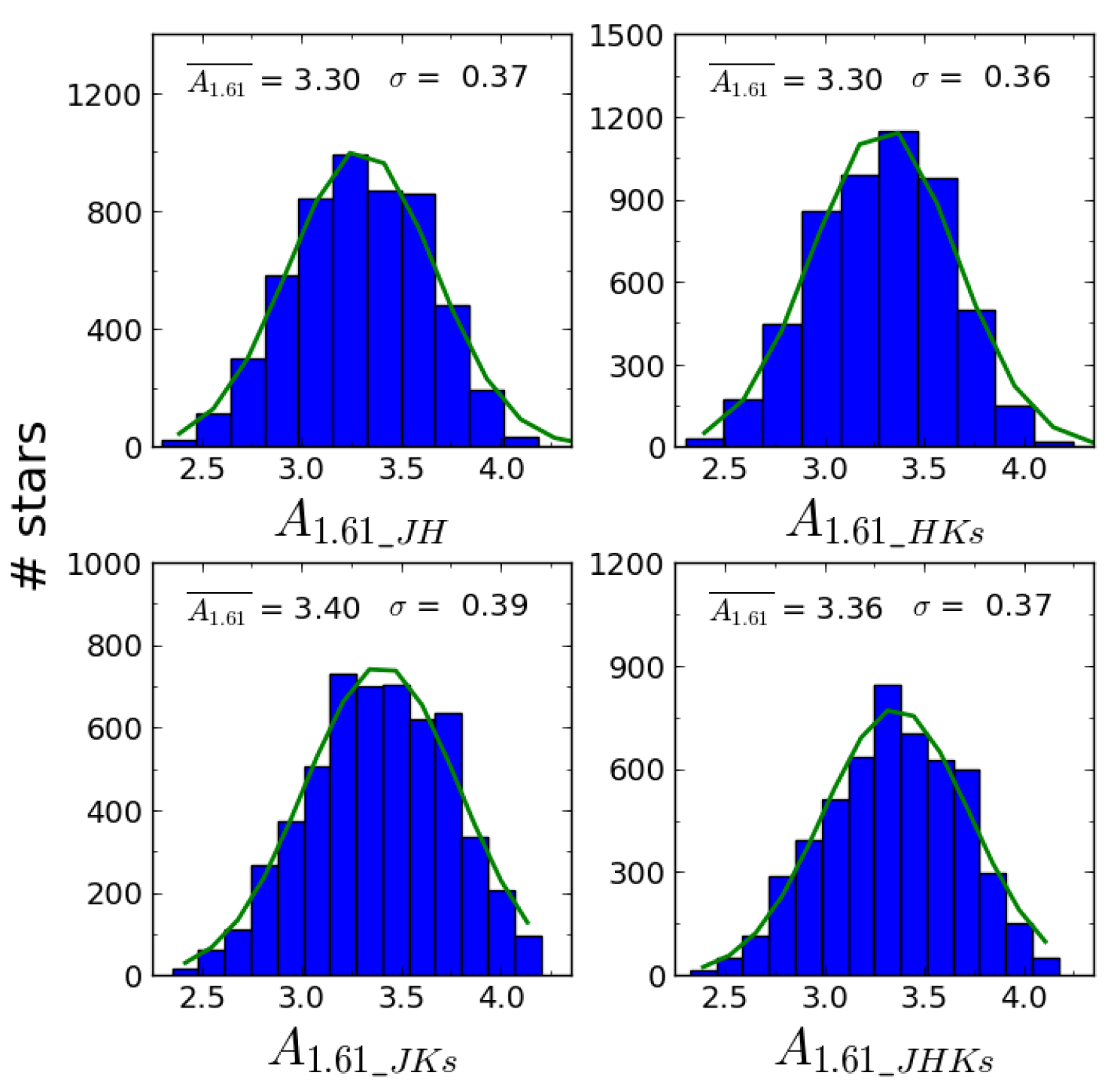}
   \caption{Histograms of $A_{1.61}$ computed with the grid
     method for the low-extinction group of RC stars. Gaussian fits are overplotted as green lines, with the mean and standard deviations annotated in the plots.}
   
   \label{A_H_grid_1}
    \end{center}   
    \end{figure}

Figures \ref{alpha_grid_1} and \ref{A_H_grid_1} show the results obtained applying the methodology described in Sect. \ref{group1} to the low extinguished RC stars.

    \subsection{Grid method. All RC stars.}
  \label{plots_all_ext}

   \begin{figure}
      \begin{center}
   \includegraphics[scale=0.36]{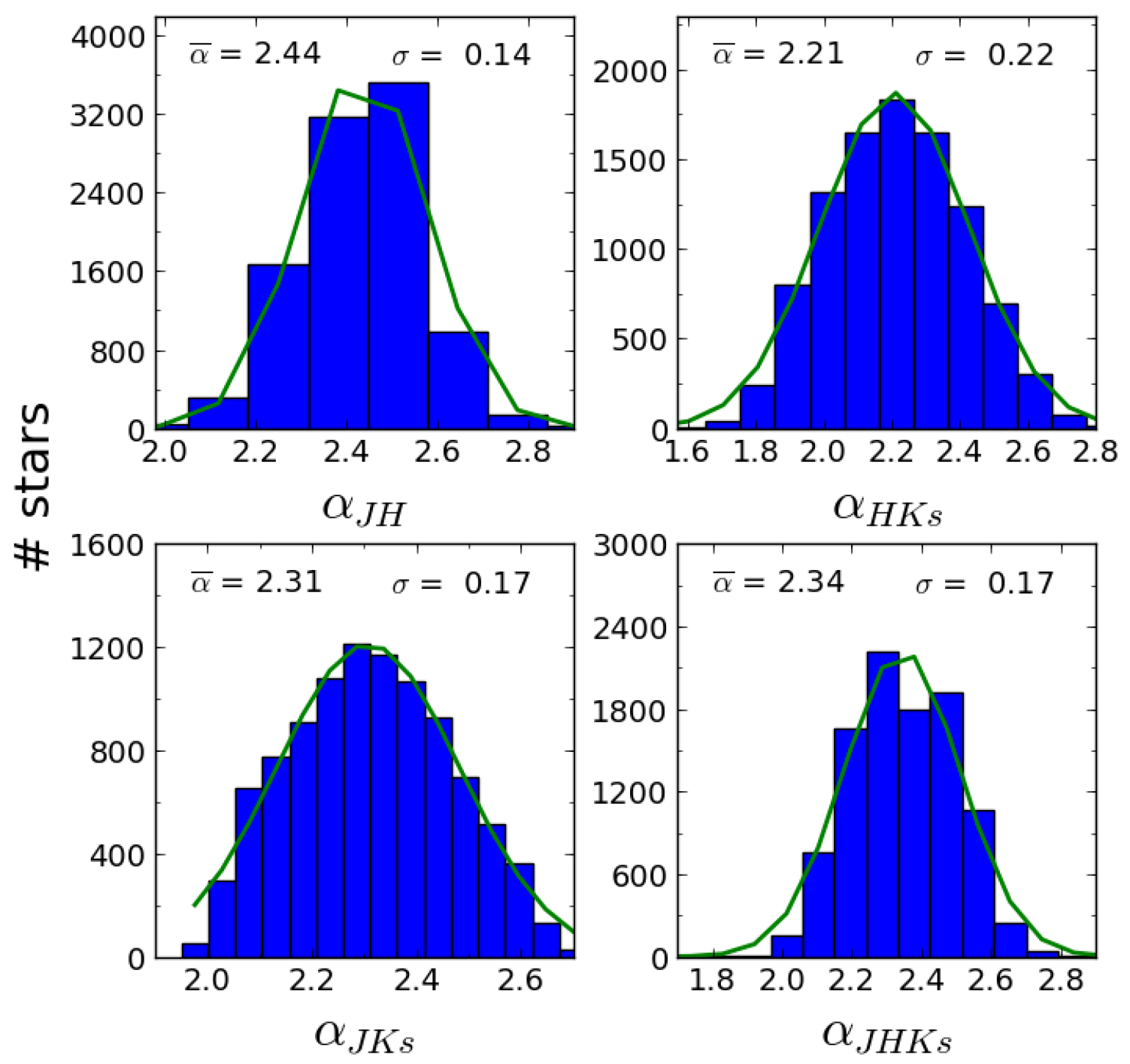}
   \caption{Histograms of $\alpha$ computed with the grid
     method for all the RC stars. Gaussian fits are overplotted as green lines, with the mean and standard deviations annotated in the plots.}
   
   \label{alpha_grid_gaussian1}
       \end{center}
    \end{figure}

   \begin{figure}
      \begin{center}
   \includegraphics[scale=0.35]{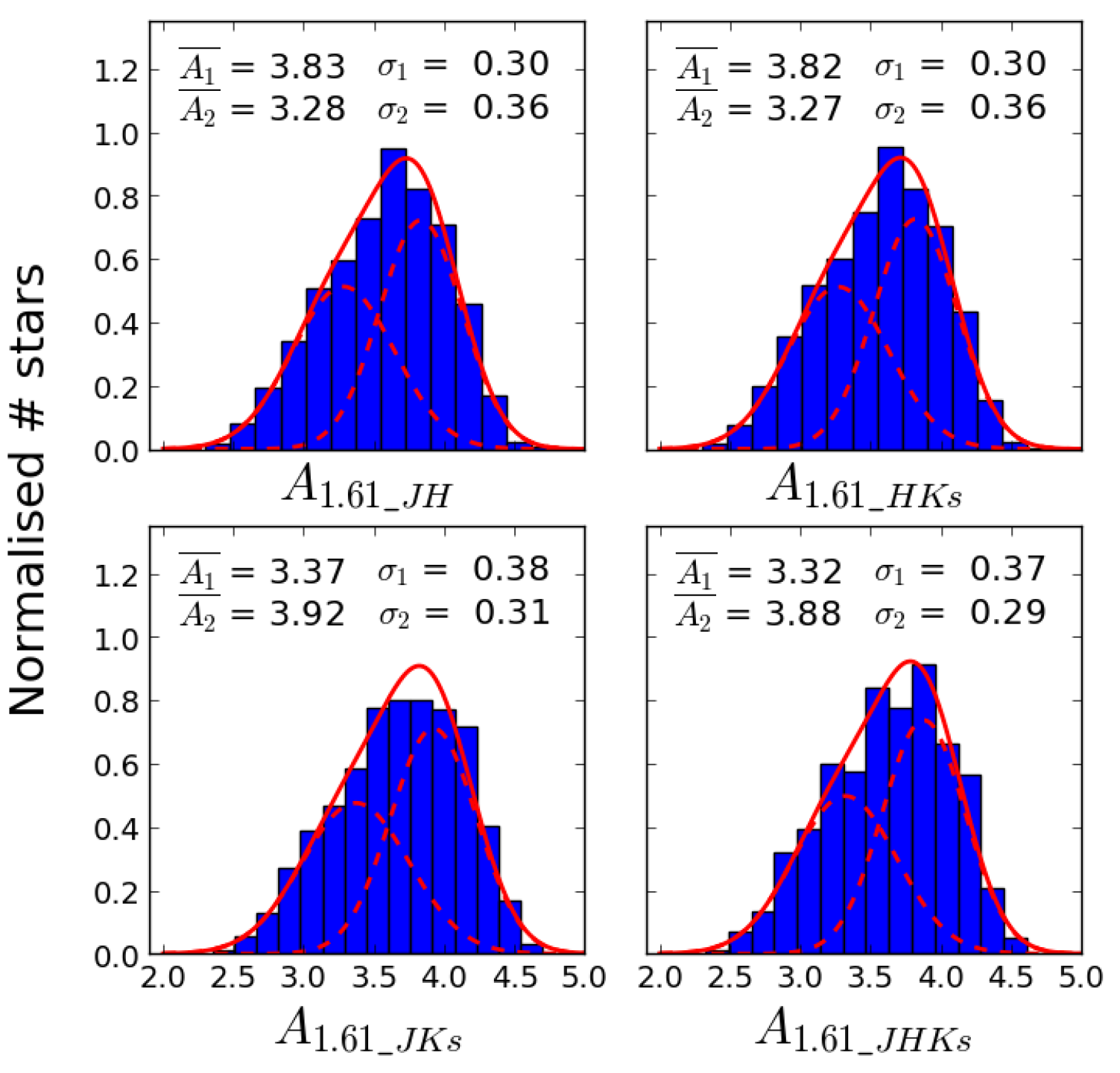}
   \caption{Normalised histograms of $A_{1.61}$ computed with the grid
     method for all the RC stars. The red continuous line shows two-Gaussian fits and the red dashed lines show the individual Gaussians of the fit. The mean
     and standard deviations of the fits are annotated in the plots.}
   
   \label{A_H_grid_gaussian1}
       \end{center}
    \end{figure}

\section{Effective wavelength}
\label{wav}
To compute the effective wavelength,  $\lambda_{\rm eff}$, we used Eq. (A3) of \citet{Tokunaga:2005jw}. We used the transmission curves for the HAWK-I $J$, $H,$ and $K_s$ filters from the instrument website and the atmospheric transmission from the Gemini telescope web site. Because the  $\lambda_{\rm eff}$ was computed for several stellar types (RC stars, late-type stars with known metallicity and temperature, and young stars), we used the more adequate Kurucz model for every case. We took the extinction index and $A_{1.61}$ obtained with the grid method described in Sect. \ref{grid},  being the value specified in every section accordingly.  

To calculate $\alpha$ we used an iterative approach in some of the methods. In those cases, we recomputed $\lambda_{\rm eff}$ using the $\alpha$ that was obtained in the first step and repeated the process updating the value until reaching convergence. The uncertainties for every $\lambda_{\rm eff}$ were estimated varying the parameters that affect them in their uncertainty ranges. In that way, we computed $\lambda_{\rm eff}$ varying the same parameter and keeping the rest constant. We considered:

\begin{itemize}

 \item The temperature of the model. For the RC stars, we used 4500, 4750, and 5000 K. For the  known young stars we used 20000, 30000, and 40000 K.

\item The metallicity of the model. We varied for all the cases from -1 dex to +1 dex in steps of 0.5.

\item We took three different values for the $\log g$ that we used in the Kurucz model: for RC, 2.0, 2.5 and 3.0, and for the young known stars 3.5, 4.0, and 4.5.

\item The amount of precipitable water vapour. We used 1.0, 1.6, and 3.0 mm.

\item The extinction index and $A_{1.61}$. We varied them according to the value and uncertainties obtained in Sect. \ref{grid}.

\end{itemize}

The final uncertainty was computed adding quadratically all the uncertainties. The final value was computed using the central parameters that were described for the uncertainties. Then,  \mbox{4750 K} and 30000 K, solar metallicity, and 1.6 mm of precipitable water vapour were used for the final values of the $\lambda_{\rm eff}$ for RC stars and young stars, respectively. Tables \ref{lambeffrc} and \ref{lambeffy} summarise the effective wavelength and the uncertainties for the high extinguished RC stars and young stars. The largest error came from $\alpha$ and $A_{1.61}$. In the case of late-type stars with known metallicity, we applied an individual model for each of them.

\begin{table}
\begin{center}
\caption{Effective wavelength and uncertainties for RC stars.}
\label{lambeffrc} 
\begin{tabular}{ccccc}
 &  &  &  & \tabularnewline
\hline 
\hline 
 &  & $J$ & $H$ & $K_s$\tabularnewline
 \hline 
$\lambda_{eff}$ &  & 1.2685 & 1.6506 & 2.1629\tabularnewline

\hline 
 & T & 0.0004 & 0.0003 & 0.0001\tabularnewline
 & met & 0.0011 & 0.0002 & 0.0003\tabularnewline
  & $\log g$ & 0.00007 & 0.00002 & 0.00003\tabularnewline
$\Delta\lambda_{eff}$ & hum & 0.0003 & 0.0001 & 0.0003\tabularnewline
 & $\alpha$ & 0.0007 & 0.0013 & 0.0002\tabularnewline
 & $A_{1.61}$ & 0.0013 & 0.0014 & 0.0008\tabularnewline
 &  Total & 0.0015 & 0.0020 & 0.0009\tabularnewline
\hline 
\end{tabular}
\end{center}
\vspace{0.35cm}
\footnotesize
\textbf{Notes.} Table shows the computed effective wavelength, the uncertainties associated to every parameter (from top to bottom: temperature of the model, metallicity, water vapour column density, extinction index, and extinction at 1.61 $\mu$m) and the final uncertainty for high extinguished RC stars.

 \end{table}

\begin{table}
\begin{center}
\caption{Effective wavelength and uncertainties for known young stars.}
\label{lambeffy} 
\begin{tabular}{ccccc}
 &  &  &  & \tabularnewline
\hline 
\hline 
 &  & $J$ & $H$ & $K_s$\tabularnewline
 \hline 
$\lambda_{eff}$ &  & 1.2687 & 1.6498 & 2.1637\tabularnewline

\hline 
 & T & 0.0012 & 0.0007 & 0.00014\tabularnewline
 & met & 0.00005 & 0.00005 & 0.00004\tabularnewline
  & $\log g$ & 0.00002 & 0.00003 & 0.000003\tabularnewline
$\Delta\lambda_{eff}$ & hum & 0.0003 & 0.00011 & 0.0003\tabularnewline
 & $\alpha$ & 0.0009 & 0.0014 & 0.0002\tabularnewline
 & $A_{1.61}$ & 0.0016 & 0.0016 & 0.0009\tabularnewline
 &  Total & 0.0022 & 0.0022 & 0.0009\tabularnewline
\hline 
\end{tabular}
\end{center}
\vspace{0.35cm}
\footnotesize
\textbf{Notes.} The uncertainties are shown independently depending on every parameter, from top to bottom: temperature of the model, metallicity, water vapour column density, extinction index, and extinction at 1.61 $\mu$m. 

 \end{table}

 \section{Intrinsic colour calculation \label{intrinsic}}

The intrinsic colours $(J-H)_0$ and $(H-K_s)_0$ were computed using the appropriate Kurucz models for each type of star. We took into account the transmission curves for the HAWK-I $J$, $H,$ and $K_s$ filters from the instrument web site and the atmospheric transmission from the Gemini telescope web site. We computed the total magnitude for every band and then normalised it using a Vega reference model (Kurucz model).  As in the calculation of $\lambda_{\rm eff}$, and taking the same values, we estimate the uncertainty varying the parameters that affect the intrinsic colour calculation (temperature of the model, metallicity, $\log g,$ and precipitable water vapour). The values for RC stars and young known stars are presented in tables \ref{irc} and \ref{iy}. For RC stars we used a model with solar metallicity, 4750 K, and 1.6 mm of precipitable water vapour. For the young known stars, we only changed the temperature of the model to 30000 K. As in Appendix \ref{wav}, the final uncertainty was computed adding quadratically all the individual uncertainties. In this way, we obtained an upper limit for the uncertainty. It can be seen that the most important factor is the temperature of the model. In the case of known late-type stars, we computed for each of them the corresponding value, depending on the metallicity and temperature.

 \begin{table}
 \begin{center}
\caption{Intrinsic colour for RC stars.}
\label{irc} 
 \begin{tabular}{cccc}
 &  &  & \tabularnewline
\hline 
\hline 
 &  & $(J-H)_0$ & $(H-K_s)_0$\tabularnewline
\hline 
Value &  & 0.495 & 0.089\tabularnewline
\hline 
 & T & 0.074 & 0.012\tabularnewline
 & met & 0.025 & 0.004\tabularnewline
Uncertainty & $\log g$ & 0.002 & 0.001\tabularnewline
 & hum & 0.0006 & 0.0003\tabularnewline
  & Total & 0.078 & 0.013\tabularnewline
\hline 
\end{tabular}

\vspace{0.35cm}
\end{center}
\footnotesize
\textbf{Notes.} The uncertainties are shown independently depending on every parameter. From top to bottom: temperature of the model, metallicity, and water vapour column density. 

 \end{table}

\begin{table}
\begin{center}
\caption{Intrinsic colour for known young stars.}
\label{iy} 
\begin{tabular}{cccc}
 &  &  & \tabularnewline
\hline 
\hline 
 &  & $(J-H)_0$ & $(H-K_s)_0$\tabularnewline
\hline 
Value &  & -0.090 & -0.117\tabularnewline
\hline 
 & T & 0.018 & 0.02\tabularnewline
 & met & 0.003 & 0.003\tabularnewline
Uncertainty & $\log g$ & 0.002 & 0.001\tabularnewline
 & hum & 0.0002 & 0.00003\tabularnewline
  & Total & 0.019 & 0.02\tabularnewline
\hline 
\end{tabular}

\vspace{0.35cm}
 \end{center}
 \footnotesize
\textbf{Notes.} The uncertainties are shown independently depending on every parameter. From top to bottom: temperature of the model, metallicity, and water vapour column density. 

 \end{table}

\bibliography{../../../BibGC.bib}

\begin{thebibliography}{53}
\expandafter\ifx\csname natexlab\endcsname\relax\def\natexlab#1{#1}\fi

\bibitem[{Akaike(1974)}]{Akaike:1974aa}
Akaike, H. 1974, Automatic Control, IEEE Transactions on, 19, 716

\bibitem[{{Aparicio} \& {Gallart}(2004)}]{Aparicio:2004fk}
{Aparicio}, A. \& {Gallart}, C. 2004, \aj, 128, 1465

\bibitem[{{Arsenault} {et~al.}(2014){Arsenault}, {Paufique}, {Kolb}, {Madec},
  {Kiekebusch}, {Argomedo}, {Jost}, {Tordo}, {Donaldson}, {Suarez},
  {Conzelmann}, {Kuntschner}, {Siebenmorgen}, {Kirchbauer}, {Rissmann}, \&
  {Schimpelsberger}}]{Arsenault:2014fv}
{Arsenault}, R., {Paufique}, J., {Kolb}, J., {et~al.} 2014, The Messenger, 156,
  2

\bibitem[{{Bertelli} {et~al.}(1994){Bertelli}, {Bressan}, {Chiosi}, {Fagotto},
  \& {Nasi}}]{Bertelli:1994aa}
{Bertelli}, G., {Bressan}, A., {Chiosi}, C., {Fagotto}, F., \& {Nasi}, E. 1994,
  \aaps, 106

\bibitem[{{Bovy} {et~al.}(2014){Bovy}, {Nidever}, {Rix}, {Girardi}, {Zasowski},
  {Chojnowski}, {Holtzman}, {Epstein}, {Frinchaboy}, {Hayden}, {Rodrigues},
  {Majewski}, {Johnson}, {Pinsonneault}, {Stello}, {Allende Prieto}, {Andrews},
  {Basu}, {Beers}, {Bizyaev}, {Burton}, {Chaplin}, {Cunha}, {Elsworth},
  {Garc{\'{\i}}a}, {Garc{\'{\i}}a-Her{\'n}andez}, {Garc{\'{\i}}a P{\'e}rez},
  {Hearty}, {Hekker}, {Kallinger}, {Kinemuchi}, {Koesterke},
  {M{\'e}sz{\'a}ros}, {Mosser}, {O'Connell}, {Oravetz}, {Pan}, {Robin},
  {Schiavon}, {Schneider}, {Schultheis}, {Serenelli}, {Shetrone}, {Silva
  Aguirre}, {Simmons}, {Skrutskie}, {Smith}, {Stassun}, {Weinberg}, {Wilson},
  \& {Zamora}}]{2014ApJ...790..127B}
{Bovy}, J., {Nidever}, D.~L., {Rix}, H.-W., {et~al.} 2014, \apj, 790, 127

\bibitem[{{Cardelli} {et~al.}(1989){Cardelli}, {Clayton}, \&
  {Mathis}}]{Cardelli:1989kx}
{Cardelli}, J.~A., {Clayton}, G.~C., \& {Mathis}, J.~S. 1989, \apj, 345, 245

\bibitem[{{Chaplin} \& {Miglio}(2013)}]{Chaplin:2013kx}
{Chaplin}, W.~J. \& {Miglio}, A. 2013, \araa, 51, 353

\bibitem[{{Diolaiti} {et~al.}(2000){Diolaiti}, {Bendinelli}, {Bonaccini},
  {Close}, {Currie}, \& {Parmeggiani}}]{Diolaiti:2000qo}
{Diolaiti}, E., {Bendinelli}, O., {Bonaccini}, D., {et~al.} 2000, \aaps, 147,
  335

\bibitem[{{Do} {et~al.}(2009){Do}, {Ghez}, {Morris}, {Lu}, {Matthews}, {Yelda},
  \& {Larkin}}]{Do:2009tg}
{Do}, T., {Ghez}, A.~M., {Morris}, M.~R., {et~al.} 2009, \apj, 703, 1323

\bibitem[{Doane(1976)}]{doi:10.1080/00031305.1976.10479172}
Doane, D.~P. 1976, The American Statistician, 30, 181

\bibitem[{{Dong} {et~al.}(2017{\natexlab{a}}){Dong}, {Lacy}, {Sch{\"o}del},
  {Nogueras-Lara}, {Gallego-Calvente}, {Mauerhan}, {Wang}, {Cotera}, \&
  {Gallego-Cano}}]{Dong:2017aa}
{Dong}, H., {Lacy}, J.~H., {Sch{\"o}del}, R., {et~al.} 2017{\natexlab{a}},
  \mnras, 470, 561

\bibitem[{{Dong} {et~al.}(2017{\natexlab{b}}){Dong}, {Sch{\"o}del}, {Williams},
  {Nogueras-Lara}, {Gallego-Cano}, {Gallego-Calvente}, {Wang}, {Rich},
  {Morris}, \& {Do}}]{Dong:2017ab}
{Dong}, H., {Sch{\"o}del}, R., {Williams}, B.~F., {et~al.} 2017{\natexlab{b}},
  \mnras, 471, 3617

\bibitem[{{Dong} {et~al.}(2011){Dong}, {Wang}, {Cotera}, {Stolovy}, {Morris},
  {Mauerhan}, {Mills}, {Schneider}, {Calzetti}, \& {Lang}}]{Dong:2011ff}
{Dong}, H., {Wang}, Q.~D., {Cotera}, A., {et~al.} 2011, \mnras, 417, 114

\bibitem[{{Draine}(1989)}]{Draine:1989eq}
{Draine}, B.~T. 1989, in ESA Special Publication, Vol. 290, Infrared
  Spectroscopy in Astronomy, ed. E.~{B{\"o}hm-Vitense}, 93--98

\bibitem[{{Feldmeier-Krause} {et~al.}(2017){Feldmeier-Krause}, {Kerzendorf},
  {Neumayer}, {Sch{\"o}del}, {Nogueras-Lara}, {Do}, {de Zeeuw}, \&
  {Kuntschner}}]{Feldmeier-Krause:2017kq}
{Feldmeier-Krause}, A., {Kerzendorf}, W., {Neumayer}, N., {et~al.} 2017,
  \mnras, 464, 194

\bibitem[{{Feldmeier-Krause} {et~al.}(2015){Feldmeier-Krause}, {Neumayer},
  {Sch{\"o}del}, {Seth}, {Hilker}, {de Zeeuw}, {Kuntschner}, {Walcher},
  {L{\"u}tzgendorf}, \& {Kissler-Patig}}]{Feldmeier-Krause:2015}
{Feldmeier-Krause}, A., {Neumayer}, N., {Sch{\"o}del}, R., {et~al.} 2015, \aap,
  584, A2

\bibitem[{Freedman \& Diaconis(1981)}]{Freedman1981}
Freedman, D. \& Diaconis, P. 1981, Probability Theory and Related Fields, 57,
  453

\bibitem[{{Fritz} {et~al.}(2011){Fritz}, {Gillessen}, {Dodds-Eden}, {Lutz},
  {Genzel}, {Raab}, {Ott}, {Pfuhl}, {Eisenhauer}, \&
  {Yusef-Zadeh}}]{Fritz:2011fk}
{Fritz}, T.~K., {Gillessen}, S., {Dodds-Eden}, K., {et~al.} 2011, \apj, 737, 73

\bibitem[{{Genzel} {et~al.}(2010){Genzel}, {Eisenhauer}, \&
  {Gillessen}}]{Genzel:2010fk}
{Genzel}, R., {Eisenhauer}, F., \& {Gillessen}, S. 2010, Reviews of Modern
  Physics, 82, 3121

\bibitem[{{Girardi}(2016)}]{Girardi:2016fk}
{Girardi}, L. 2016, \araa, 54, 95

\bibitem[{{Gosling} {et~al.}(2009){Gosling}, {Bandyopadhyay}, \&
  {Blundell}}]{Gosling:2009kl}
{Gosling}, A.~J., {Bandyopadhyay}, R.~M., \& {Blundell}, K.~M. 2009, \mnras,
  394, 2247

\bibitem[{{Kissler-Patig} {et~al.}(2008){Kissler-Patig}, {Pirard}, {Casali},
  {Moorwood}, {Ageorges}, {Alves de Oliveira}, {Baksai}, {Bedin}, {Bendek},
  {Biereichel}, {Delabre}, {Dorn}, {Esteves}, {Finger}, {Gojak}, {Huster},
  {Jung}, {Kiekebush}, {Klein}, {Koch}, {Lizon}, {Mehrgan}, {Petr-Gotzens},
  {Pritchard}, {Selman}, \& {Stegmeier}}]{Kissler-Patig:2008fr}
{Kissler-Patig}, M., {Pirard}, J.-F., {Casali}, M., {et~al.} 2008, \aap, 491,
  941

\bibitem[{{Kurucz}(1993)}]{Kurucz:1993fk}
{Kurucz}, R.~L. 1993, VizieR Online Data Catalog, 6039, 0

\bibitem[{{Landsman}(1993)}]{Landsman:1993aa}
{Landsman}, W.~B. 1993, in Astronomical Society of the Pacific Conference
  Series, Vol.~52, Astronomical Data Analysis Software and Systems II, ed.
  R.~J. {Hanisch}, R.~J.~V. {Brissenden}, \& J.~{Barnes}, 246

\bibitem[{{Launhardt} {et~al.}(2002){Launhardt}, {Zylka}, \&
  {Mezger}}]{Launhardt:2002nx}
{Launhardt}, R., {Zylka}, R., \& {Mezger}, P.~G. 2002, \aap, 384, 112

\bibitem[{{Lejeune} {et~al.}(1997){Lejeune}, {Cuisinier}, \&
  {Buser}}]{Lejeune:1997aa}
{Lejeune}, T., {Cuisinier}, F., \& {Buser}, R. 1997, \aaps, 125
  [\eprint{astro-ph/9701019}]

\bibitem[{{Malkin}(2013)}]{Malkin:2013fk}
{Malkin}, Z. 2013, in IAU Symposium, Vol. 289, IAU Symposium, ed. R.~{de
  Grijs}, 406--409

\bibitem[{{Marigo} {et~al.}(2017){Marigo}, {Girardi}, {Bressan}, {Rosenfield},
  {Aringer}, {Chen}, {Dussin}, {Nanni}, {Pastorelli}, {Rodrigues}, {Trabucchi},
  {Bladh}, {Dalcanton}, {Groenewegen}, {Montalb{\'a}n}, \&
  {Wood}}]{Marigo:2017aa}
{Marigo}, P., {Girardi}, L., {Bressan}, A., {et~al.} 2017, \apj, 835, 77

\bibitem[{{Massari} {et~al.}(2016){Massari}, {Fiorentino}, {McConnachie},
  {Bellini}, {Tolstoy}, {Turri}, {Andersen}, {Bono}, {Stetson}, \&
  {Veran}}]{Massari:2016uq}
{Massari}, D., {Fiorentino}, G., {McConnachie}, A., {et~al.} 2016, \aap, 595,
  L2

\bibitem[{{Minniti} {et~al.}(2010){Minniti}, {Lucas}, {Emerson}, {Saito},
  {Hempel}, {Pietrukowicz}, {Ahumada}, {Alonso}, {Alonso-Garcia}, {Arias},
  {Bandyopadhyay}, {Barb{\'a}}, {Barbuy}, {Bedin}, {Bica}, {Borissova},
  {Bronfman}, {Carraro}, {Catelan}, {Clari{\'a}}, {Cross}, {de Grijs},
  {D{\'e}k{\'a}ny}, {Drew}, {Fari{\~n}a}, {Feinstein}, {Fern{\'a}ndez
  Laj{\'u}s}, {Gamen}, {Geisler}, {Gieren}, {Goldman}, {Gonzalez}, {Gunthardt},
  {Gurovich}, {Hambly}, {Irwin}, {Ivanov}, {Jord{\'a}n}, {Kerins}, {Kinemuchi},
  {Kurtev}, {L{\'o}pez-Corredoira}, {Maccarone}, {Masetti}, {Merlo},
  {Messineo}, {Mirabel}, {Monaco}, {Morelli}, {Padilla}, {Palma}, {Parisi},
  {Pignata}, {Rejkuba}, {Roman-Lopes}, {Sale}, {Schreiber}, {Schr{\"o}der},
  {Smith}, {}, {Soto}, {Tamura}, {Tappert}, {Thompson}, {Toledo}, {Zoccali}, \&
  {Pietrzynski}}]{Minniti:2010fk}
{Minniti}, D., {Lucas}, P.~W., {Emerson}, J.~P., {et~al.} 2010, \na, 15, 433

\bibitem[{{Nagayama} {et~al.}(2003){Nagayama}, {Nagashima}, {Nakajima},
  {Nagata}, {Sato}, {Nakaya}, {Yamamuro}, {Sugitani}, \&
  {Tamura}}]{Nagayama:2003fk}
{Nagayama}, T., {Nagashima}, C., {Nakajima}, Y., {et~al.} 2003, in \procspie,
  Vol. 4841, Instrument Design and Performance for Optical/Infrared
  Ground-based Telescopes, ed. M.~{Iye} \& A.~F.~M. {Moorwood}, 459--464

\bibitem[{{Nishiyama} {et~al.}(2006{\natexlab{a}}){Nishiyama}, {Nagata},
  {Kusakabe}, {Matsunaga}, {Naoi}, {Kato}, {Nagashima}, {Sugitani}, {Tamura},
  {Tanab{\'e}}, \& {Sato}}]{Nishiyama:2006tx}
{Nishiyama}, S., {Nagata}, T., {Kusakabe}, N., {et~al.} 2006{\natexlab{a}},
  \apj, 638, 839

\bibitem[{{Nishiyama} {et~al.}(2006{\natexlab{b}}){Nishiyama}, {Nagata},
  {Sato}, {Kato}, {Nagayama}, {Kusakabe}, {Matsunaga}, {Naoi}, {Sugitani}, \&
  {Tamura}}]{Nishiyama:2006ai}
{Nishiyama}, S., {Nagata}, T., {Sato}, S., {et~al.} 2006{\natexlab{b}}, \apj,
  647, 1093

\bibitem[{{Nishiyama} {et~al.}(2008){Nishiyama}, {Nagata}, {Tamura}, {Kandori},
  {Hatano}, {Sato}, \& {Sugitani}}]{Nishiyama:2008qa}
{Nishiyama}, S., {Nagata}, T., {Tamura}, M., {et~al.} 2008, \apj, 680, 1174

\bibitem[{{Paufique} {et~al.}(2010){Paufique}, {Bruton}, {Glindemann}, {Jost},
  {Kolb}, {Jochum}, {Le Louarn}, {Kiekebusch}, {Hubin}, {Madec}, {Conzelmann},
  {Siebenmorgen}, {Donaldson}, {Arsenault}, \& {Tordo}}]{Paufique:2010zl}
{Paufique}, J., {Bruton}, A., {Glindemann}, A., {et~al.} 2010, in \procspie,
  Vol. 7736, Adaptive Optics Systems II, 77361P

\bibitem[{{Paumard} {et~al.}(2006){Paumard}, {Genzel}, {Martins}, {Nayakshin},
  {Beloborodov}, {Levin}, {Trippe}, {Eisenhauer}, {Ott}, {Gillessen}, {Abuter},
  {Cuadra}, {Alexander}, \& {Sternberg}}]{Paumard:2006xd}
{Paumard}, T., {Genzel}, R., {Martins}, F., {et~al.} 2006, \apj, 643, 1011

\bibitem[{Pedregosa {et~al.}(2011)Pedregosa, Varoquaux, Gramfort, Michel,
  Thirion, Grisel, Blondel, Prettenhofer, Weiss, Dubourg, Vanderplas, Passos,
  Cournapeau, Brucher, Perrot, \& Duchesnay}]{Pedregosa:2011aa}
Pedregosa, F., Varoquaux, G., Gramfort, A., {et~al.} 2011, Journal of Machine
  Learning Research, 12, 2825

\bibitem[{{Petr} {et~al.}(1998){Petr}, {Coude Du Foresto}, {Beckwith},
  {Richichi}, \& {McCaughrean}}]{Petr:1998vn}
{Petr}, M.~G., {Coude Du Foresto}, V., {Beckwith}, S.~V.~W., {Richichi}, A., \&
  {McCaughrean}, M.~J. 1998, \apj, 500, 825

\bibitem[{{Pfuhl} {et~al.}(2011){Pfuhl}, {Fritz}, {Zilka}, {Maness},
  {Eisenhauer}, {Genzel}, {Gillessen}, {Ott}, {Dodds-Eden}, \&
  {Sternberg}}]{Pfuhl:2011uq}
{Pfuhl}, O., {Fritz}, T.~K., {Zilka}, M., {et~al.} 2011, \apj, 741, 108

\bibitem[{{Primot} {et~al.}(1990){Primot}, {Rousset}, \&
  {Fontanella}}]{Primot:1990fk}
{Primot}, J., {Rousset}, G., \& {Fontanella}, J.~C. 1990, Journal of the
  Optical Society of America A, 7, 1598

\bibitem[{{Reid} \& {Brunthaler}(2004)}]{Reid:2004ph}
{Reid}, M.~J. \& {Brunthaler}, A. 2004, \apj, 616, 872

\bibitem[{{Rieke} \& {Lebofsky}(1985)}]{Rieke:1985fq}
{Rieke}, G.~H. \& {Lebofsky}, M.~J. 1985, \apj, 288, 618

\bibitem[{{Saito} {et~al.}(2012){Saito}, {Minniti}, {Dias}, {Hempel},
  {Rejkuba}, {Alonso-Garc{\'{\i}}a}, {Barbuy}, {Catelan}, {Emerson},
  {Gonzalez}, {Lucas}, \& {Zoccali}}]{Saito:2012fk}
{Saito}, R.~K., {Minniti}, D., {Dias}, B., {et~al.} 2012, \aap, 544, A147

\bibitem[{{Sch{\"o}del} {et~al.}(2014){Sch{\"o}del}, {Feldmeier}, {Neumayer},
  {Meyer}, \& {Yelda}}]{Schodel:2014bn}
{Sch{\"o}del}, R., {Feldmeier}, A., {Neumayer}, N., {Meyer}, L., \& {Yelda}, S.
  2014, Classical and Quantum Gravity, 31, 244007

\bibitem[{{Sch{\"o}del} {et~al.}(2010){Sch{\"o}del}, {Najarro}, {Muzic}, \&
  {Eckart}}]{Schodel:2010fk}
{Sch{\"o}del}, R., {Najarro}, F., {Muzic}, K., \& {Eckart}, A. 2010, \aap, 511,
  A18+

\bibitem[{{Sch{\"o}del} {et~al.}(2013){Sch{\"o}del}, {Yelda}, {Ghez}, {Girard},
  {Labadie}, {Rebolo}, {P{\'e}rez-Garrido}, \& {Morris}}]{Schodel:2013fk}
{Sch{\"o}del}, R., {Yelda}, S., {Ghez}, A., {et~al.} 2013, \mnras, 429, 1367

\bibitem[{Schwarz(1978)}]{Schwarz:1978aa}
Schwarz, G. 1978, The Annals of Statistics, 6, 461

\bibitem[{Scott(1979)}]{SCOTT:1979aa}
Scott, D.~W. 1979, Biometrika, 66, 605

\bibitem[{{Scoville} {et~al.}(2003){Scoville}, {Stolovy}, {Rieke},
  {Christopher}, \& {Yusef-Zadeh}}]{Scoville:2003la}
{Scoville}, N.~Z., {Stolovy}, S.~R., {Rieke}, M., {Christopher}, M., \&
  {Yusef-Zadeh}, F. 2003, \apj, 594, 294

\bibitem[{{Stead} \& {Hoare}(2009)}]{Stead:2009uq}
{Stead}, J.~J. \& {Hoare}, M.~G. 2009, \mnras, 400, 731

\bibitem[{Sturges(1926)}]{doi:10.1080/01621459.1926.10502161}
Sturges, H.~A. 1926, Journal of the American Statistical Association, 21, 65

\bibitem[{{Tokunaga} \& {Vacca}(2005)}]{Tokunaga:2005jw}
{Tokunaga}, A.~T. \& {Vacca}, W.~D. 2005, \pasp, 117, 421

\bibitem[{{Zasowski} {et~al.}(2009){Zasowski}, {Majewski}, {Indebetouw},
  {Meade}, {Nidever}, {Patterson}, {Babler}, {Skrutskie}, {Watson}, {Whitney},
  \& {Churchwell}}]{Zasowski:2009ys}
{Zasowski}, G., {Majewski}, S.~R., {Indebetouw}, R., {et~al.} 2009, \apj, 707,
  510

\end{thebibliography}

\end{document}